\documentclass[preprint,pra,aps,prfluids,showkeys,superscriptaddress,longbibliography]{revtex4-1}

\usepackage{amsmath,amssymb,amsfonts} %
\usepackage{graphicx}

\usepackage[caption=false]{subfig}% Support for small, `sub' figures and tables
\begin{document}

%\articletype{ARTICLE TEMPLATE}% Specify the article type or omit as appropriate

\title{Structure and fluctuations of the premelted liquid film of ice at the
triple point.}

\author{Jorge Benet, Pablo Llombart, Eduardo Sanz, Luis G.
MacDowell$^\dag$}
\affiliation{$^\dag$Departamento de Qu\'{\i}mica-F\'{\i}sica, Facultad de
Ciencias Qu\'{\i}micas, Universidad Complutense de Madrid, 28040 Madrid, Spain}

\email[]{lgmac@quim.ucm.es}

\begin{abstract}
In this paper we study the structure of the ice/vapor interface in the
neighborhood of the triple point for the TIP4P/2005 model. We probe the
fluctuations of the ice/film and film/vapor surfaces that separate the liquid
film from the coexisting bulk phases at basal, primary prismatic and secondary
prismatic planes. The results are interpreted using a coupled sine Gordon plus
Interface Hamiltonian model.   At large length-scales, the two bounding surfaces
are correlated and behave as a single complex ice/vapor interface. For  small
length, on the contrary, the ice/film and film/vapor surfaces behave very much
like independent ice/water and water/vapor interfaces. The study suggests that
the basal facet of the TIP4P/2005 model is smooth, the prismatic facet is close
to a roughening transition, and the secondary prismatic facet is rough. For the
faceted basal face,  our fluctuation analysis allows us to estimate the step
free energy in good agreement with experiment. Our results allow for a
quantitative characterization of the extent to which the adsorbed quasi-liquid
layer behaves as water, and explains experimental observations which  reveal
similar activation energies for crystals grown in bulk vapor or bulk water.
\end{abstract}

The Version of Record of this article has been published and is available
is {\em Molelcular Physics}, 05 March 2019,
{http://www.tandfonline.com/10.1080/00268976.2019.1583388}

\keywords{Roughening; Premelting, Layering; Surface Melting; Quasi-liquid layer;
Capillary Waves; Sine Gordon model.}

\maketitle

\section{Introduction}

Snowflakes offer the opportunity to observe the beautiful structure of ice
micro-crystals.\cite{libbrecht05} The striking symmetry  that is revealed,
is related to the stability of well defined crystal facets, which intersect at
the edges making well defined angles.\cite{pfalzgraf10} As temperature is increased, some crystal
facets altogether disappear, while the edges and sides gradually blur and eventually become
rounded.

Whereas ice crystals in snowflakes grow under kinetic
control,\cite{libbrecht05,ball16,demange17}  the
process described is an  illustration of a thermodynamic surface phase
transition.\cite{burton51,vanbeijeren77,abraham80,lipowsky82,fisher83,rommelse87,ball88,lipowsky90,evans92} These transitions, which are
best characterized for independent well defined facets at
equilibrium, determine overall the equilibrium crystal shape of a crystalline
solid.\cite{herring51,rottman84,chaikin95,kuroda82}

In surface physics there are two main types of phase transitions, roughening and
surface melting, that have been
characterized on the basis of well understood solvable
models.\cite{vanbeijeren77,abraham80,lipowsky82,fisher83,lipowsky90}

The roughening transition characterizes the thermal disorder that occurs on
facets of a two phase system well away from the triple point. This transition
separate smooth facets, with a low number density of defects, and
small finite perpendicular fluctuations, from rough surfaces, exhibiting a large
number of defects and diverging height fluctuations that do not differ at a
coarse scale from those found in fluid-fluid interfaces. It is believed that
when the correlation length of parallel height fluctuations becomes larger than
the size of the crystal, the facet disappears and becomes
round.\cite{jayaprakash83}

Surface melting  characterizes a different type of transition that involves
three, rather than two phases. It occurs as the
crystal is heated close to  the melting point. In such cases, it is often found that
the metastable liquid phase that is approached from below builds a small surface layer,
of finite
thickness.\cite{conde08,yang13,limmer14,kling18,pickering18,mohandesi18,qiu18} 
This process is known as premelting. Surface melting occurs
in those other instances where the thickness of the premelting layer diverges as
the  melting point is approached, very much as in a wetting
transition.\cite{schick90}  For the  the ice/vapor
interface, surface melting can only occur at the triple point, and it is defined
as a divergence of the premelting film as the triple point is approached along
the sublimation line.
Unfortunately, the relation of premelting with the corresponding equilibrium
crystal shape remains unclear, and it is a matter of concern whether the
premelting transition could round off the edges of a crystal shape as in
roughening.\cite{dosch95}

Whereas these prototypical transitions serve as a benchmark to assess surface
induced disorder in real systems, they may be well insufficient to interpret the
variety of complex surface phenomena that are found in real substances. In the
case of a roughening transition, for example, one sometimes finds roughening can
occur on surfaces which preserve the crystalline
structure (the vertical displacements remain congruent with the lattice spacing),
as in surfaces of nickel,\cite{cao90} while other reports refer to a
surface disordering transition with complete loss of the translational order on the
surface, as in gold.\cite{gibbs88} Already for such simple atomic crystals, it is
possible to find surfaces that neither premelt, nor roughen, that roughen
without premelting, or exhibit both roughening and premelting  before the triple
point is reached.\cite{mei07,yang13} 

A key issue that could be missing in the conventional picture is the
interplay of roughening and premelting with the related phenomenon
of layering.\cite{ball88,evans92} 
This is a sequence of layer by layer transitions that can occur
on top of a substrate and lead to the discontinuous growth of an
adsorbed film. At the mean field level, typically
surface melting is preceded by a large number of such transitions. 
However,  surface phase transitions related to short-range molecular 
correlations are fluctuation
dominated processes,\cite{jasnow84} and the mean filed picture is often
considerably transformed after renormalization of surface
capillary waves.\cite{chernov88,chernov89,henderson94}

Given this variety of surface phenomena, it is not unexpected to find how
difficult and controversial the characterization of equilibrium surface
properties is in such a common and important molecular crystal as
ice.\cite{dash06,elbaum91,elbaum91b,elbaum93,lied94,dosch95,bluhm02,sazaki12,
asakawa15,asakawa16,murata16,sanchez17,slater19}

There is currently ample experimental evidence indicating that the ice/vapor crystal surface exhibits
premelting at both the basal and prismatic planes close to the triple
point.\cite{elbaum93,lied94,dosch95,bluhm02,sadtchenko02,sanchez17,smit17} However, several
features of this transition layer are still a matter of debate. Firstly, the
temperature at which the premelting transition occurs, which varies from -50 C,
to a few Celsius below the triple point depending on the experimental
source.\cite{dash06} Secondly, there is also no consensus on the thickness of the
premelting layer, which varies between 10 to several hundred Angstroms close to
the triple point.\cite{dash06,michaelides17}  Thirdly, the nature of
the transition itself, which appear likely to be a continuous process as
reported recently.\cite{pickering18,qiu18} Finally, It had been
apparently unresolved for a long time whether
the premelting layer remains finite,\cite{elbaum91,elbaum91b,elbaum93}
or diverges (surface melting),\cite{lied94,dosch95} at the triple point.
However, recent advances in confocal microscopy have allowed the direct
visual inspection of ice surfaces which clearly confirm previous
hints of the appearance of water droplets on the ice surface very close
to the triple point.\cite{elbaum93,gonda99} Such observations seem to confirm that the ice
surface remains only partially wet at the triple point, whence, does not
exhibit surface melting.\cite{sazaki12,asakawa15,asakawa16,murata16}
Notice, however, that there is some evidence that contamination of  water 
or air largely increases the size of the premelted
film,\cite{elbaum93,wettlaufer99,bluhm02,mitsui19}, an indication that could reconcile to some extent
the conflicting results from different
laboratories.\cite{elbaum91,elbaum91b,elbaum93} Finally, very recent
experiments show indications of  layering  on the
basal ice surface,\cite{sanchez17,michaelides17} though the interpretation
of Frequency Sum Generation experiments is difficult and often consistent
with alternative explanations.\cite{smit17}
A layer-wise packing of adsorbed liquid onto a solid surface is certainly
not unexpected by liquid state theory,\cite{ball88,evans92} but as noticed 
earlier,
this is a fluctuation dominated phenomena and it is still to be determined
whether this structural feature is a proper thermodynamic surface phase
transition in this case.\cite{chernov88}

Further hints on the ice surface may be obtained from crystal growth
experiments.\cite{libbrecht13} Controlled growth of ice
crystals  at  about 10-20 ÂC below zero, usually produce prisms, with flat basal facets and
hexagonal shape, whether as grown from the vapor,\cite{elbaum91}
or the liquid phase.\cite{maruyama97} After the appearance of a thin premelting
film, the hexagonal shape of such crystallites is
observed to round as the triple point approaches, possibly indicating a roughening
transition of the prismatic facets .\cite{elbaum91,maruyama97,gonda99,asakawa15} In such studies,
the basal plane is found to remain smooth, indicating no roughening of the basal
orientation. Such observations are consistent with an ice surface exhibiting
premelting before roughening for the prismatic face, and no roughening at all
for the basal face.

On the other hand, there are claims that the equilibrium crystal shape of ice
should be completely rounded above -6 C,\cite{gonda78,colbeck83,dosch95} an
observation which would indicate fully roughened planes. In
fact, x-ray reflectivity experiments have reported observation of a fully rough
surface of the basal plane before the advent of premelting at about -13 C.\cite{lied94,dosch95}
These observations seem reasonable on theoretical grounds, since theoretical
estimates of the surface free energy,\cite{libbrecht13} as well as computer
simulations of different ice models, indicate a very small anisotropy of
the surface free energy,\cite{benet14c} whence, the expectation of a quasi-spherical equilibrium
crystal shape.

Accordingly, it would appear that not only the thickness of the premelting film is
unknown. Even the relative order of the premelting and roughening transitions, or
the occurrence of the latter altogether, are still a matter of debate.

Simulation studies could be a very useful tool for the study of the ice/vapor
interface, since direct observation at atomic scale is possible. Studies up to
date have confirmed the presence of a premelting layer, but it is difficult to
determine whether these models exhibit surface melting, for problems of limited
system sizes and inaccuracy in the location of the melting
point.\cite{furukawa97,conde08,bishop08,pereyra09,pan11}
Recently, Limmer and Chandler performed extensive simulations of the ice/vapor
interface, and observed a logarithmic divergence of the premelting layer
consistent with a surface melting transition on the
basal plane; as well
as a rough solid surface of diverging correlation length.\cite{limmer14} This is
a very careful study of surface melting, but it is arguable whether the results
may be extrapolated to describe a real substance such as water. Clearly, the
presence of the surface melting transition is an extremely subtle property
which is likely to require a very fine molecular model to describe reliably.
Indeed, it has been suggested that the absence of surface melting in water is
the result of complicated many-body interactions which require to take into
account the time dependent dielectric response of solid and liquid phases, as
well as  retardation effects.\cite{elbaum91b} These are fine features that are well beyond the coarse scale of simple non-polarizable point
charge models (let alone models that altogether ignore dispersion forces and the electric interactions
of water). 

In this work we extend our study of the ice/vapor
interface,\cite{benet15b,benet16} with ice described using the  
TIP4P/2005 model of water.\cite{abascal05} This force field is
apparently very close to the best rigid-point charge model of
water,\cite{abascal11} and
is therefore a good starting point for the study of short range contributions to
 surface premelting.  By introducing a convenient order parameter, we are able
to resolve the liquid from the solid, and study the fluctuations of the
resulting solid/film and film/vapor surfaces. This offers us a unique opportunity
to study the interplay between premelting and roughening of
the solid/film surfaces. Our results show that both basal and prismatic faces
exhibit premelting, with hints of a roughening transition on the prismatic
facet.  Unfortunately, a conclusive statement is still not possible, because
some limitations of the model that are unimportant in the study of bulk
properties turn out to be major concerns whenever two or more phases are
involved, as is the case in our study. Firstly, for reasons of numerical
convenience, the dispersive interactions are cut-off at a finite distance.
Secondly, the non-polarizable model is known to exhibit a static dielectric
constant that is smaller for the solid than for the liquid phase, at odds with
real water.\cite{macdowell10} Finally,  because of the absence of
polarizability, no retardation effects are incorporated at all. Fortunately, it
is expected that the range where these effects are important is beyond the film
thickness observed in our simulations.\cite{elbaum91b}

\section{Theory}

\subsection{Roughening transition of an interface}

The most significant feature of a roughening transition is a divergence of the
parallel and perpendicular correlations of the interface. Whereas this is
essentially a general feature in all roughening processes, the detailed physics
of the considered interfaces may be quite different. Here we briefly review
how such transition comes about in two important cases, namely, and adsorbed
premelted film, and a solid surface.

\subsubsection{Complete roughening of an adsorbed liquid-vapor interface}

Consider a system with two bulk phases in coexistence (such as a solid and vapor
phases), and a third metastable phase (such as a liquid) that is adsorbed
between the solid and vapor.  The
premelted liquid film exhibits a quasi-liquid-vapor interface, which may be described
in terms of its local height above the solid phase, $h({\bf x})$, where ${\bf
x}$ is a point on the plane of the substrate. In the capillary wave
approximation, the free energy $H[h]$ of a given film profile is given as:
\begin{equation}\label{eq:Hlv}
  H_{l/v} = \int d{\bf x} \left ( g(h) +  \gamma_{lv} \sqrt{ 1 + (\nabla h)^2 } \right )
\end{equation}
The first term,
$g(h)$ is a binding potential describing the effective interaction of the
liquid-vapor interface with the underlying solid substrate. In principle, the
binding or interface potential for an adsorbed liquid on an inert substrate
 may be calculated from computer simulations.\cite{macdowell14,benet14b}
However, the extension to premelting in a single component system seems difficult.
Here, it suffices to assume that there exists a binding potential whose minimum
sets the equilibrium film thickness of the system.
The second term
is governed by the liquid-vapor surface tension, $\gamma_{lv}$, and penalizes
increments of the surface area.
Expanding the Hamiltonian to second order in the fluctuations of $h({\bf x})$
away from the equilibrium film height, one finds:
\begin{equation}
 \Delta H_{l/v} =  \frac{1}{2}  \int d{\bf x} \left ( g''  h^2 + \gamma_{lv}
(\nabla h)^2 \right )
\end{equation}
where the primes in $g''$ denote differentiation with respect to $h$.
The partition function for this Hamiltonian may be worked out by expanding
$h({\bf x})$  in Fourier modes. This yields  the following result for the spectrum
of film height fluctuations:
\begin{equation}
  \langle  h^2(q) \rangle = \frac{k_BT}{A(g''+\gamma_{lv} q^2 )}
\end{equation}
This spectrum is the signature of a film with finite roughness. For
wave-lengths  that are smaller than
a parallel correlation length $\xi=(\gamma_{lv}/g'')^{1/2}$, the
fluctuations correspond to a rough interface, with a spectrum  characterized 
by a
$q^{-2}$ power law divergence. As the wave-vector becomes small, however, $g''$ damps the
fluctuations, which become smooth for wavelengths larger than $\xi$. In the limit where
$g''\to0$, however, the correlation length becomes infinite, the power law follows down to zero wave-vectors,
and the  interface becomes rough on all length scales, indicating complete
roughness of the interface.

\subsubsection{Roughening transition of a solid's surface}

We now consider a roughening transition that does not correspond to the
unbinding of a fluid film from a solid substrate, but rather, to the unbinding of the
solid-liquid interface from its own underlying bulk solid substrate. Traditionally,
this process has been described using so called Solid on Solid models, which
describe the solid as made of prismatic columns, of discrete heights, $h_i$,
that are multiples of the inter-plane spacing, $b$.\cite{chaikin95,nelson04} At 0 K, a high symmetry
surface is completely smooth, such that  all columns are of equal height. Rising
a column by one lattice spacing creates a defect of energy $J$, which, however,
increases the surface entropy. The energy of a given realization of column
heights may be described qualitatively using the SOS Gaussian model:
\begin{equation}\label{eq:gsos}
 H_{s/l} = \frac{J}{b^2}  \sum_{i,j} ( h_i - h_j)^2
\end{equation}
where the sum runs over all neighboring lattices, and it is understood that the
column heights are multiples of the lattice spacing.

Whereas this model offers a
rather clear description of the roughening process, it is difficult to solve analytically. For this
reason, it is convenient to resort to a  somewhat more abstract continuum
model, known as the sine-Gordon model, which has the advantage of being solvable.
In this way, the energy is now written as:
\begin{equation}\label{eq:Hsl}
 H_{s/l} =  \int d{\bf x} \left ( \frac{1}{2}
\tilde\gamma_{sl} (\nabla h)^2 -u \cos(\frac{2\pi}{b} h) \right )
\end{equation}
Here, the discrete column heights are now transformed into a continuum surface
height, $h({\bf x})$, so that the squared differences of Eq \ref{eq:gsos} are
transformed into a squared gradient. Together with this contribution, a bulk
pining field is added in order to favor surface heights that are multiples of the lattice spacing.
The parameter $\tilde \gamma_{sl}$ is the interface stiffness,
\cite{fisher83,chaikin95,nelson04} which penalizes
deviations from the planar configuration, while $u$ is a bulk-surface coupling
parameter which dictates the strength of the bulk pining field.

This model exhibits a roughening transition, at a temperature
$T_R=2\tilde\gamma_{sl}/b^2$, where the
bulk-surface coupling constant effectively vanishes. Above this temperature, the
bulk pining field is absent,  and the Hamiltonian becomes exactly as the
capillary wave Hamiltonian of  fluid-fluid interfaces. For highly symmetric
crystals with low anisotropy, the stiffness does not show large differences among
different facets, and the roughening temperature is mainly governed by the
distance between equivalent planes, $b$. Accordingly, high symmetry faces, with
small inter-plane spacing, are usually those with highest roughening
temperature.

It is instructive to expand the sine-Gordon Hamiltonian to quadratic order in
the surface profile. Up to an additive irrelevant constant, we obtain:\cite{nelson04,safran94}
\begin{equation}
 \Delta H_{s/l} = \frac{1}{2}  \int d{\bf x} \left ( \frac{4\pi^2 u}{b^2}\, 
 h^2 + \tilde \gamma_{sl}\,
(\nabla h)^2 \right )
\end{equation}
Whence, to quadratic order in the surface height, the sine-Gordon Hamiltonian
for column fluctuations is essentially equal to the Capillary Wave Hamiltonian
for adsorbed films, with an effective pining strength $\upsilon=\frac{4\pi^2
u}{b^2}$ in place of $g''$. Accordingly, the spectrum of fluctuations is:
\begin{equation}
  \langle  h^2(q) \rangle = \frac{k_BT}{A(\upsilon + \tilde
\gamma_{sl}\, q^2 )}
\end{equation}
As long as the pining coefficient $u$ remains finite, the surface height
fluctuations are bound, and the surface is said to be smooth. If, however, $u\to
0$, the fluctuations diverge as $q\to 0$ and the surface becomes rough on all
length scales. According to the theory of equilibrium crystal shapes, crystal
facets of length smaller than the correlation length
$(\tilde\gamma_{sl}/\upsilon)^{1/2}$ disappear and become round.\cite{dash06}

\subsection{Model for coupled interface fluctuations}
\label{modelo}
We now attempt to provide a phenomenological description of interface
fluctuations of a premelted solid-vapor interface. Consider, to be specific, a
premelted water film (f), that is adsorbed on top of a bulk ice phase (i) separating
it from a bulk vapor phase (v). Overall, the ice/vapor interface may be
described in terms of two different dividing surfaces, one, separating solid ice
from the water film, ({\em if}), and other, separating the film from the vapor phase
({\em fv}). In our phenomenological model, we describe the fluctuations of the
ice/film  surface using the sine-Gordon Hamiltonian, and the film/vapor surface
using the Capillary Wave Hamiltonian, as follows:
\begin{equation}
\renewcommand{\arraystretch}{2.2}
%\begin{array}{l l}
 \Delta H_{s/f/v} =  \int d{\bf x} \left ( 
\frac{1}{2}\tilde\gamma_{iw} (\nabla h_{if})^2 -u \cos(\frac{2\pi}{b} h_{if}) 
 + g(h_{fv}-h_{if}) + \gamma_{wv} \sqrt{1+(\nabla h_{fv})^2}
\right )
%\end{array}
\label{eq:Hmodelo}
\end{equation}
where $h_{if}$ and $h_{fv}$ are the local positions of the {\em i/f} and {\em f/v}
surfaces, respectively; $\tilde\gamma_{iw}$ is the stiffness of the {\em i/w}
interface, $\gamma_{wv}$ is the surface tension of the {\em w/v} interface and
$g(x)$ is a local interface potential which binds the film of premelted ice to
the bulk ice phase.

This Hamiltonian may be simplified by expanding to quadratic order in $h_{if}$
and $h_{fv}$, as noted previously. This results in a total energy which is
essentially the sum of Eq \ref{eq:Hsl} and Eq \ref{eq:Hlv}, with $h_{if}$ and $h_{fv}$ coupled via
the interface potential. The Hamiltonian may be worked out as before, by
writing the film heights in Fourier modes, yielding:
\begin{equation}
\renewcommand{\arraystretch}{2.2}
%\begin{array}{l l}
 \Delta H_{s/f/v} = \frac{1}{2} \sum_{\bf q} \left \{
 [ \upsilon + g'' + \tilde\gamma_{iw} q^2 ] |h_{if}^2({\bf q})| +
 [  g'' + \gamma_{wv} q^2 ] |h_{fv}^2({\bf q})| -
 2 g''  |h_{if}({\bf q}) h_{fv}^{*}({\bf q})|
\right \} 
%\end{array}
\end{equation}
where we have introduced $\upsilon=4\pi^2 u/b^2$ as the effective bulk crystal field strength, for short.

The statistical weight of this Hamiltonian, $exp(-\Delta H/k_BT)$ yields
a Gaussian bivariate distribution for the surface modes analogous to that found for coupled fluid-fluid interfaces under gravity.\cite{li01,fukuto06,pershan12} This can be solved
immediately, providing the following result for the spectrum of fluctuations:
\begin{equation}\label{eq:h2surf}
\begin{array}{ccc}
 \langle |h_{if}^2({\bf q})| \rangle & = &
 \displaystyle{     \frac{k_BT}{A} \frac{g'' + \gamma_{wv} q^2}{[ \upsilon + g'' +
\tilde\gamma_{iw} q^2 ][  g'' + \gamma_{wv} q^2 ] - g''^2} } \\
 & & \\
 \langle |h_{fv}^2({\bf q})| \rangle & = &
  \displaystyle{      \frac{k_BT}{A} \frac{\upsilon + g'' + \tilde\gamma_{iw} q^2}{[ \upsilon + g'' +
\tilde\gamma_{iw} q^2 ][  g'' + \gamma_{wv} q^2 ] - g''^2}  } \\
& & \\
 \langle h_{if}({\bf q}) h_{fv}^*({\bf q}) \rangle & = &
   \displaystyle{      \frac{k_BT}{A} \frac{g''}{[ \upsilon + g'' +
\tilde\gamma_{iw} q^2 ][  g'' + \gamma_{wv} q^2 ] - g''^2}  }
\end{array}
\end{equation}
These set of equations for the compound ice/vapor fluctuating interface is
denoted henceforth as the sine Gordon plus Capillary Wave model (SG-CW).

In order to assess the significance of these equations, it is
convenient to introduce the effective surface stiffness, as:
\begin{equation}\label{eq:defg}
  \Gamma_{\alpha-\beta}({\bf q}) =
    \frac{k_BT}{A}\frac{1}{ \langle h_{\alpha}({\bf q}) h_{\beta}^*({\bf q})
\rangle q^2}
\end{equation}   
where the sub-indexes $\alpha$  and $\beta$ denote here  either the $if$ or 
$fv$ surfaces. Notice
$\Gamma_{\alpha-\beta}$ has dimensions of a surface free energy 
and corresponds exactly
to the surface tension for rough and isotropic interfaces
in the limit of vanishing wave-vector. In the
event that one or both of the bulk phases involved are anisotropic,
it corresponds rather to the surface stiffness coefficient, which
effectively diverges to infinity for a  smooth interface as the
wave-vector vanishes.

In order to analyze the rich and complex behavior afforded by the coupled
sine Gordon+capillary wave model, we introduce  parallel correlation lengths 
characteristic of isolated ice/water and water/vapor interfaces,
$\xi_{iw}^2=\tilde\gamma_{iw}/\upsilon$, and $\xi_{wv}^2=\gamma_{wv}/g''$,
respectively.

\begin{table}[h]
\centering
\footnotesize
\begin{tabular}{c c c c c}
\hline
\hline
  $\Gamma_{\alpha-\beta}({\bf q})$/q range      &    $(\xi_{iw}^2+\xi_{wv}^2) q^2 \ll 1$  &  $\xi_{wv}^2 q^2 \ll 1 \ll \xi_{iw}^2 q^2$ &  $\xi_{iw}^2 q^2 \ll 1 \ll \xi_{wv}^2 q^2$ & $(\xi_{iw}^2+\xi_{wv}^2) q^2 \gg 1$  \\
\hline
   $\Gamma_{if-if}({\bf q})$          &      $\upsilon q^{-2}$             &   $(\tilde\gamma_{iw}+\gamma_{wv})$        &   $ \upsilon q^{-2} $      & $ \tilde\gamma_{iw} $          \\
   $\Gamma_{fv-fv}({\bf q})$          &      $\frac{\upsilon g''}{\upsilon +  g''}q^{-2} $  &   $(\tilde\gamma_{iw}+\gamma_{wv})$ & $\gamma_{wv} $  & $ \gamma_{wv} $          \\
   $\Gamma_{if-fv}({\bf q})$          &      $\upsilon q^{-2}$ &   $(\tilde\gamma_{iw}+\gamma_{wv})$ &  $ \frac{\upsilon \gamma_{wv}}{g''} $    &
   $\frac{\tilde\gamma_{iw}\gamma_{wv}}{g''} q^2  $         \\
\hline
 Comment                                        &       Smooth and pinned         &      Rough and pinned                      &    Smooth and depinned                  & Rough and depinned   \\
  
\hline
\hline
\end{tabular}
\caption{
Summary of limiting behavior of the effective stiffness
$\Gamma_{\alpha-\beta}({\bf q})$ as a function of wave-vector. The different
ranges are given relative to the length-scales 
$\xi_{iw}^2=\tilde\gamma_{iw}/\upsilon$ and $\xi_{wv}^2 = \gamma_{wv}/g''$.
Roughness/smoothness describes the behavior of the ice/film surface; while
pinned/depinned refers to the behavior of the film/vapor surface relative to
that of the ice/film surface.
}
\label{limites}
\end{table}

Depending on the relative value of $q$ with respect to the
parallel correlation lengths, we can identify four different
regimes. In the very small wave-vector  regime, all three effective
stiffness coefficients $\Gamma_{if-if}$, $\Gamma_{fv-fv}$ and
$\Gamma_{if-fv}$ diverge. It corresponds to the case where
the {\em if} surface is smooth, and the {\em fv} surface is pinned
to the solid. On the contrary,  for length-scales that are
small compared to both $\xi_{iw}$ and $\xi_{wv}$, 
the {\em if} and {\em fv} surfaces become  
uncorrelated, while the effective stiffness become finite
and adopt the value corresponding to independent rough interfaces. 
For small wave-vectors, one of either two intermediate regimes can occur 
in the event that the
parallel correlation lengths of the ice/water or water/vapor
surfaces are very different. If $\xi_{iw}\gg\xi_{wv}$,   and
$\xi_{wv}^2 q^2 \ll 1 \ll \xi_{iw}^2 q^2$, then {\em if} and {\em fv}
surfaces become strongly correlated and behave as one single
rough interface with a stiffness coefficient that is the sum
of the independent stiffness coefficients, $\tilde\gamma_{iw}+\gamma_{wv}$.
If, on the other hand, $\xi_{wv}\gg\xi_{iw}$, but
$\xi_{iw}^2 q^2 \ll 1 \ll \xi_{wv}^2 q^2$, then
the {\em if} surface remains smooth, but the {\em fv} surface depins
and effectively becomes rough.

In practice, since we are actually interested in the behavior
of ice covered by a premelting film of finite thickness, $g''$ is
finite, and only the first two low wave-vector regimes are of
interest.  Most significantly, the model allows for
a clear distinction between two different possible
scenarios, namely, 
1) the case where the film/vapor unbinds (surface melting), 
before the roughening transition is reached.  
Then $\upsilon$ remains finite, and the low wave-vector 
stiffness coefficients become effectively infinite. This
is the smooth-and-pinned scenario.
2) the case where a
roughening transition occurs before the unbinding of
the {\em fv} surface from the {\em if} surface. 
In this case, $\upsilon = 0$, and  the effective stiffness of both the 
ice/fluid and fluid/vapor surfaces, as well as their 
coupling  attain a finite value equal
to the sum of $\tilde\gamma_{iw}$ and $\gamma_{wv}$. This
is the rough-and-pinned scenario. The Table \ref{limites} summarizes the 
limits of $\Gamma$ in each of the four possible situations described above.

\section{Methods}

We have used the TIP4P/2005 model of water. Our systems consist of an
ice slab placed in the middle of the $z$ direction of a rectangular simulation
box of sides $L_x$, $L_y$ and $L_z$ with the interface placed at the $x,y$
plane. Surrounding this slab there are water molecules if an ice/water system
is simulated, or vacuum for an ice/vapor system.

In order to deal with a reasonable number of molecules we set  $L_x>>L_y$. In
this way our systems present an elongated interfacial area which allow us
to study capillary waves propagating along the $x$ direction.
The values of the box sides and the number of molecules of
each system are shown in Table \ref{sizes-ice-water}.

In order to analyze our systems we need to find a discrete function $h(\bf{x})$
describing the interface. To manage this
we make use of the order parameter
$\bar{q}_6$ of Lechner and Dellago \cite{lechner08}, which allows us
to distinguish between solid and liquid molecules.

When analyzing ice/water systems we follow the same procedure as in Ref.
\cite{benet14}. In this procedure we get rid of fluid--like molecules
by making use of the order parameter and the biggest solid cluster is
found.

However, for the ice/vapor system the procedure is not always the same, as we
can find two different surfaces as mentioned in the introduction: the ice/film
surface and the film/vapor surface. For the ice/film surface the procedure
is exactly the same as for the ice/water surface, since we are interested only
in solid--like molecules. On the contrary, for the film/vapor surface we are
interested in the interface between a liquid--like film and a vapor. For this
reason we do not make use of any order parameter and we just look for the
biggest cluster, regardless it is solid or fluid. By doing this we get rid
of any vapor particle.

Finally, once we have isolated the molecules which concern us for each surface
we can define a discretised interface profile along the $x$ direction,
$h(\bf{x})$. Due to the anisotropy of the ice facets, the surface
fluctuations are characterized not only by the chosen facet, but also
by the direction of the $x$ axis along which the fluctuations are measured.
To distinguish different realizations of the fluctuations, we denote
the surface in round parenthesis, and the direction perpendicular
to the propagation of the
fluctuations in squared parenthesis, as described in detail
in Ref.\cite{benet14,benet15b,benet16}  For example, the primary prismatic
facet is denoted here as (pI). Propagation of waves along the direction
perpendicular to the basal or secondary prismatic planes are non-equivalent.
We specify this by indicating in square brackets the crystal direction
perpendicular to the direction of wave propagation. Whence,
(pI)[pII] corresponds to propagation of surface waves on the pI plane
that run along the Basal direction.

We prepare our systems by equilibrating an ice Ih configuration at T=248K and
1 bar, about 2~K below the triple point of the model \cite{abascal11,rozmanov11,conde17}. We then
re-scale the simulation box to the average value of $L_x$, $L_y$ and $L_z$ to
avoid any stress. The solid is then placed next to a liquid or vacuum,
for ice/water and ice/vapor interfaces respectively, and equilibrated in the
NVT ensemble until the energy of the systems remains stable.

Then we perform production runs of about $0.5\mu s$ in the NVT ensemble with
the time step for the Velocity-Verlet integrator fixed to $0.003 ps$.  The
cut-off distance for Lennard-Jones interactions was set at 0.9~nm and
standard Ewald summations were used.
Snapshots were saved every $75ps$, resulting in a total of about 6500
snapshots. The temperature if the system was fixed by using the
velocity-rescaling thermostat of Bussy, Donadio and Parrinello \cite{bussi07}.

\begin{table}[h]
\centering
\footnotesize
\begin{tabular}{c c c c c}
\hline
\hline
Orientation & ice/water interface  & Molecules &  ice/vapour interface  &    Molecules \\
            & $L_xxL_yxL_z (nm^3)$ &  &   $L_xxL_yxL_z (nm^3)$ &     \\
\hline
(Basal)[pII]& 18.7696x1.8039x9.3319 & 10112 & 18.7696x1.8039x9.3319 & 4632 \\
(pI)[Basal] & 18.0134x2.1991x8.0808 & 10240 & 18.0577x2.2045x9.5000 & 5520 \\
(pI)[pII]   & 17.6430x2.3491x7.8227 & 10368 & 17.6430x2.3491x9.0000 & 5760 \\
(pII)[Basal]& 17.9927x2.2047x8.3875 & 10670 & 18.0596x2.2063x9.0000 & 5760 \\
(pII)[pI]   & 18.3690x1.8037x8.3928 & 8896  & 18.3661x1.8035x9.0000 & 4800 \\
(pI)[Basal] &                       &       & 36.1163x2.2062x19.0000 & 23040 \\
(pI)[pII]   &                       &       & 36.7309x1.8034x19.0000 & 19200 \\
\hline
\hline
\end{tabular}
\caption{Box dimensions  of the different systems studied.}
\label{sizes-ice-water}
\end{table}

\section{Results}
 
Here we present results for the structure of the ice/water and ice/vapor
interface of basal, primary prismatic (pI) and secondary prismatic (pII)
planes. 
The
simulations of the ice/water interface are carried out at a temperature of
ca. T=248.5K,
while those of the ice/vapor interface are simulated at ca. T=248.7K. 

Unfortunately, the melting point is difficult to determine with state of
the art simulations to a precision better than $\pm 0.5$~K, and the literature
reports data scattered between about 249 and
253~K.\cite{abascal11,rozmanov11,conde17}
The melting point seems to depend not only on system size, as discussed
recently,\cite{conde17} but also on the Lennard-Jones cutoff-distance
(c.f. Ref.\cite{mastny07}).
Whereas we did not make a precise evaluation of the melting point here,
our results seem consistent with results reported for systems with a 
Lennard-Jones cutoff of $1$~nm, whence, $T_t=250.5$~K. To avoid the need
for a detailed knowledge of the model properties, we refer to our simulation
henceforth somewhat loosely as being 2~K away from the triple point.

\subsection{Density profiles}

\subsubsection{Ice/water interface}

\begin{figure}[h]
\centering
\subfloat[]{%
\resizebox*{5cm}{!}{\includegraphics{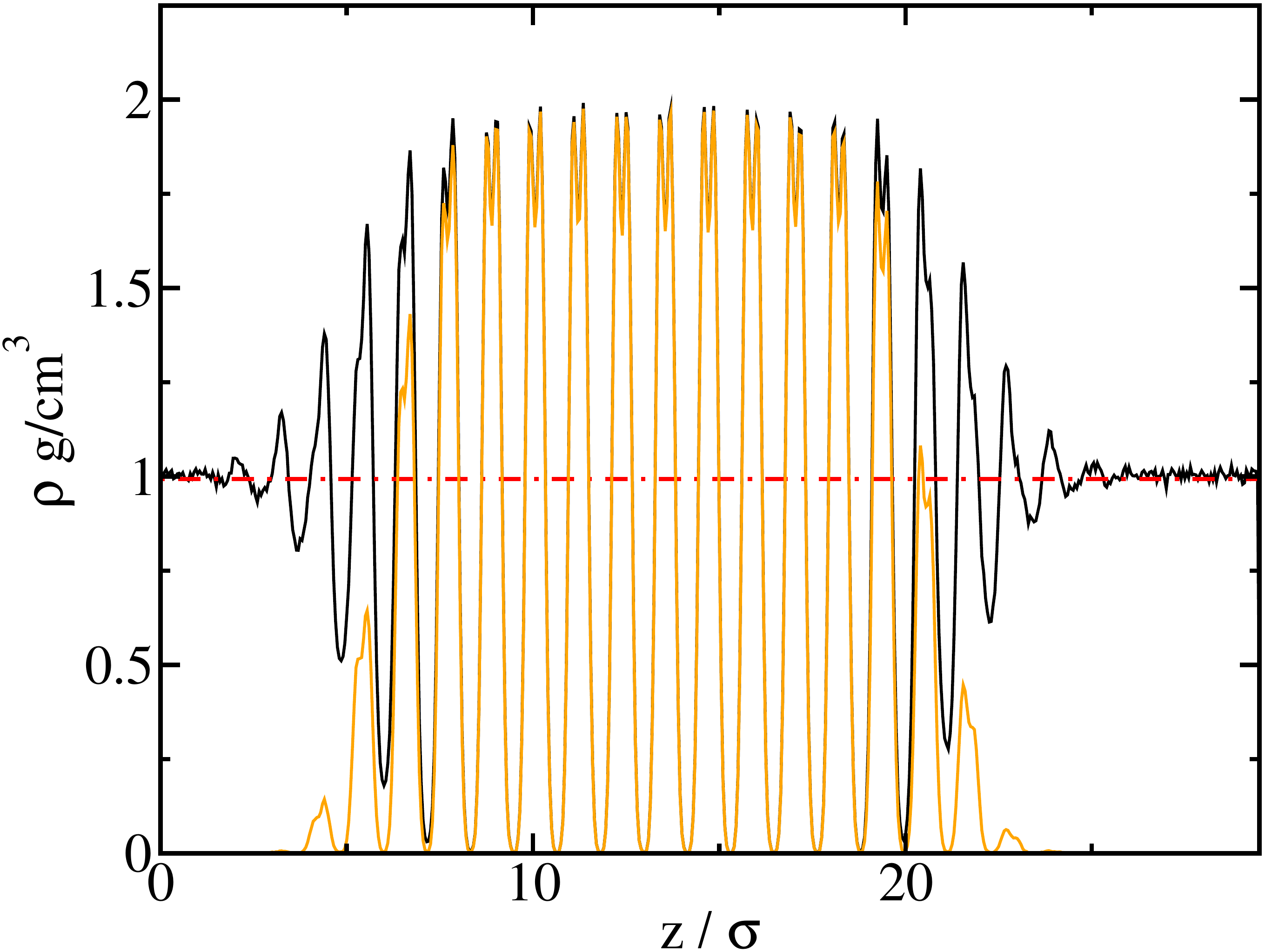}}}
\hspace{0.5cm}
\subfloat[]{%
\resizebox*{5cm}{!}{\includegraphics{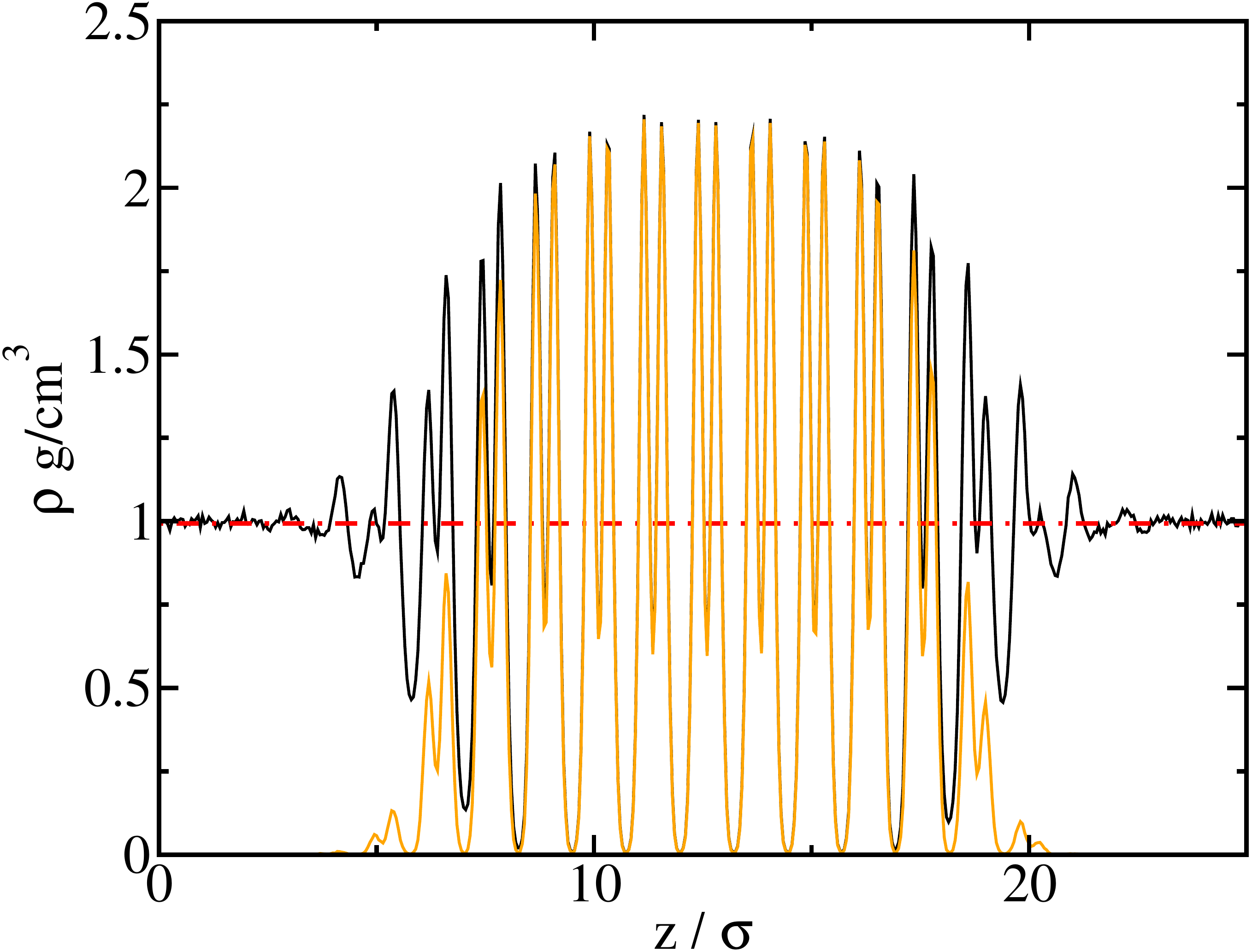}}}
\hspace{0.5cm}
\subfloat[]{%
\resizebox*{5cm}{!}{\includegraphics{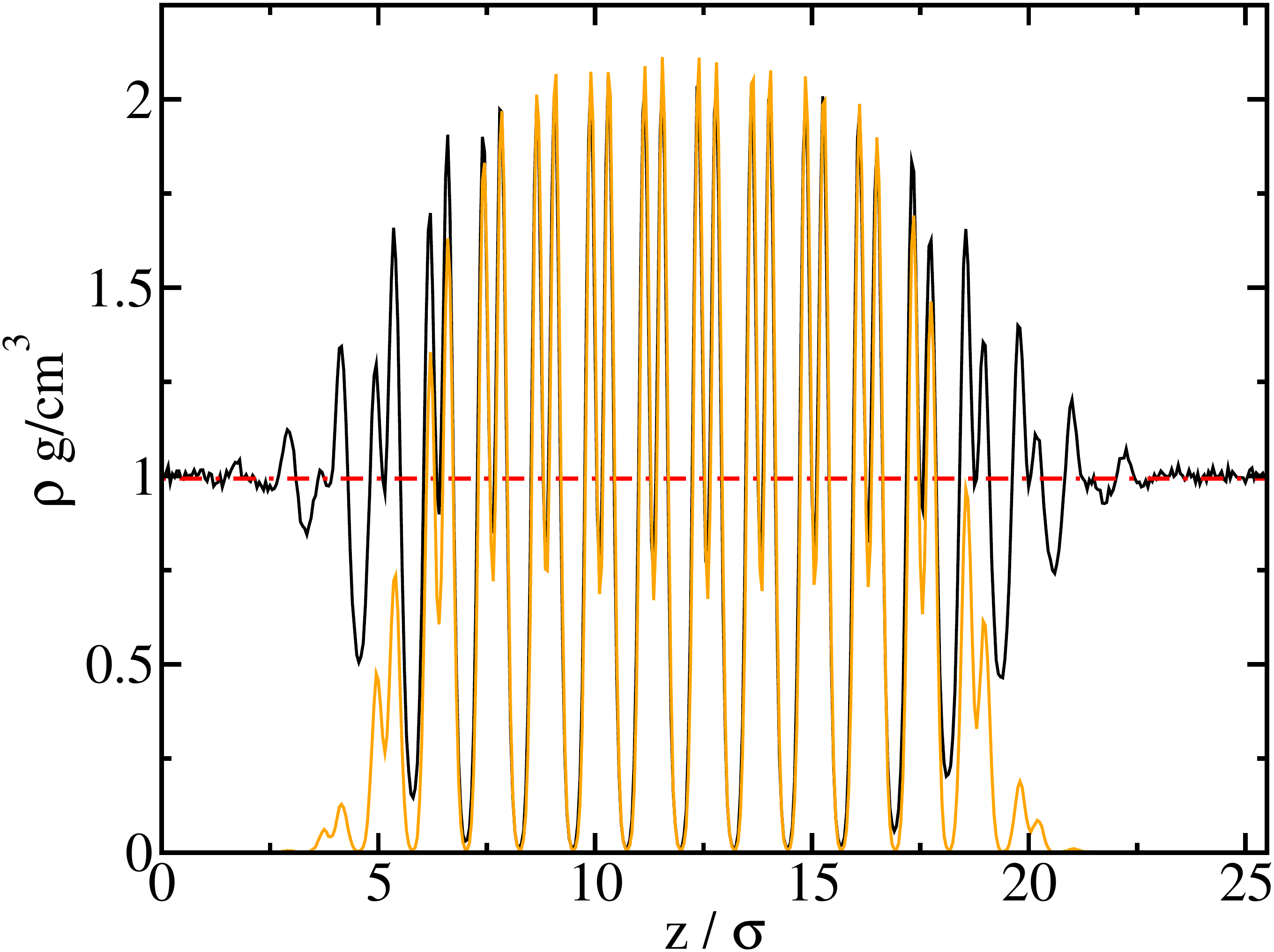}}}  \\
\subfloat[]{%
\resizebox*{5cm}{!}{\includegraphics{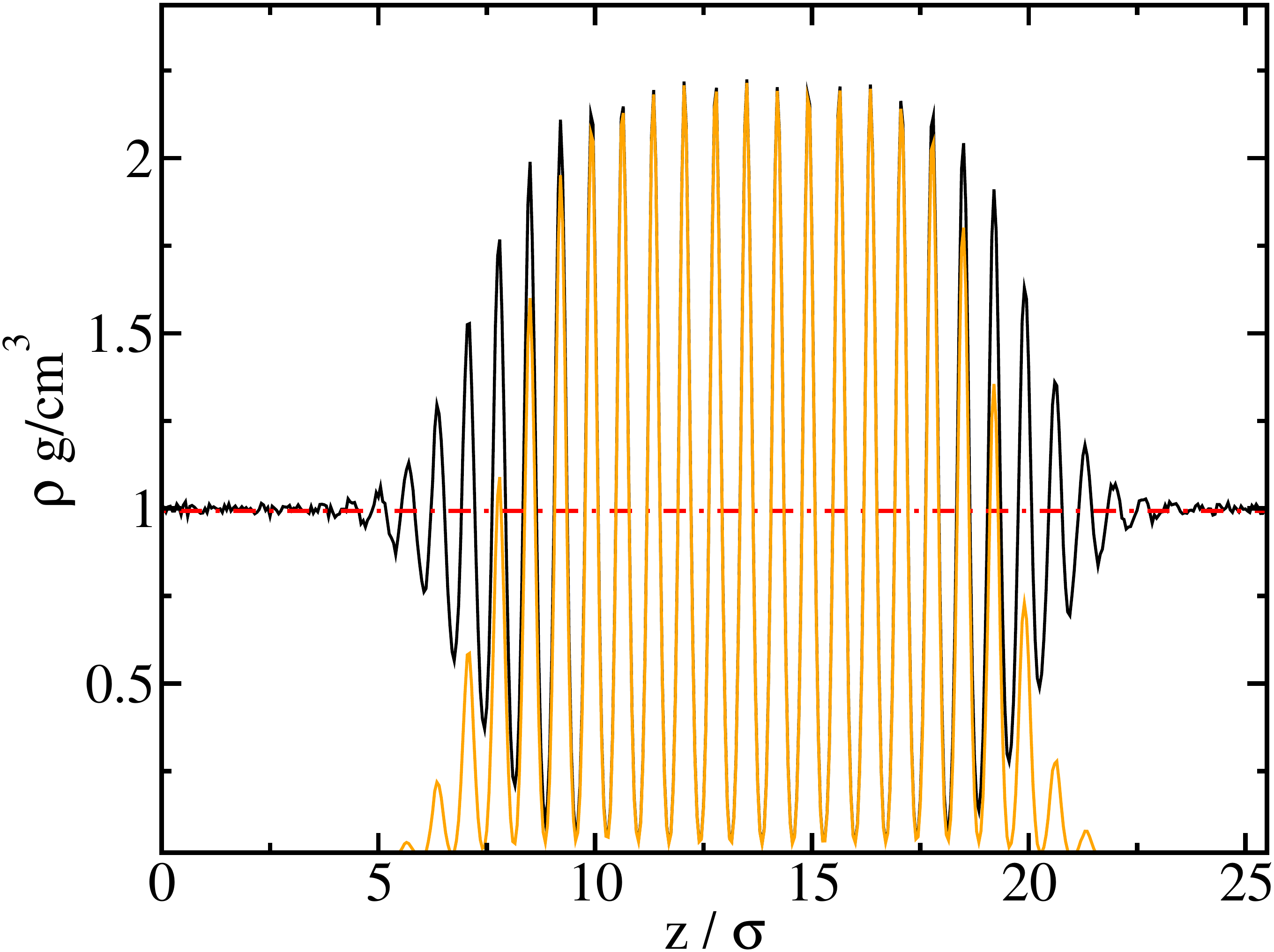}}}
\hspace{0.5cm}
\subfloat[]{%
\resizebox*{5cm}{!}{\includegraphics{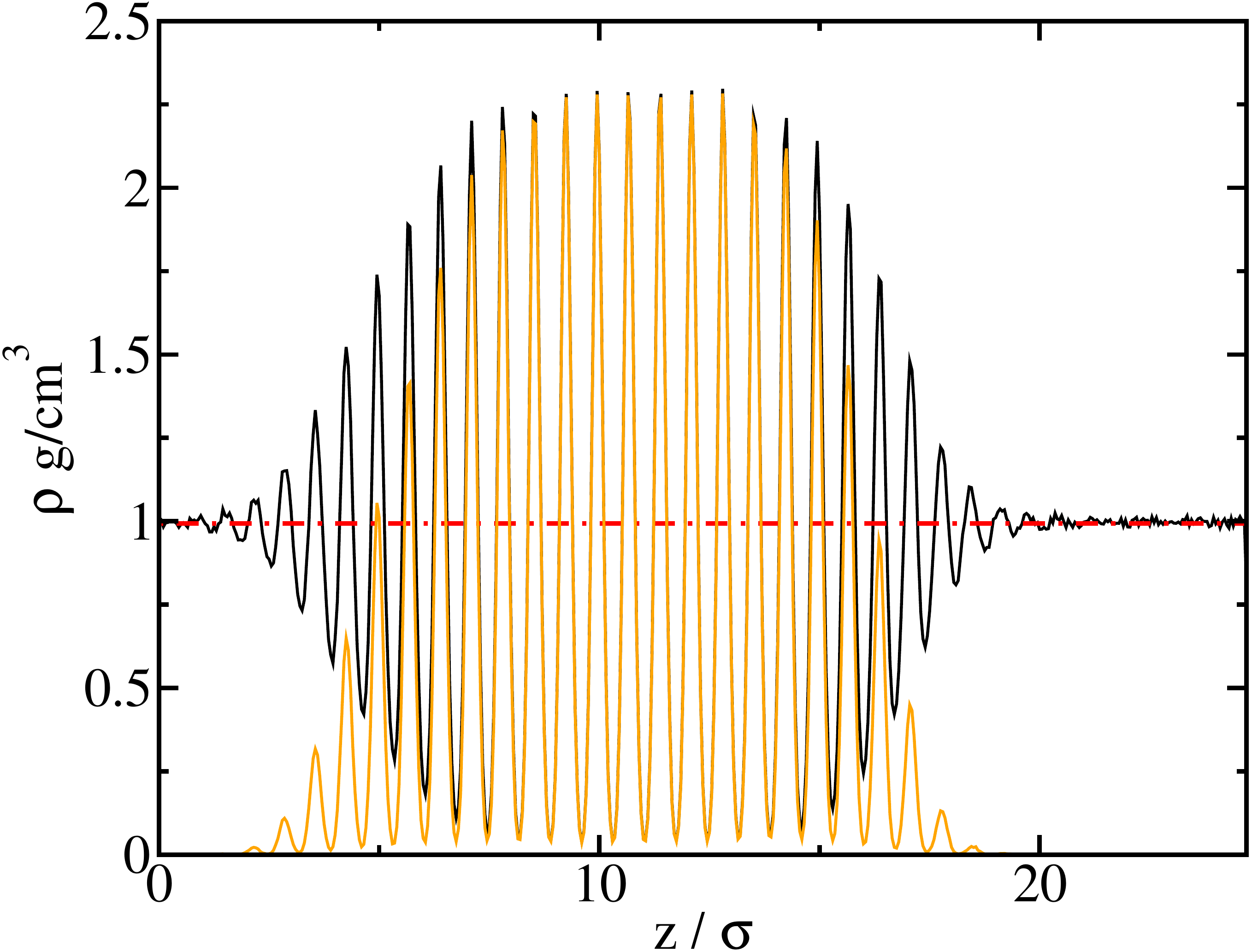}}}
\caption{Density profiles of ice--water systems along z direction.
(a) (Basal)[pII], (b) (pI)[Basal], (c) (pI)[pII], (d) (pII)[Basal],
(e) (pII)[pI]. The density profiles have been calculated with slabs of
thickness 0.05$\sigma$.
Black lines correspond to the whole system and orange lines correspond
to ice--like molecules. Horizontal dotted--dashed lines correspond
to the average bulk density of the fluid phase.}
 \label{fig:perf-sl}
\end{figure}

Fig. \ref{fig:perf-sl} shows results for the density profiles across the ice/water interface for
all three planes studied and different surface setups.\cite{benet15b} Notice that the density
profiles are not intrinsic properties of the bulk thermodynamic field, but
rather, depend also on the lateral system size. With this caution, however, we
can interpret the density profiles in the mean field sense.

 For each plane, it is
apparent the coexistence of a well equilibrated bulk solid phase, with
oscillatory behavior, and a homogeneous liquid phase of uniform density. This
can be inferred by comparing the total density profile (black lines), with the
density of molecules labeled as solid (orange lines), which are fully coincident
within a large slab several layers thick.  The
bulk solid phase acts on average as a hard wall, whereupon damped oscillations of
the liquid phase decay towards the bulk liquid phase due to packing correlations.

Although the thickness of the interface is almost the same in the three planes
studied \cite{benet14c} the number of layers involved in it
differs. The basal plane exhibits
5 distinct layers of ordered liquid before decaying to the bulk density; the pI
plane shows four bimodal oscillations, and the pII is that exhibiting a larger
number of layers with about six clear oscillations before decaying to the liquid
density. In all cases, there is a clear penetration of the solid density into
the region where the liquid is the majority phase. This indicates either a rough
interface, or the presence of terraces, such that, along the same layer, a
partially filled solid stacking is interrupted by pockets of liquid water.

As expected, the density profiles of equal planes but different geometries,
do not differ  from each other. Unlike the stiffness coefficients, the density
profiles are properties of the plane only,  not of  a privileged direction for
wave propagation within that plane. Hence, the density profiles of the (pI)[basal]
and (pI)[pII] setups are essentially identical, and similarly, those of the
(pII)[basal] and (pII)[pI] are also equal.

\subsubsection{Ice/vapor interface}

\begin{figure}[h]
\centering
\subfloat[]{%
\resizebox{5cm}{!}{\includegraphics{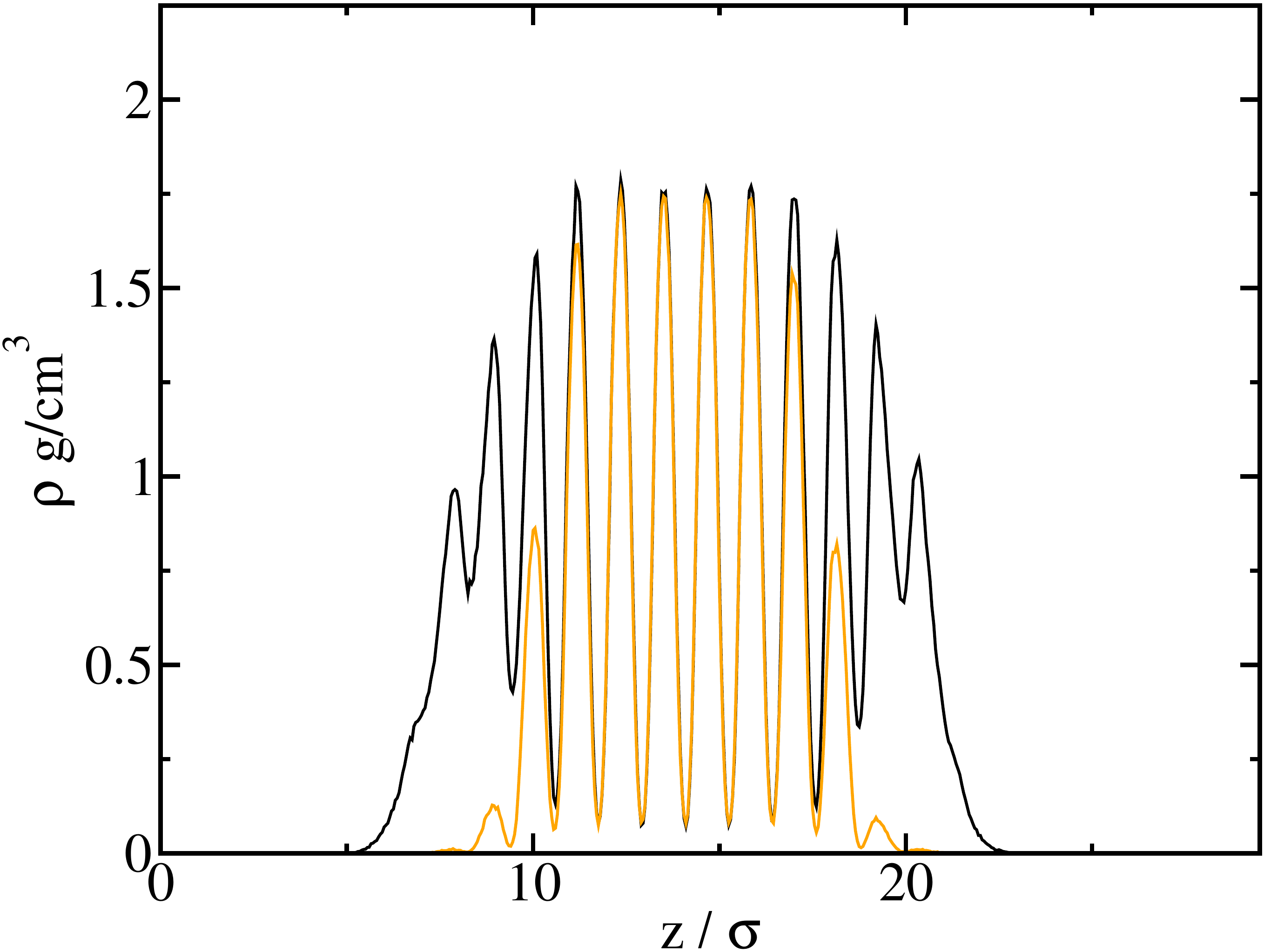}}}
\hspace{0.5cm}
\subfloat[]{%
\resizebox{5cm}{!}{\includegraphics{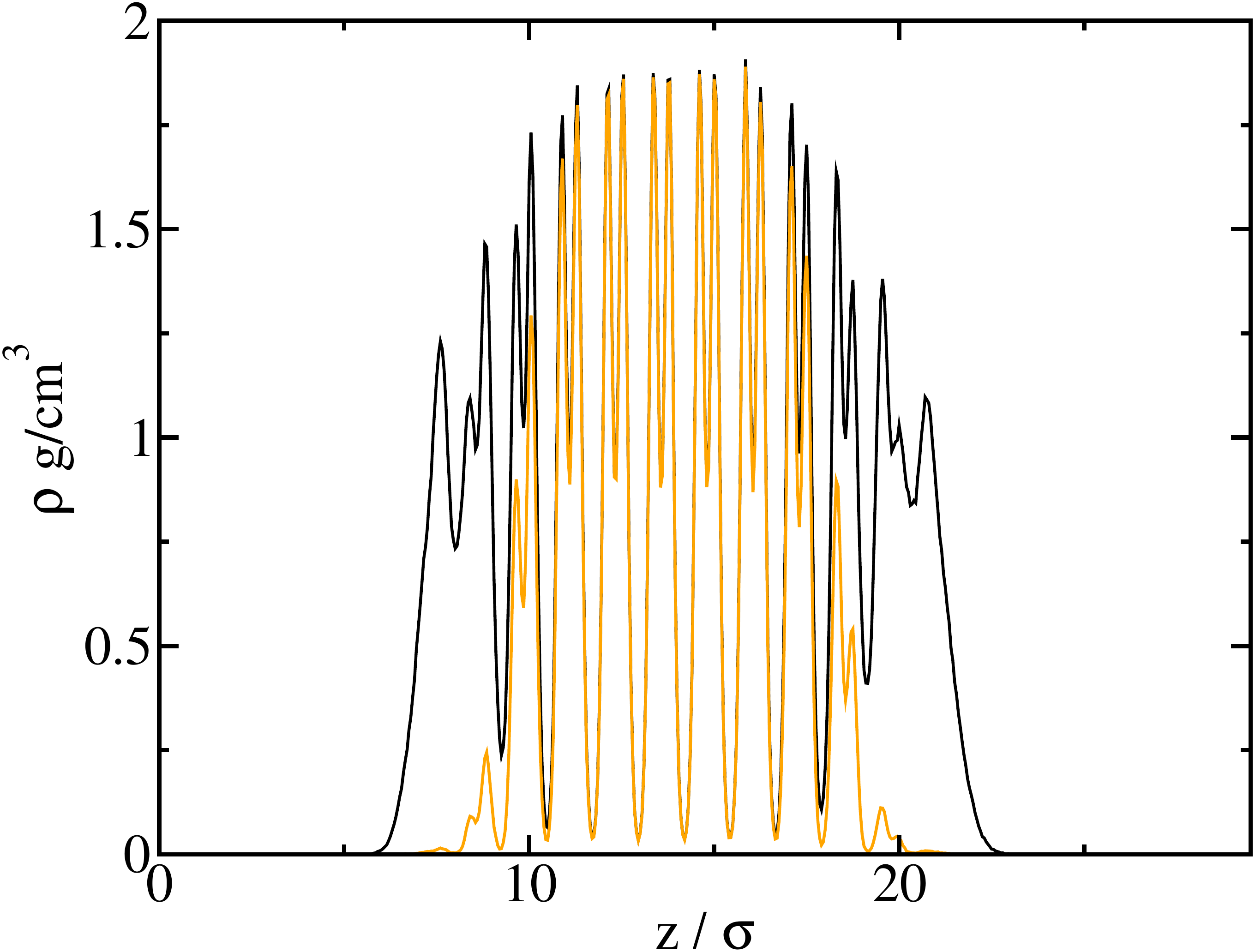}}}
\hspace{0.5cm}
\subfloat[]{%
\resizebox{5cm}{!}{\includegraphics{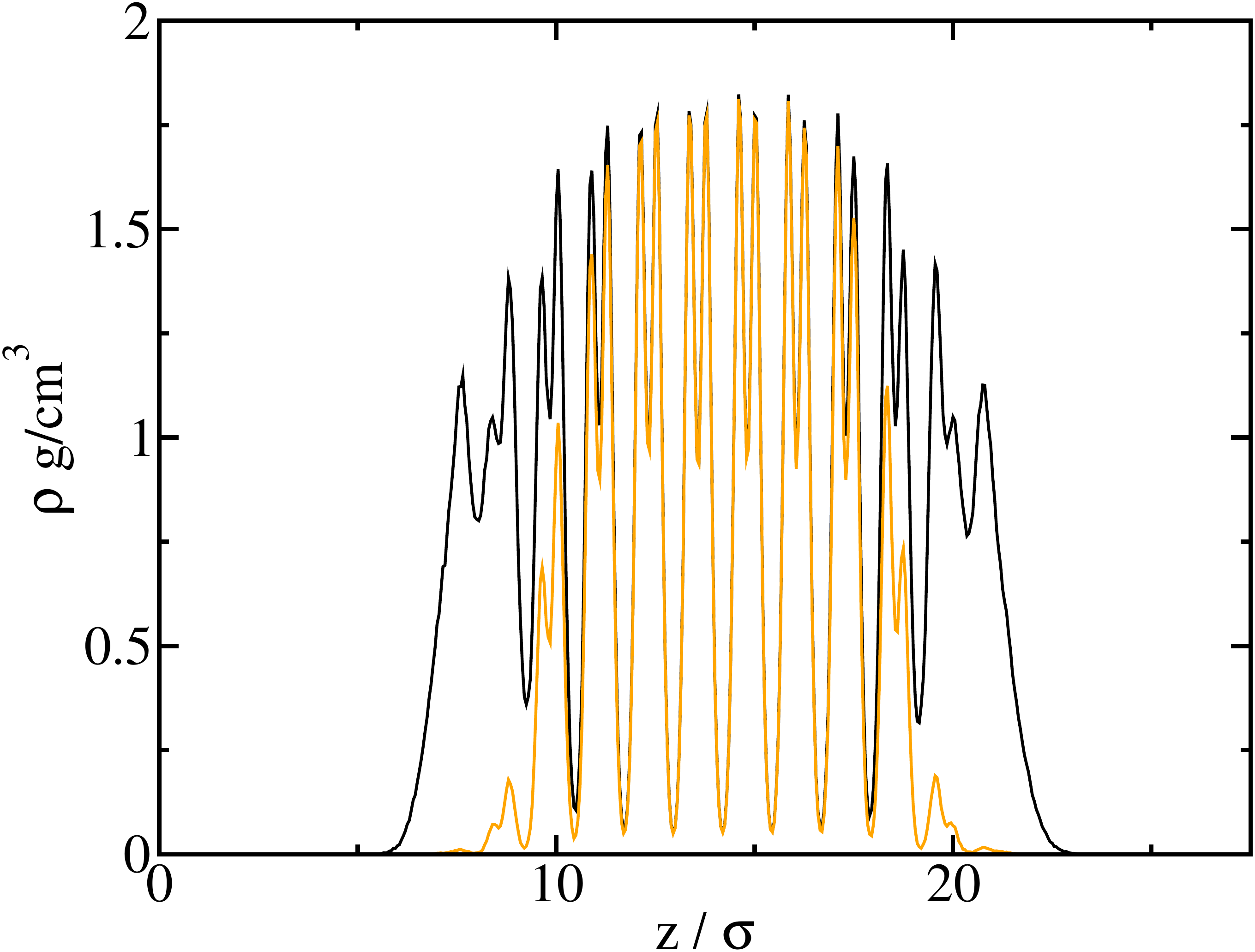}}} \\
\subfloat[]{%
\resizebox{5cm}{!}{\includegraphics{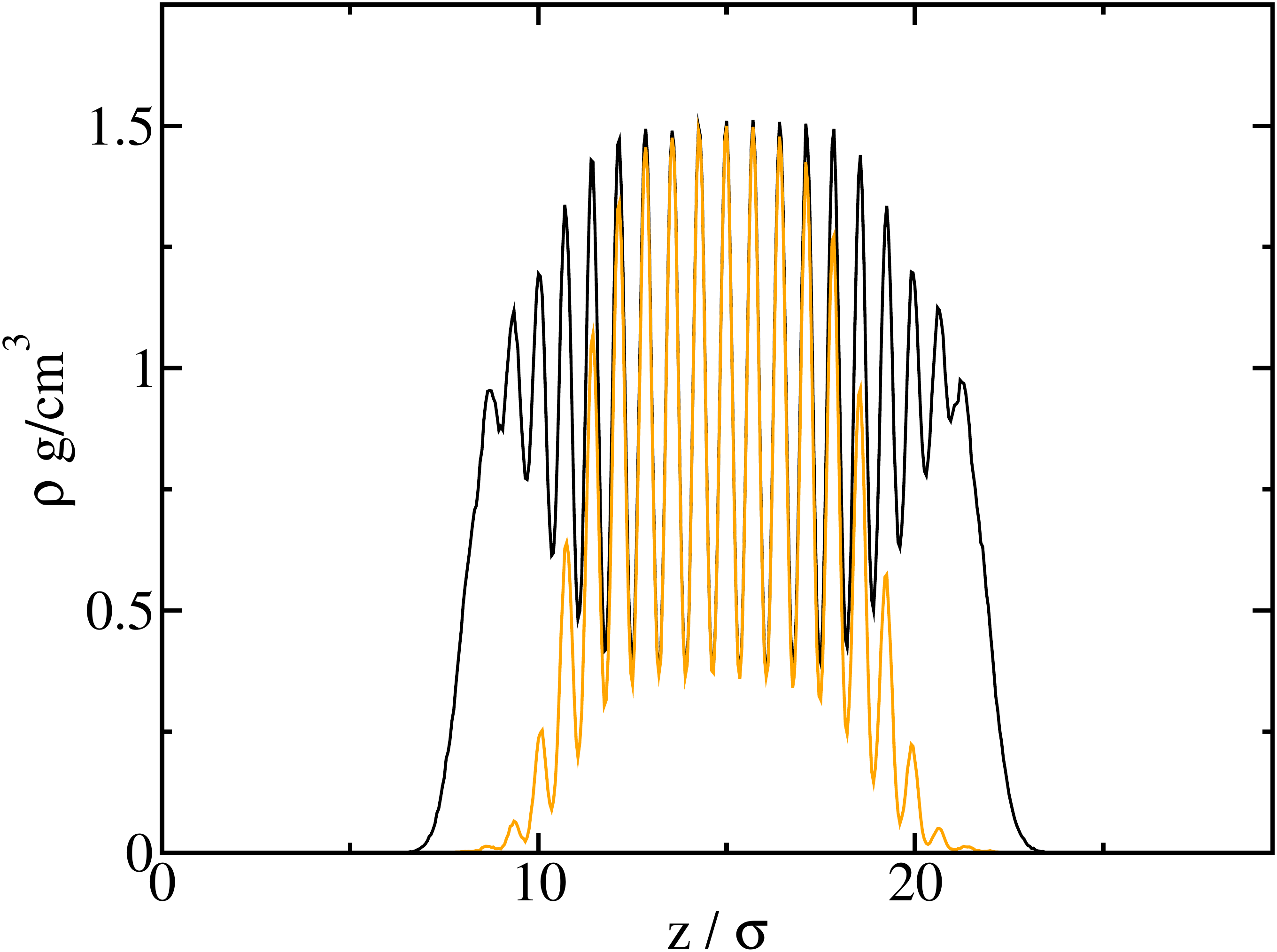}}}
\hspace{0.5cm}
\subfloat[]{%
\resizebox{5cm}{!}{\includegraphics{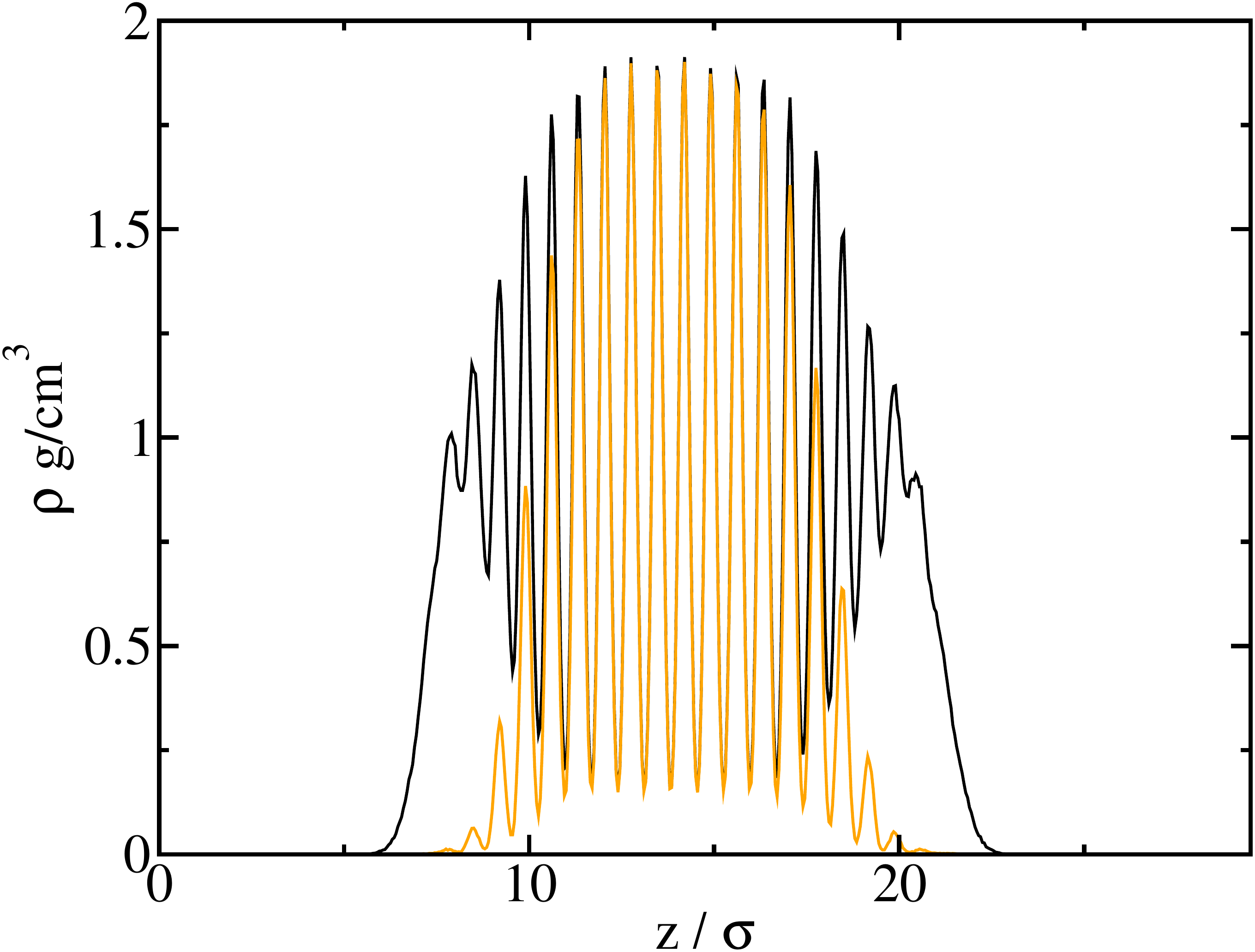}}}
\caption{Density profiles of the ice--vapour interfaces along z direction.
(a) (Basal)[pII], (b) (pI)[Basal], (c) (pI)[pII], (d) (pII)[Basal],
(e) (pII)[pI]. The density profiles have been calculated with slabs of
thickness 0.05$\sigma$.
Black lines correspond to the whole system and orange lines correspond
to ice--like molecules.}
 \label{fig:perf-sv}
\end{figure}

The density profiles of the ice/vapor interface are shown in Fig.
\ref{fig:perf-sv} for the same
planes and geometries studied previously.\cite{benet15b} Again it is possible to identify a
bulk  solid phase several layers thick, and a vapor phase of very small
density (essentially zero density in the scale of the figure). Compared to the
ice/water interface, however, the presence of a third liquid phase protruding
between the bulk solid and vapor phases is fairly apparent, as indicated by the
high density regions with damped oscillations corresponding to water molecules
labeled as liquid phase. Whence, it is concluded that the ice/vapor interface is
best described as an ice/film/vapor system, with a premelted liquid film between
the vapor and the solid. A full characterization of the fluctuating interface
then requires to distinguish between the ice/film and the film/vapor surfaces,
which could in principle, exhibit different correlations, at least at large
wave-vectors.

\begin{figure}[h]
\centering
\subfloat[]{%
\resizebox{5cm}{!}{\includegraphics{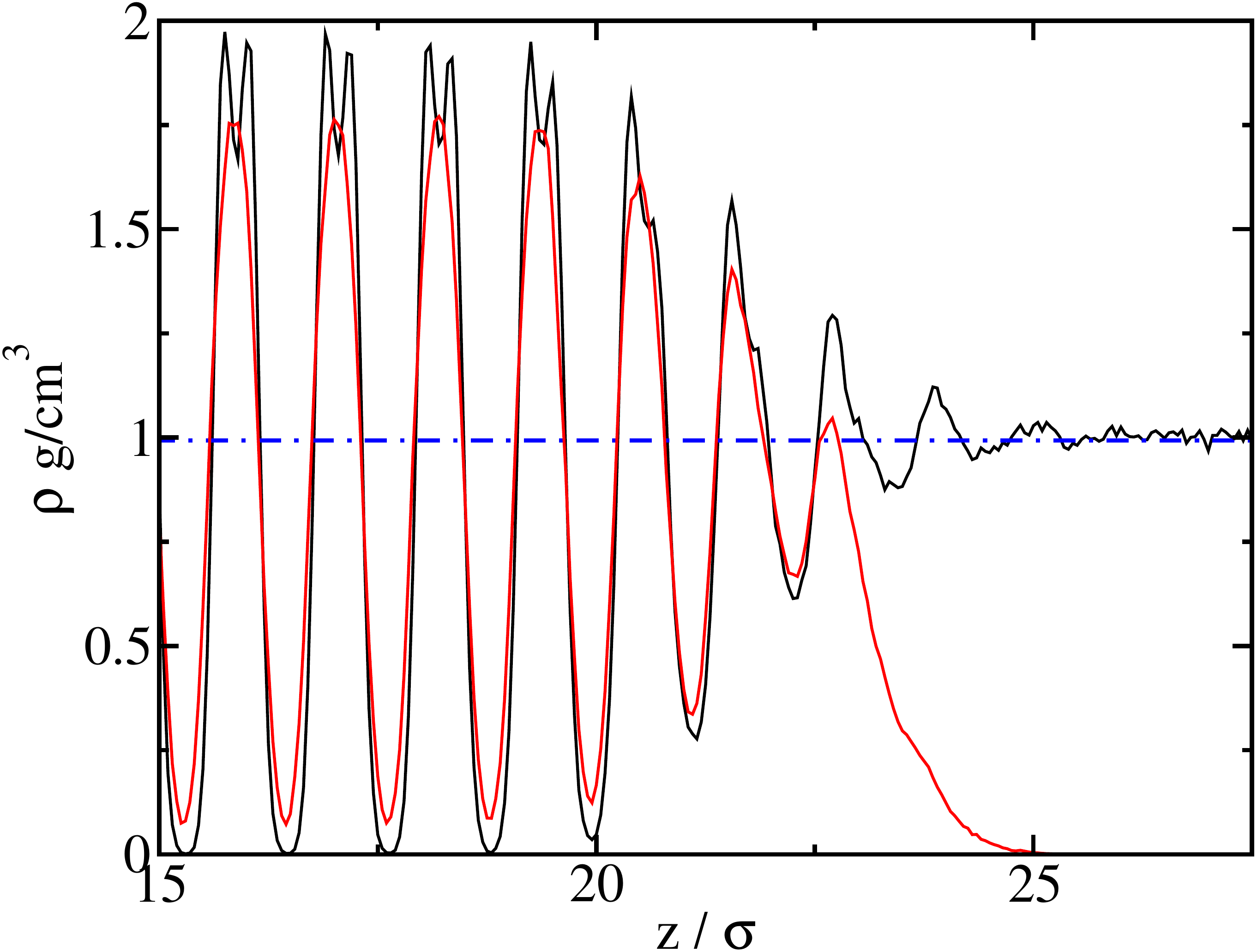}}}
\hspace{0.5cm}
\subfloat[]{%
\resizebox{5cm}{!}{\includegraphics{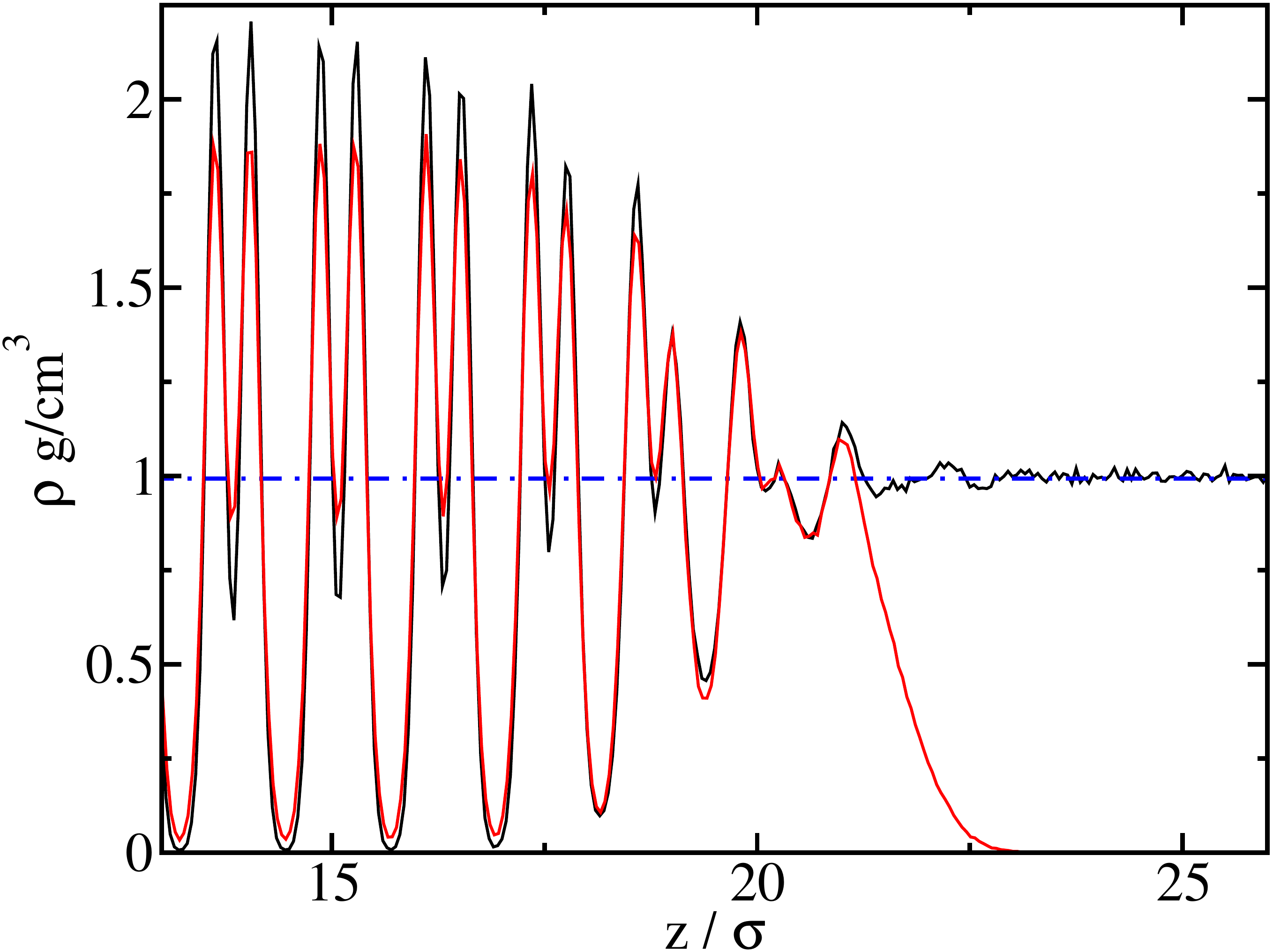}}}
\hspace{0.5cm}
\subfloat[]{%
\resizebox{5cm}{!}{\includegraphics{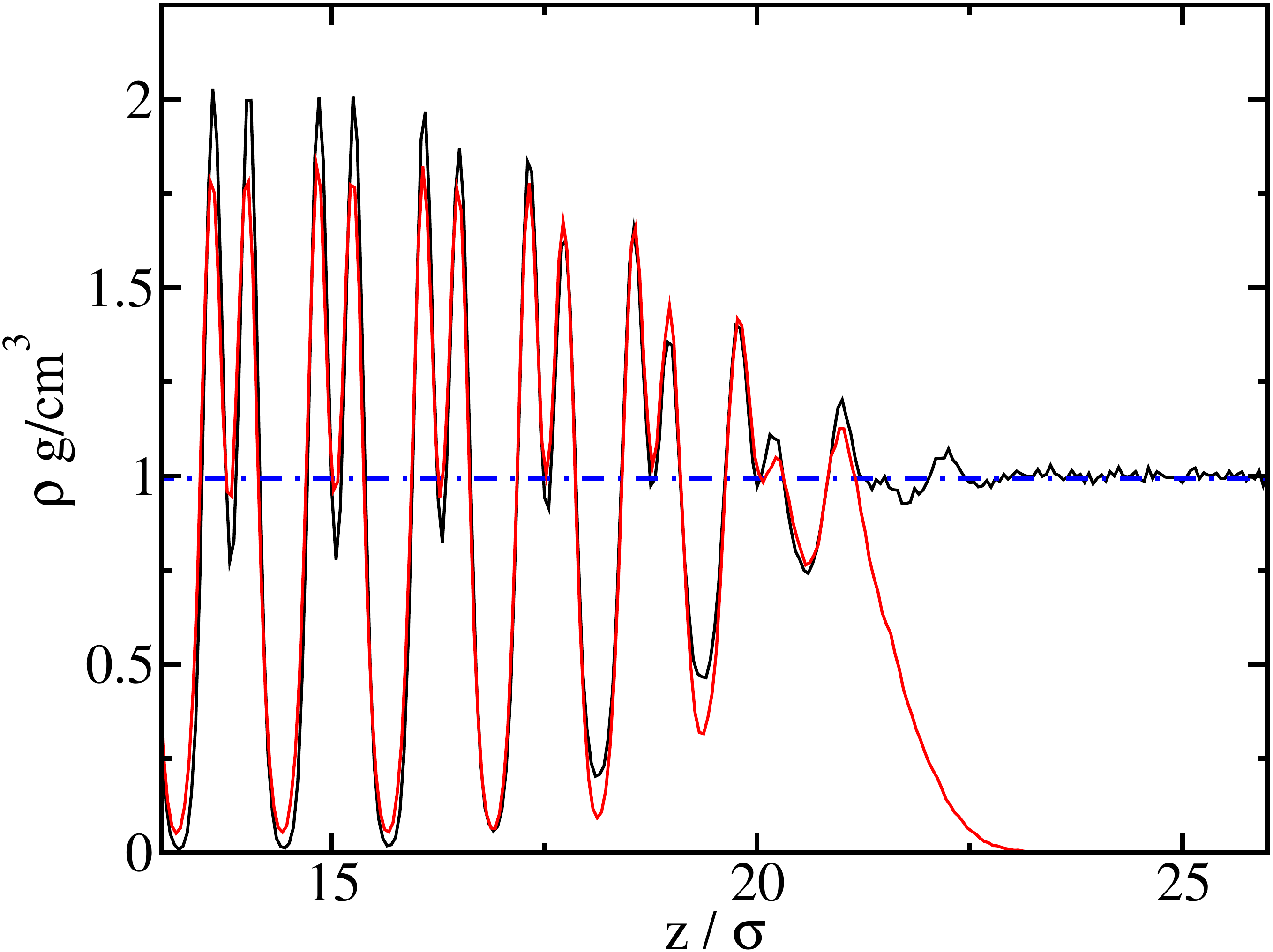}}}  \\
\subfloat[]{%
\resizebox{5cm}{!}{\includegraphics{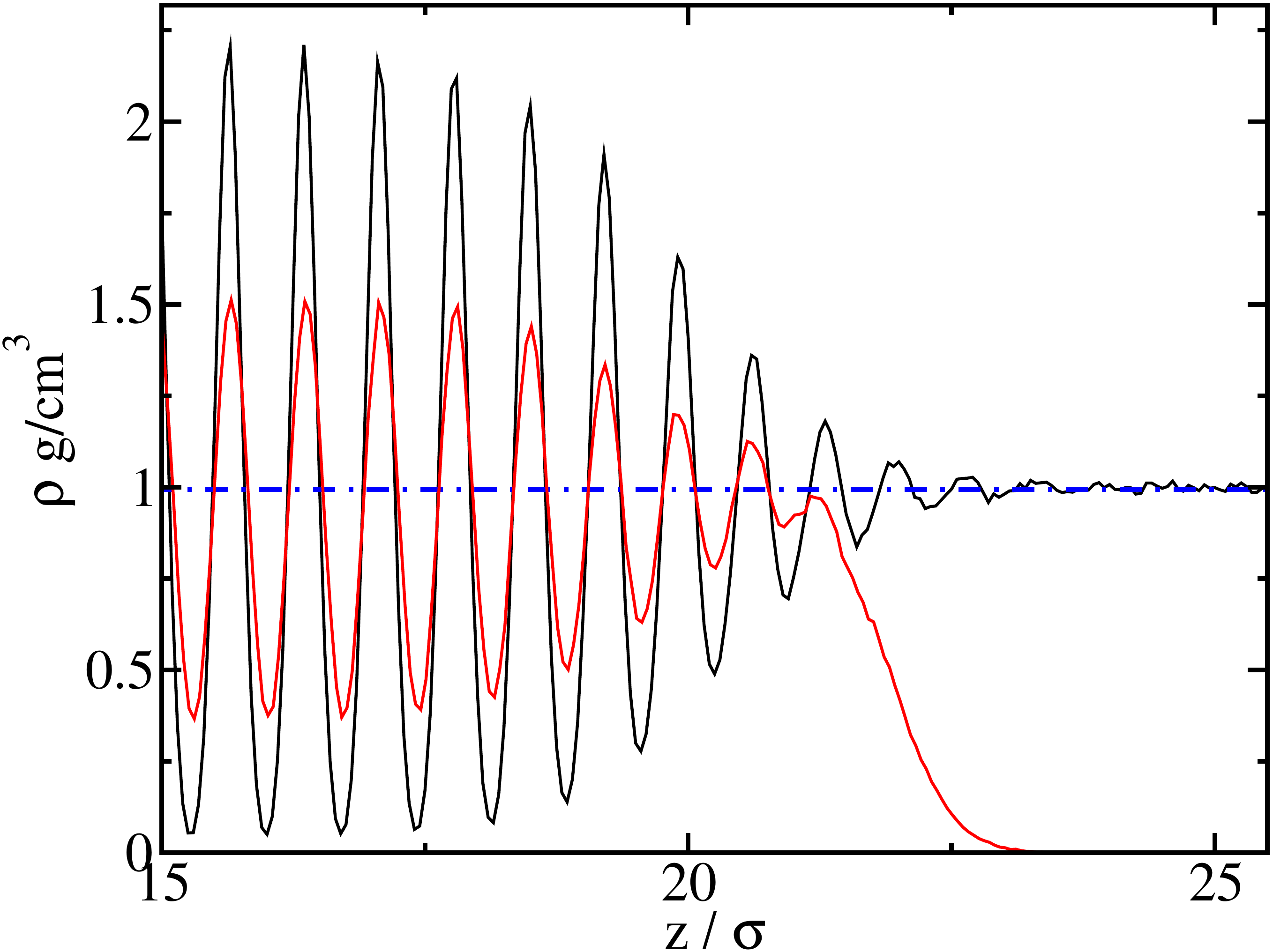}}}
\hspace{0.5cm}
\subfloat[]{%
\resizebox{5cm}{!}{\includegraphics{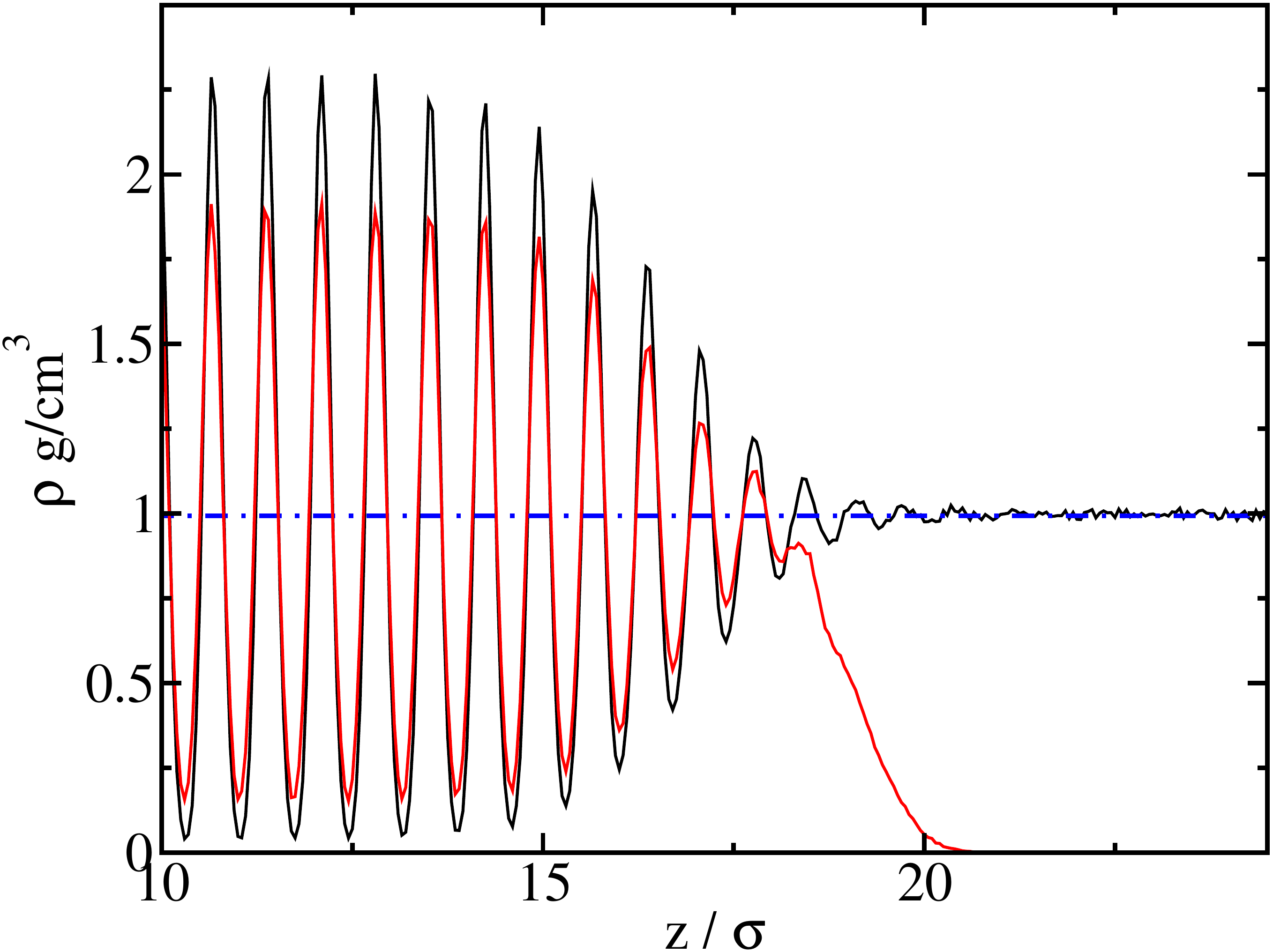}}}
\caption{Complete density profiles of the ice--water (black)
and the ice--vapour (red) systems along z direction.
(a) (Basal)[pII], (b) (pI)[Basal], (c) (pI)[pII], (d) (pII)[Basal],
(e) (pII)[pI]. Dashed and dotted blue line correspond to the average density
of the fluid phase. The density profiles have been calculated with slabs of
thickness 0.05$\sigma$.}
 \label{fig:liquid}
\end{figure}

Interestingly, a comparison of the decaying oscillations of  the
premelted film and the ice/water interface reveals a rather similar structure.
This is best seen in Fig.\ref{fig:liquid}, where the total density profile of the ice/water and
ice/vapor interfaces is compared for all planes studied. The figure clearly
shows that the ice/vapor interface is nearly equal to the ice/water interface.
Not only it follows the oscillations expected for the bulk solid phase, but also
mimics accurately the damped oscillations of the decaying water profile,
up to a point where the density suddenly falls to the bulk vapor density.

\begin{figure}[h]
\centering
\subfloat[]{%
\resizebox{5cm}{!}{\includegraphics{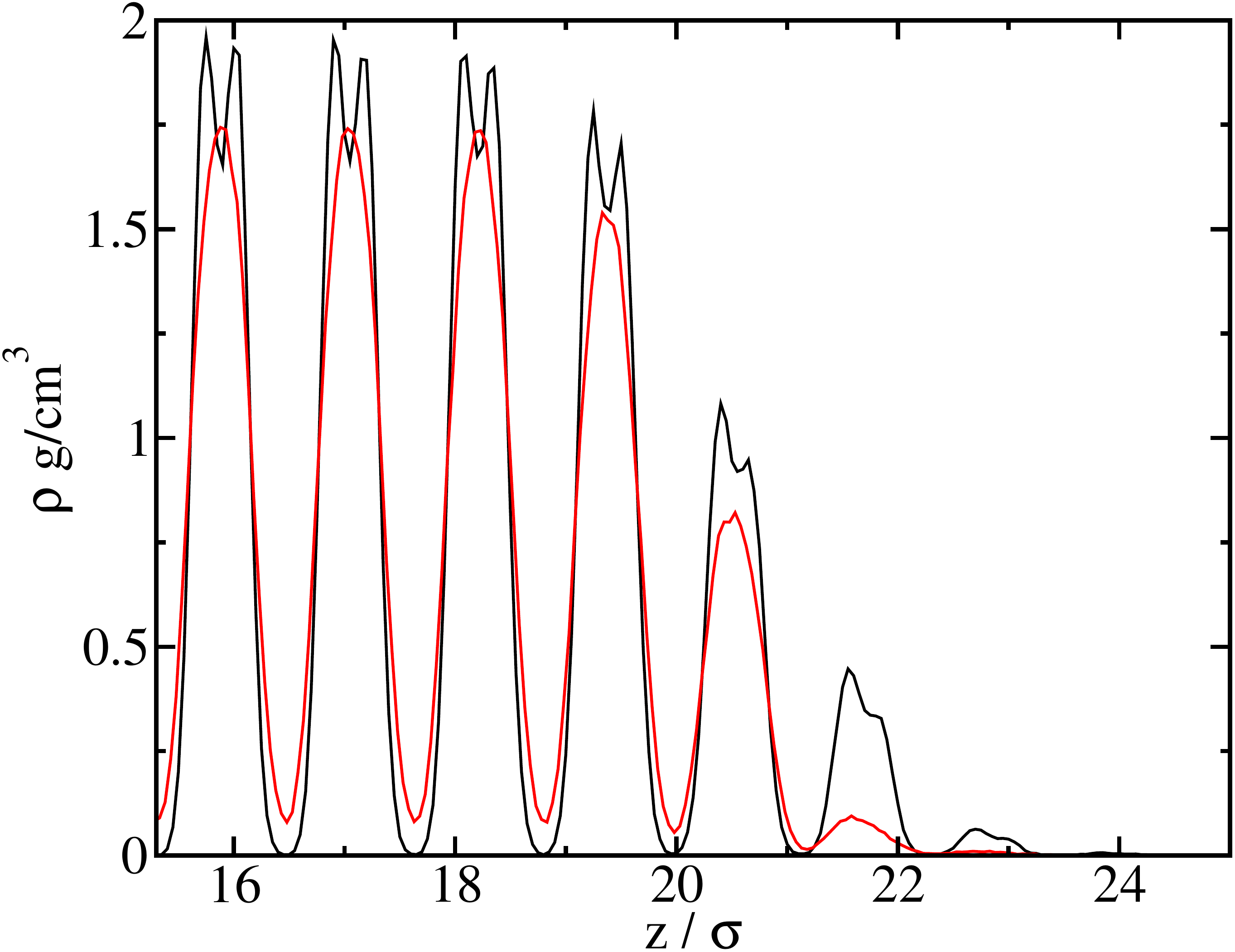}}}
\subfloat[]{%
\hspace{0.5cm}
\resizebox{5cm}{!}{\includegraphics{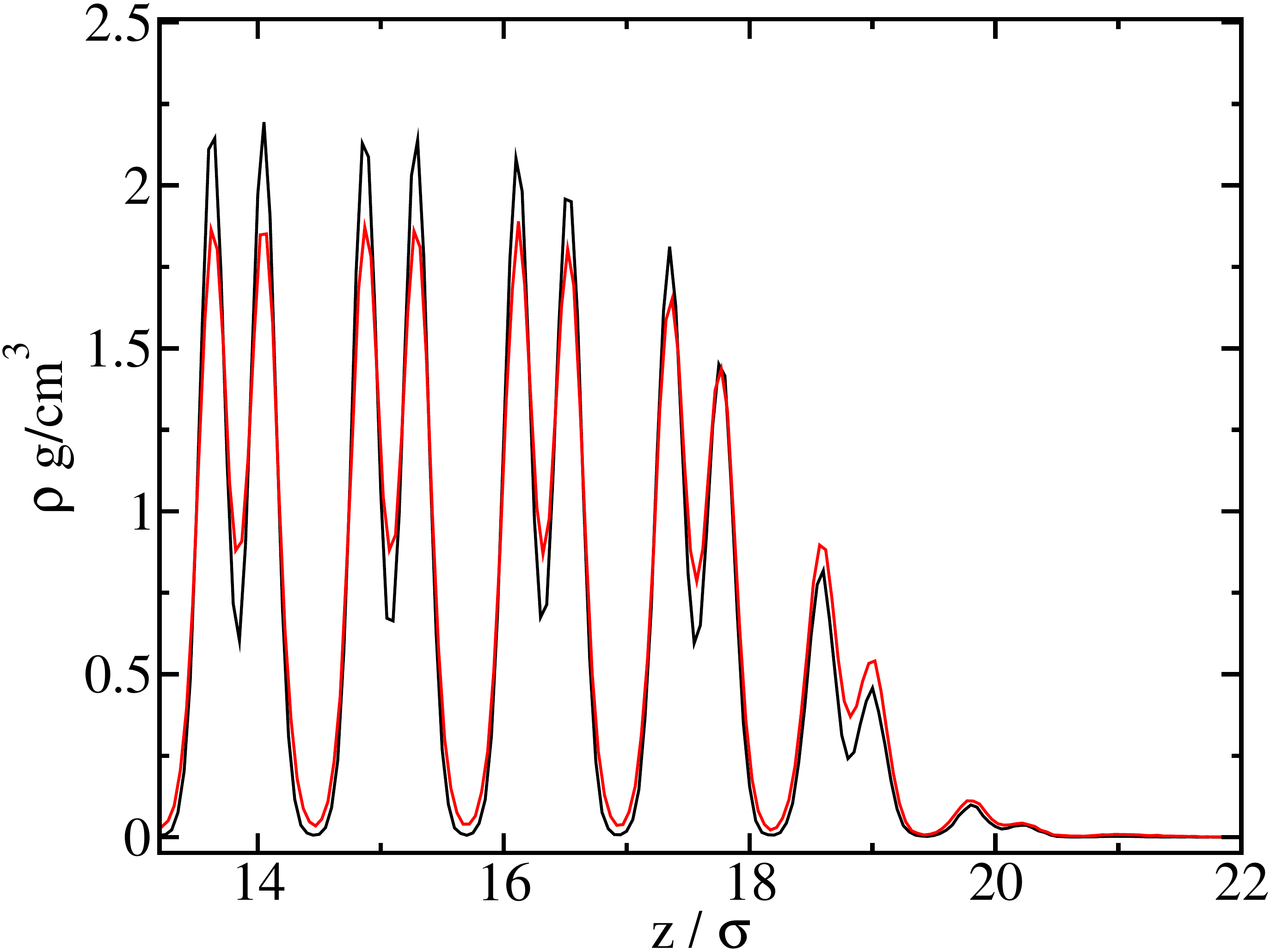}}}
\hspace{0.5cm}
\subfloat[]{%
\resizebox{5cm}{!}{\includegraphics{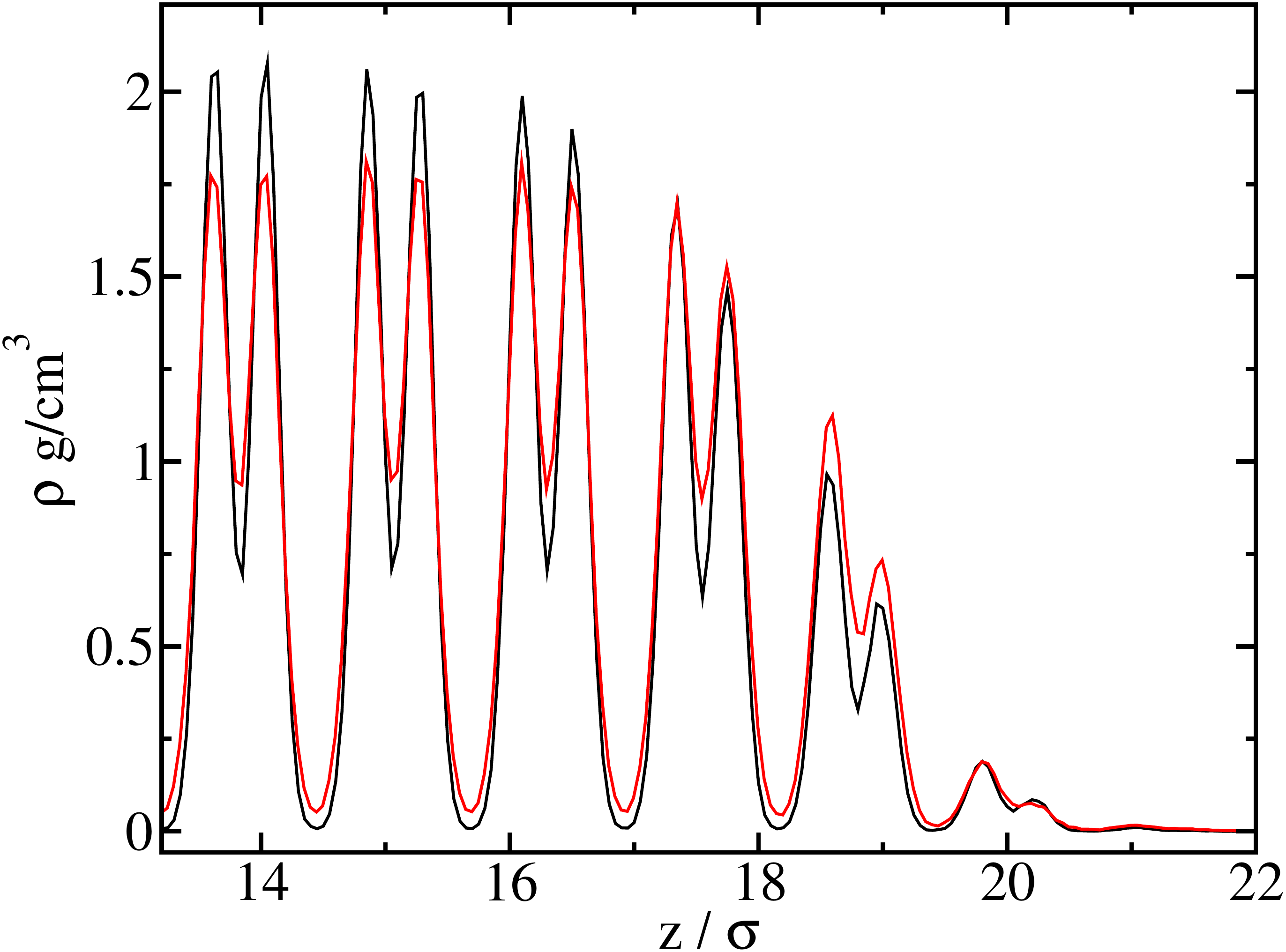}}} \\
\subfloat[]{%
\resizebox{5cm}{!}{\includegraphics{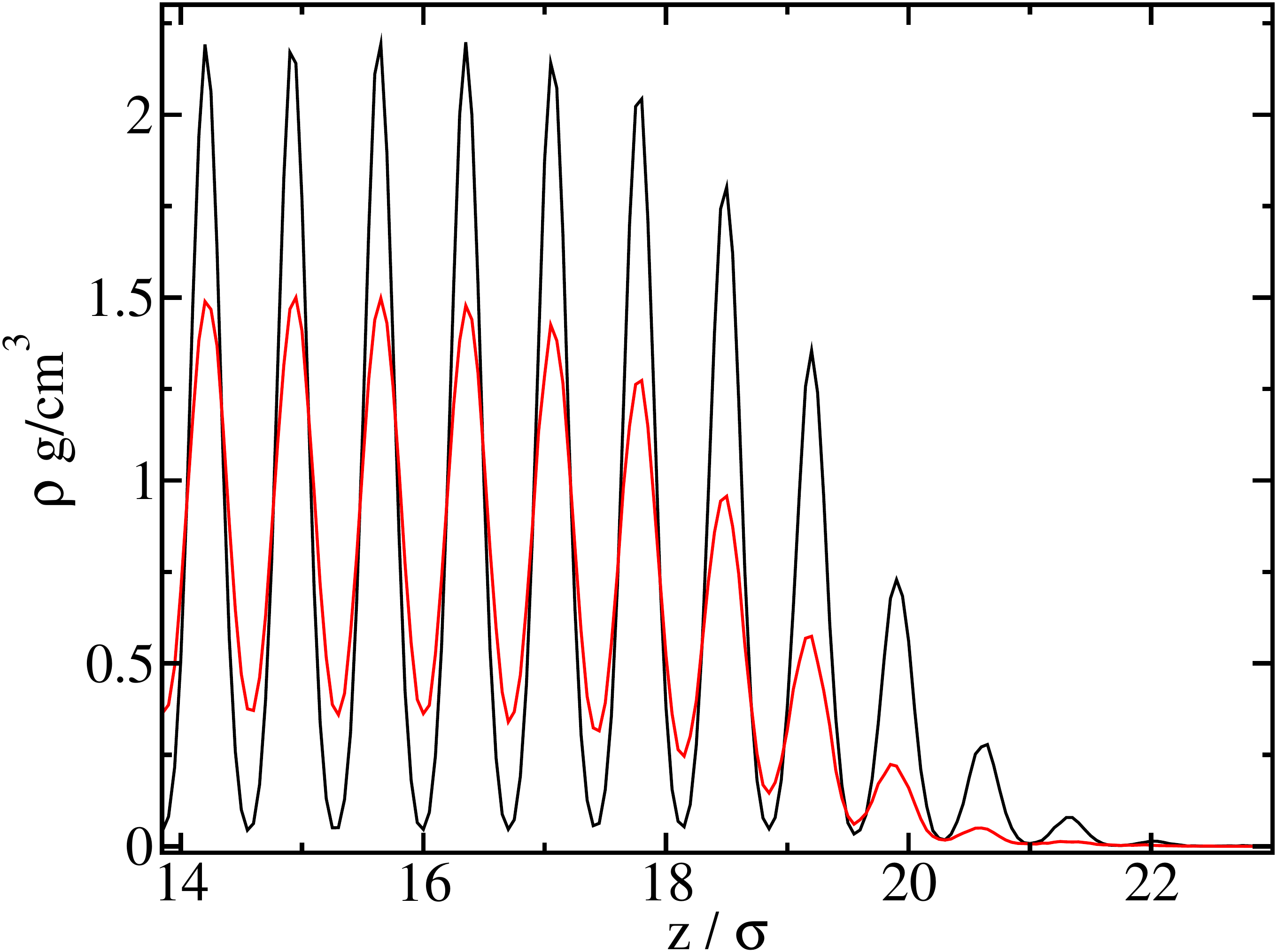}}}
\hspace{0.5cm}
\subfloat[]{%
\resizebox{5cm}{!}{\includegraphics{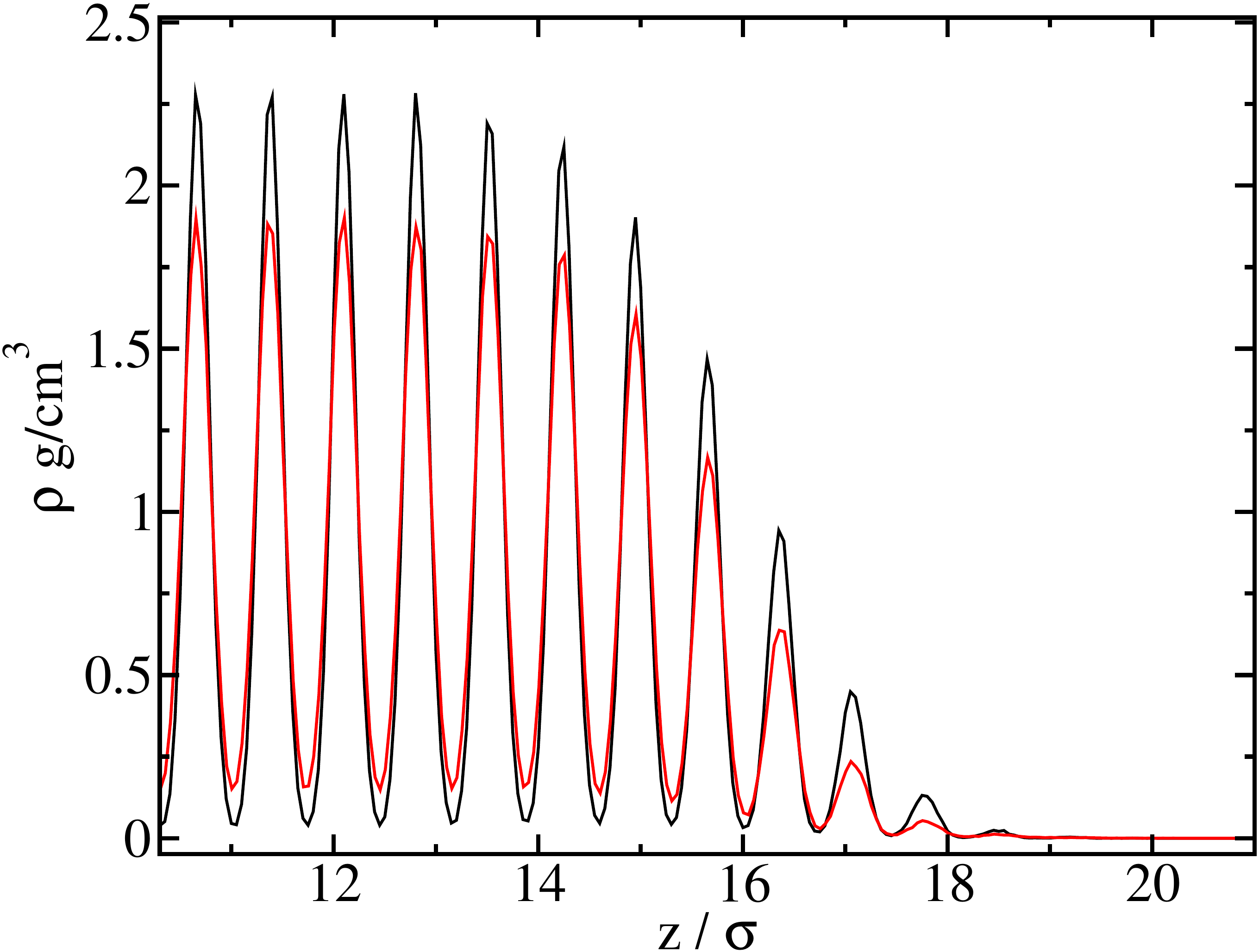}}}
\caption{Density profiles of ice--like molecules for the ice--water (black)
and the ice--vapour (red) systems along z direction.
(a) (Basal)[pII], (b) (pI)[Basal], (c) (pI)[pII], (d) (pII)[Basal],
(e) (pII)[pI]. The density profiles have been calculated with slabs of
thickness 0.05$\sigma$}
 \label{fig:solid}
\end{figure}

Similarly, the density profile of solid like atoms is the same in both the
ice/water and ice/vapor interfaces, both within the bulk solid slab, and in the
decaying density profile. We test this in  Fig.~\ref{fig:solid},
where the density profiles of solid like molecules
for the ice/water and ice/vapor interfaces are compared. Clearly, the structure
of the density oscillations is nearly equal, with only somewhat smaller
solid molecule densities in the ice/vapor system. Such differences are obviously
a result of the somewhat smaller chemical potential that is imposed along the
sublimation line, as compared to that of the melting line.

Finally, we see from Fig. \ref{fig:solid} that the structure of the film formed
at the ice/vapor interface is the same as that of the liquid phase of the
ice/water interface.

These set of figures  clearly indicate that the ice/film boundary of the ice/vapor interface is
very similar to that of the ice/water boundary, at least at temperatures a few
degrees below the triple point. Interestingly, this observation is quite consistent with recent
measurement of ice growth, which revealed an activated mechanism with
equal molecular step energies for
ice crystallites grown in water or vapor bulk phases, and supports
the hypothesis that the rate determining step of crystal growth, whether from
the liquid or the vapor phase, is the stacking of crystal  planes at
the liquid/ice boundary.\cite{libbrecht14,murata18}

%%%%%%%%%%%%%%%%%%%%%%%%%%%%%%%%%%%%%%%%%%%%%%%%%%%%%%%%%%%%%%%%%%%%%%%%%%%%%
%%%%%%%%%%%%%%%%%%%%%%%%%%%%%%%%%%%%%%%%%%%%%%%%%%%%%%%%%%%%%%%%%%%%%%%%%%%%%

%%%%%%%%%%%%%%%%%%%%%%%%%%%%%%%%%%%%%%%%%%%%%%%%%%%%%%%%%%%%%%%%%%%%%%%%%%%%%
%%%%%%%%%%%%%%%%%%%%%%%%%%%%%%%%%%%%%%%%%%%%%%%%%%%%%%%%%%%%%%%%%%%%%%%%%%%%%

\subsection{Roughness}

%%%%%%%%%%%%%%%%%%%%%%%%%%%%%%%%%%%%%%%%%%%%%%%%%%%%%%%%%%%%%%%%%%%%%%%%%%%%%
\begin{figure}[h]
\centering
\subfloat[]{%
\resizebox{7cm}{!}{\includegraphics{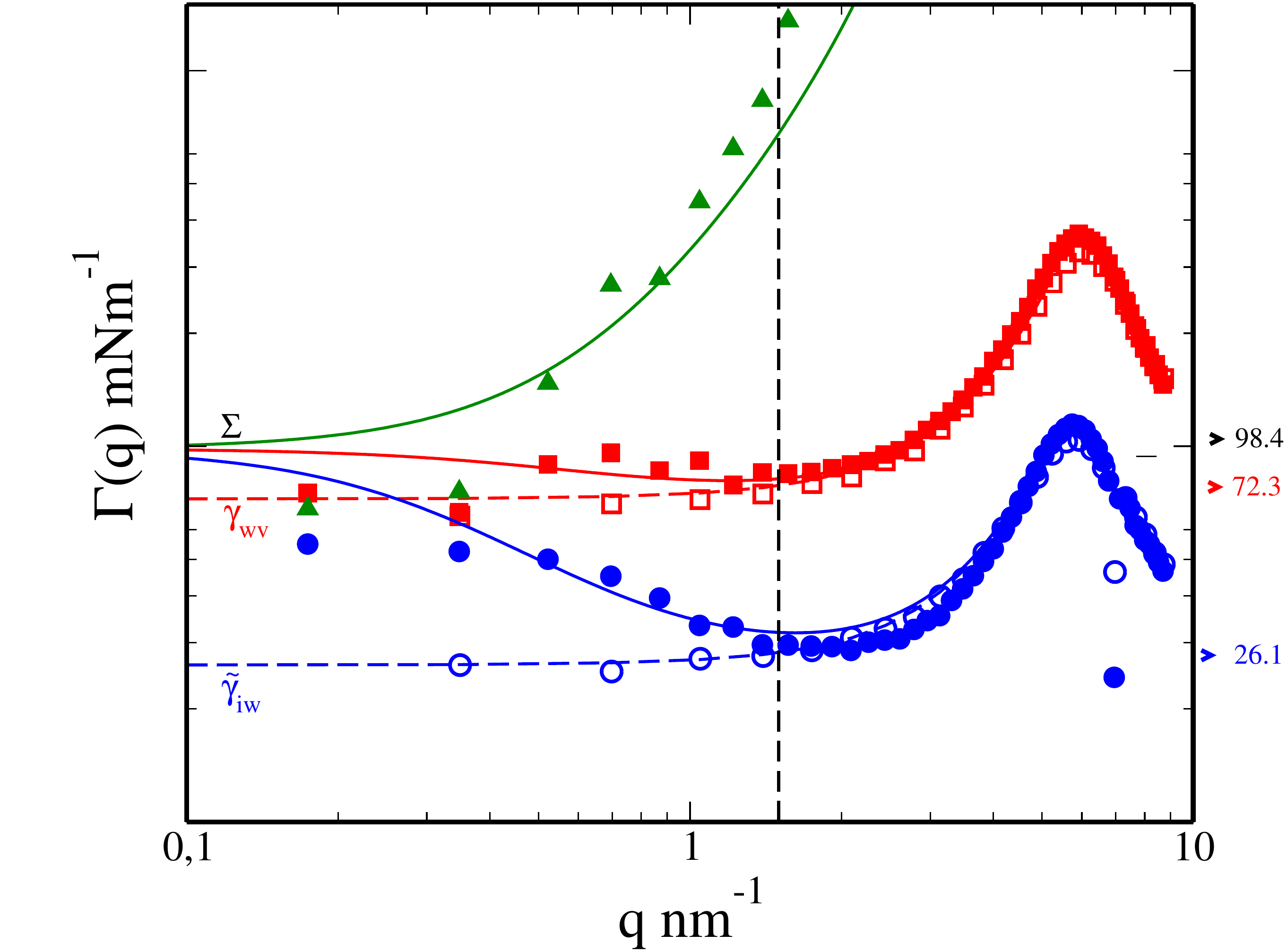}}}
%\hspace{0.5cm}
\subfloat[]{%
\resizebox{7cm}{!}{\includegraphics{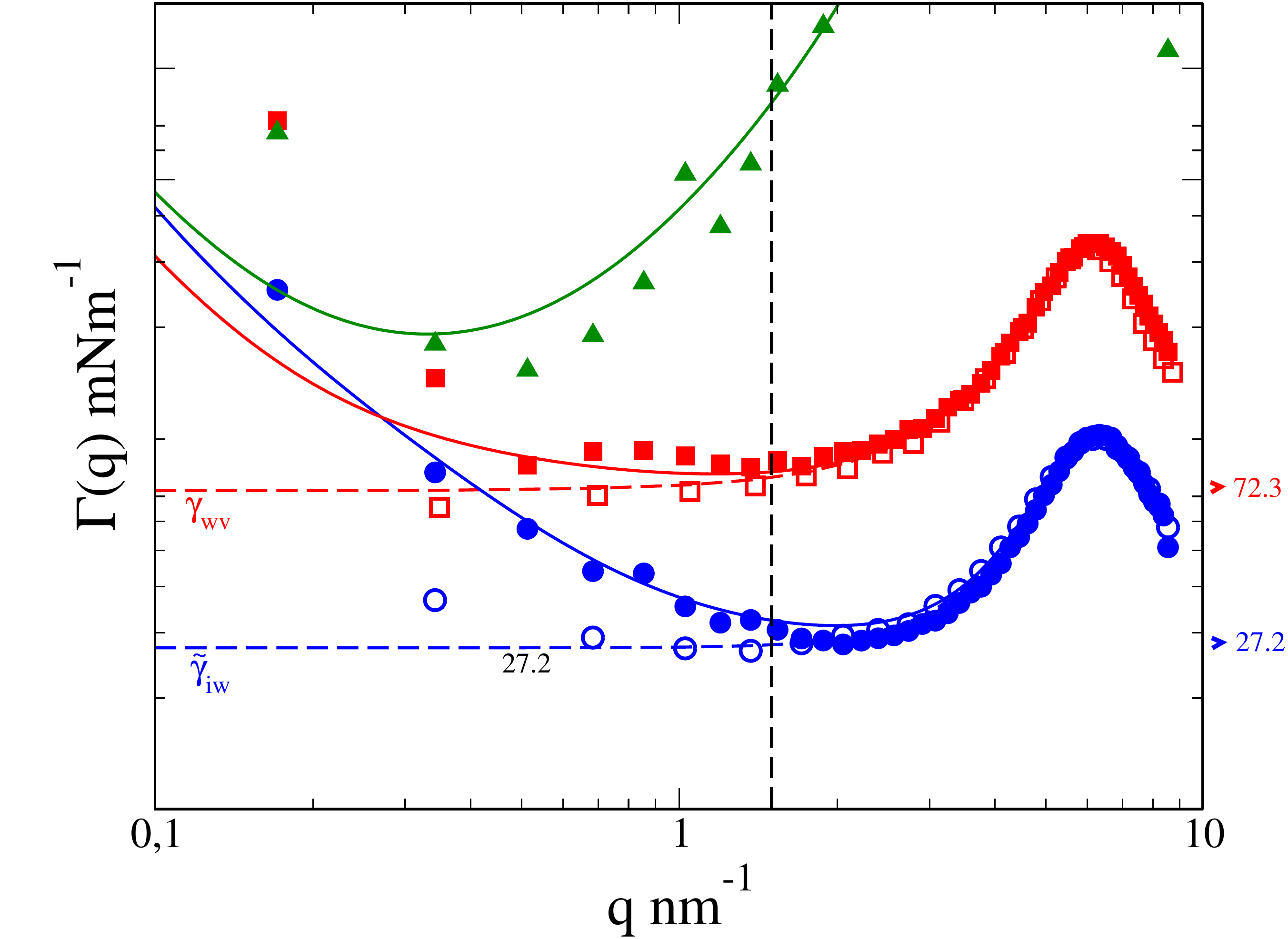}}}  \\
\subfloat[]{%
\resizebox{7cm}{!}{\includegraphics{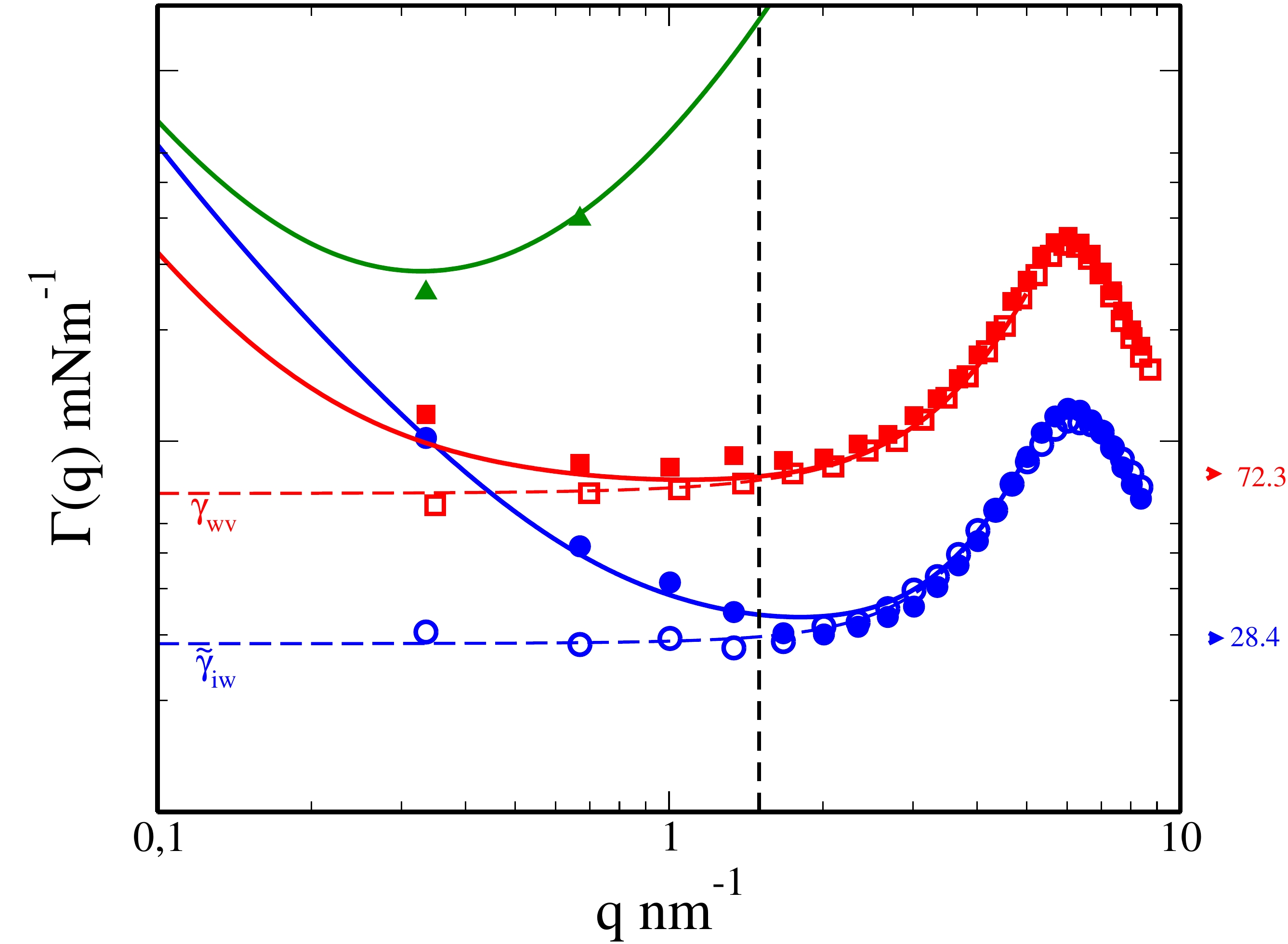}}}
%\hspace{0.5cm}
\subfloat[]{%
\resizebox{7cm}{!}{\includegraphics{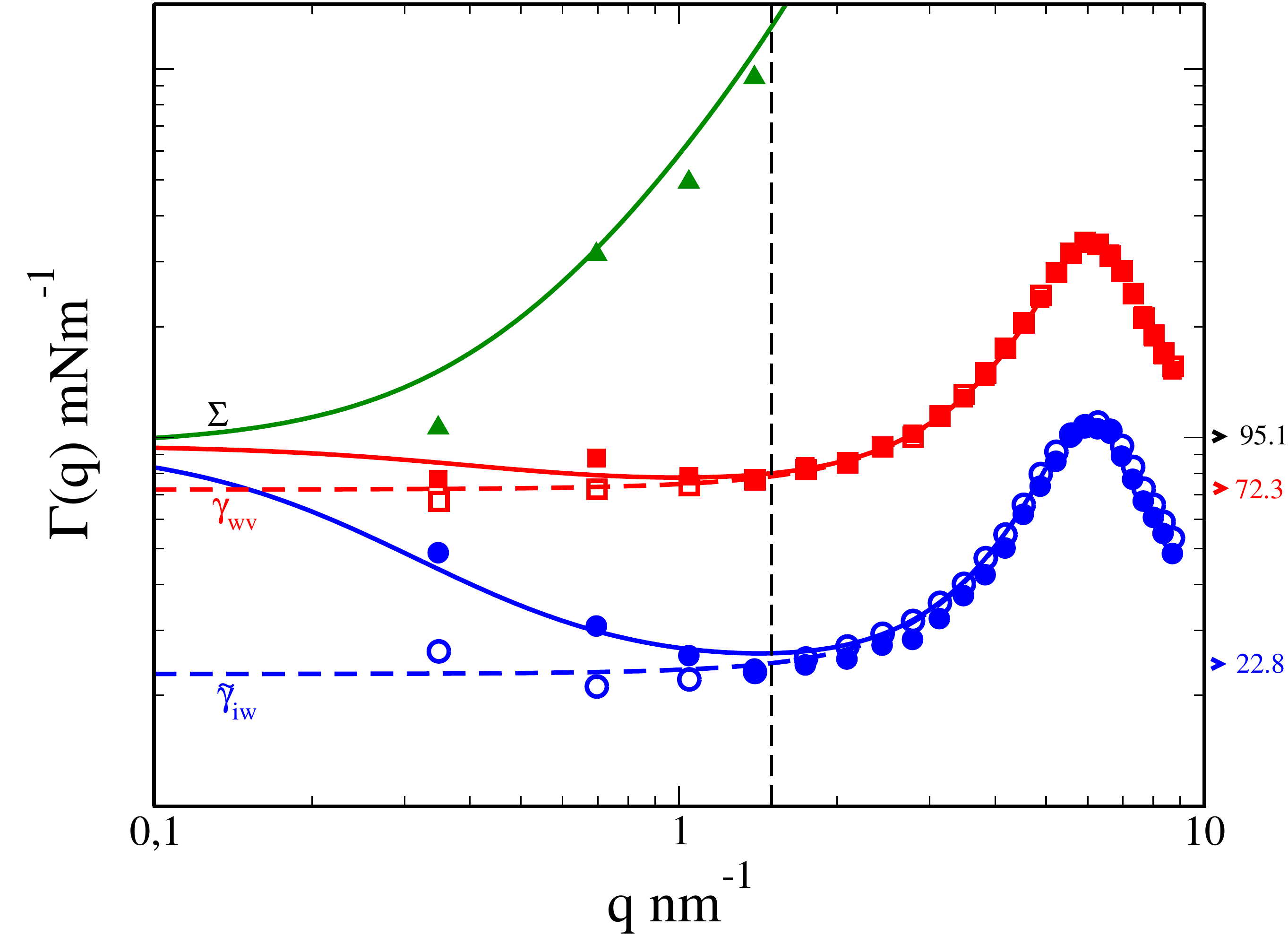}}}
%\subfloat[]{%
%\resizebox{5cm}{!}{\includegraphics{./figuras/hq2vsq-pii-pi-eps-converted-to}}}
 \caption{Plots of $\Gamma(q)$ vs q for several ice--vapour
interfaces studied. (a) (pI)[Basal], (b) (pI)[pII], (c) (Basal)[pII] and (d) (pII)[Basal].
Results pertaining to the ice/vapor interface are shown as filled symbols, with
$\Gamma_{if}(q)$ (blue circles) $\Gamma_{fv}(q)$ (red squares) and
$\Gamma_{if-fv}(q)$ (green triangles). Full lines with the corresponding colour
are fits of the simulation results to the SG-CW model, with parameters of the
fit shown in Table \ref{tab:parametrosagua}. Empty symbols represent stiffness coefficients
for the ice/water (blue circles) and water/vapor (red squares) interfaces. Dashed lines are
corresponding fits to the model of Eq.\ref{ajusteq}. Greek letters close to the y-axis
denote limiting values for the stiffnesses, with
$\Sigma=\tilde\gamma_{iw}+\gamma{wv}$.
}
 \label{fig:hq}
\end{figure}
%%%%%%%%%%%%%%%%%%%%%%%%%%%%%%%%%%%%%%%%%%%%%%%%%%%%%%%%%%%%%%%%%%%%%%%%%%%%%
%%%%%%%%%%%%%%%%%%%%%%%%%%%%%%%%%%%%%%%%%%%%%%%%%%%%%%%%%%%%%%%%%%%%%%%%%%%%%
%%%%%%%%%%%%%%%%%%%%%%%%%%%%%%%%%%%%%%%%%%%%%%%%%%%%%%%%%%%%%%%%%%%%%%%%%%%%%

As discussed previously,  the density profiles of the ice/water and the
ice/vapor interface clearly reveal a considerable degree of surface disorder.
This is apparent in the ice/vapor interface by the presence of a premelted film,
but also, by the inter-penetration of the liquid profile into the solid profile
across several solid layers (Fig \ref{fig:solid}).  This implies that the 
ice/film surface is either rough or has a large density of surface steps.

We note at this stage that our previous
study of the ice/water interface for the TIP4P/2005 model,\cite{benet14c} revealed that all
three basal, pI and pII planes where rough at least up to  the largest
length scale of our simulation box, i.e., about $\lambda=18$~nm.

In Fig. \ref{fig:hq} we plot the effective stiffnesses for the ice/film
and film/vapor surfaces, $\Gamma_{if-if}(q)$, $\Gamma_{fv-fv}(q)$, 
as well as the stiffness coefficient for the crossed correlations
$\Gamma_{if-fv}(q)$. For short, we name these $\Gamma_{if}(q)$,
$\Gamma_{fv}(q)$ and $\Gamma_{iv}(q)$, respectively. Results are
shown for the pI, basal and pII facets. Results for the
ice/water surface from our previous study are also shown for
comparison.\cite{benet14c}  

According to the model of section
\ref{modelo}, for small wave-vectors
the $\Gamma(q)$ remain finite for a rough interface (vanishing $\upsilon$), but diverge if
the interface is smooth (finite $\upsilon$), as indicated in
Table.\ref{limites} Our results for the primary prismatic facet (pI) 
illustrate this clearly. The stiffness coefficients
obtained from fluctuations along the basal direction (Fig.\ref{fig:hq}.a)
seem to converge to a finite constant value equal to the sum
of $\tilde\gamma_{iw}$ and $\gamma_{wv}$, as expected for a rough surface
(the rough and pinned scenario).
On the contrary, the coefficients obtained from fluctuations along
the pII direction appear to diverge, as expected for a smooth interface
in the smooth and pinned scenario (Fig.\ref{fig:hq}.b).
In our previous work,\cite{benet16} we interpreted this observation
as revealing the neighborhood of a {\em roughening} transition on the
pI face of ice very near to the triple point. The difficulty to stabilize
the temperature  to less than tenths of K over the very large simulation 
runs of about half a microsecond could explain the differences observed when studying
the fluctuations on the pI facet along different directions. We can,
however, not discard a very complicated dependence of the results
on the system size. The roughening transition is a fluctuation dominated
 process which is highly geometry dependent. The need to study
quasi-one-dimensional systems to access the low q regime at an affordable
price could potentially have an impact on the results. However, since
in experiments the pI facet has been observed to undergo a roughening
transition very close to the triple point,\cite{elbaum91,asakawa15} 
we interpret the results obtained for the (pI)[basal] direction as indicative
of a roughening transition for the TIP4P/2005 model.

Together with the results of the pI facet, we show results for
the the stiffness on the basal facet along the pII direction 
(basal)[pII] (Fig.\ref{fig:hq}.c) and the secondary prismatic
facet along the basal direction (pII)[basal] (Fig.\ref{fig:hq}.d).
The results in this case are also difficult to interpret, because
we have studied smaller system sizes which are only starting to
exhibit the low q regime. However, subject to some reservations
the results seem to indicate a divergence of the stiffness coefficients
for the basal facet (smooth and pinned scenario) and convergence to a 
constant value for the
pII facet (rough and pinned scenario). To help interpret these results, we have
indicated with an arrow in the figures
the $q=0$ limit expected from the model for a rough interface (i.e.,
$\tilde\gamma_{iw}+\gamma_{wv}$). A smooth extrapolation of the
simulated data (as performed by visual inspection), would seem to indicate that
the results for (Basal)[pII] are larger than this limit,
while those for  (pII)[Basal]  
seem to favor the hypothesis of a completely rough interface. 
This is very much consistent with experimental observations.
Indeed, indications of nucleated or spiral growth on the basal facet
have been reported  in recent years,\cite{inomata17,murata18,murata19} 
confirming
the expectations from crystal growth measurements that the basal
face of ice is smooth up to the triple point.\cite{beckmann82,libbrecht14} 
On the contrary,
the absence of pII facets in ice crystallites suggests a roughening
transition for this face well below the triple
point.\cite{herring51,rottman84,chaikin95}

\begin{figure}
\centering
\subfloat[\text{(Basal)[pII] Ice/water}]{%
\resizebox{2.7in}{!}{\includegraphics{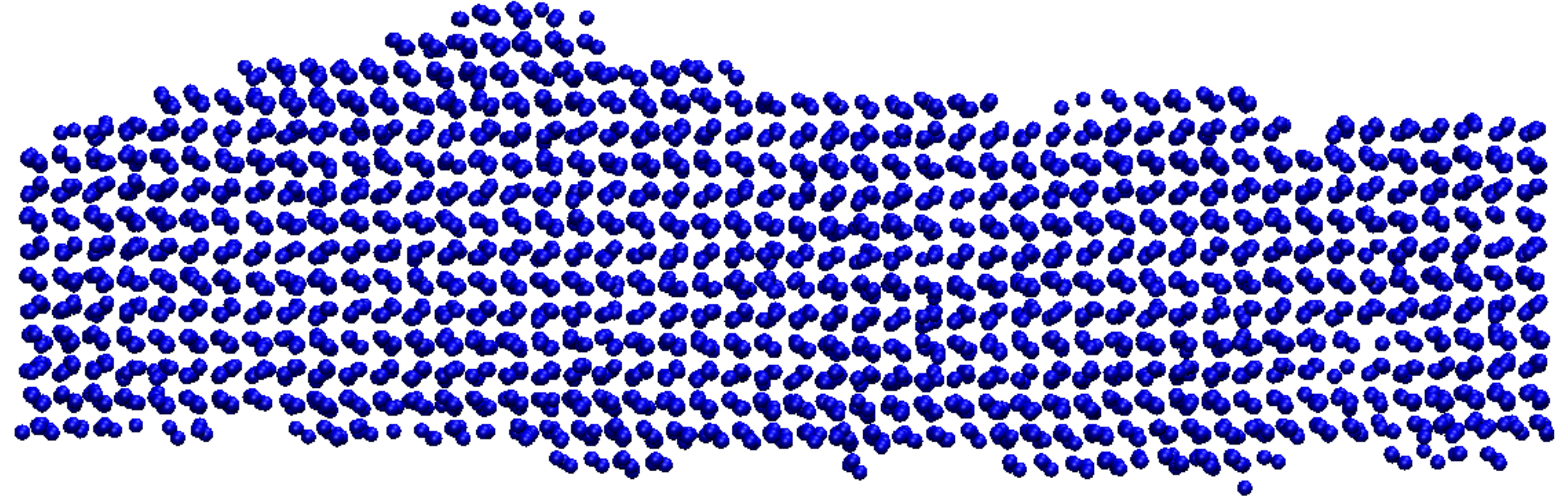}}}
\subfloat[\text{(Basal)[pII] Ice/vapor}]{%
\resizebox{2.7in}{!}{\includegraphics{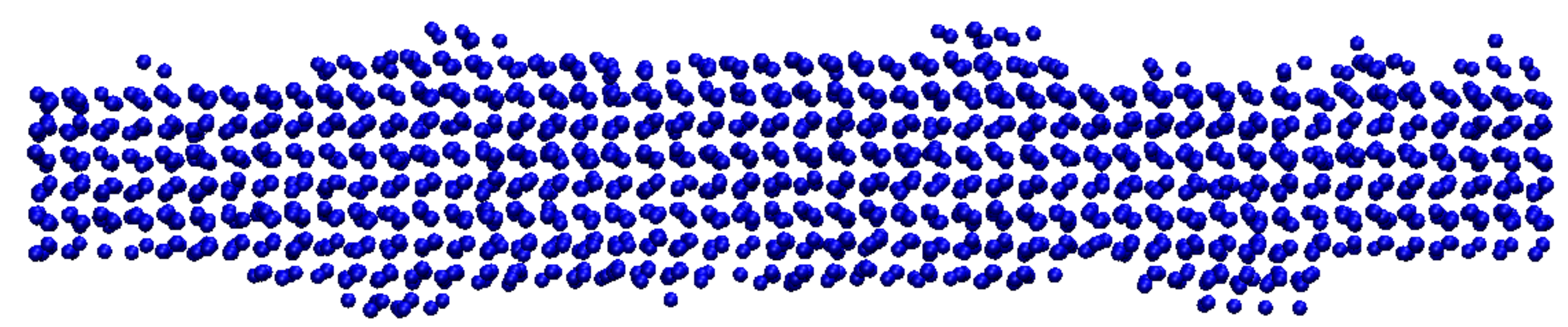}}}  \\
%\caption{\scriptsize (Basal)[pII] Ice/vapor}
\subfloat[\text{(pI)[Basal] Ice/water}]{%
\resizebox{2.7in}{!}{\includegraphics{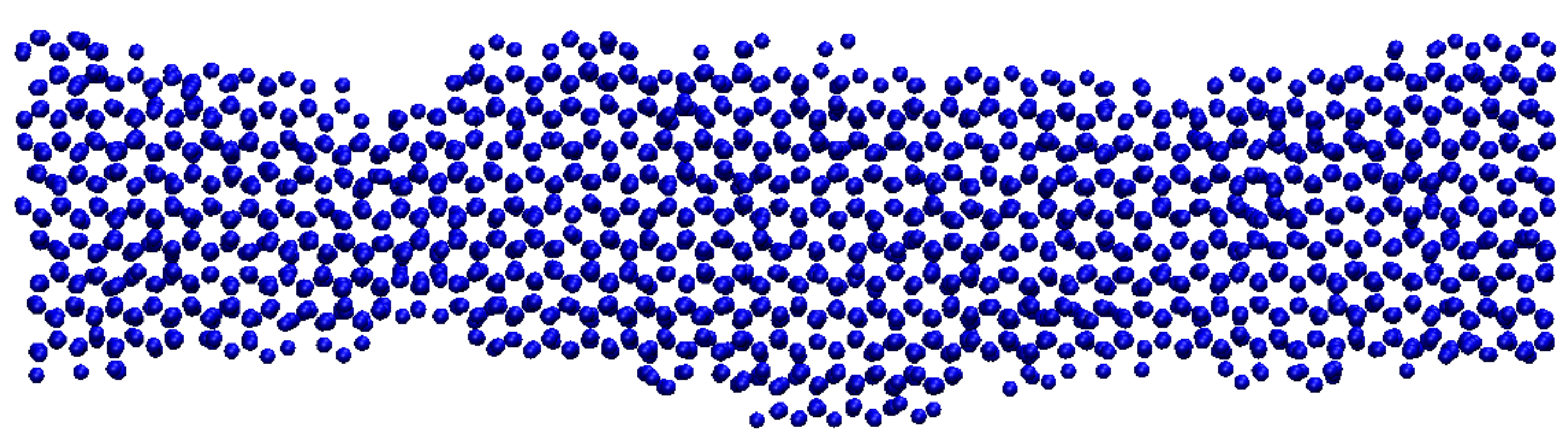}}}
%\caption{\scriptsize (pI)[Basal] Ice/water}
\subfloat[\text{(pI)[Basal] Ice/vapor}]{%
\resizebox{2.7in}{!}{\includegraphics{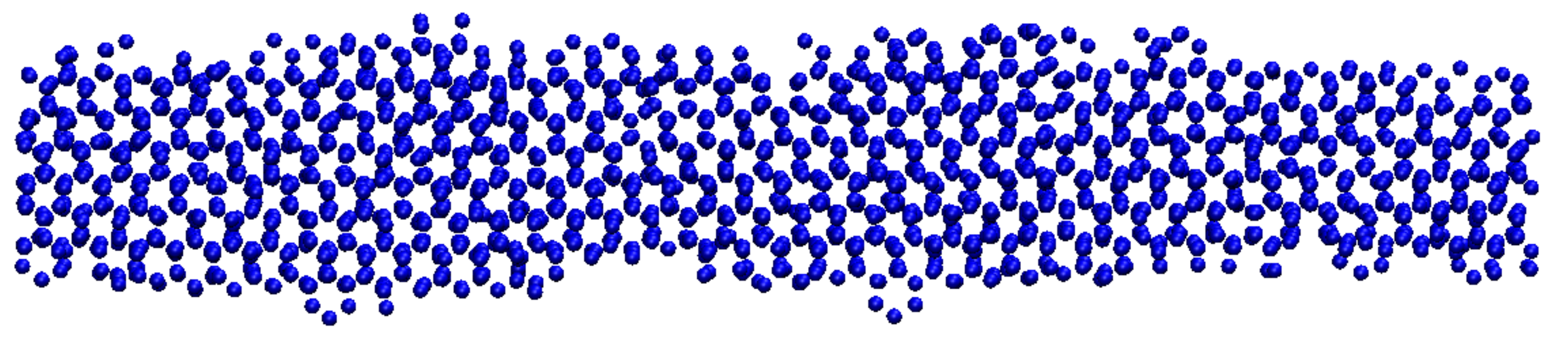}}}  \\
%\caption{\scriptsize (pI)[Basal] Ice/vapor}
\subfloat[\text{(pII)[pI] Ice/water}]{%
\resizebox{2.7in}{!}{\includegraphics{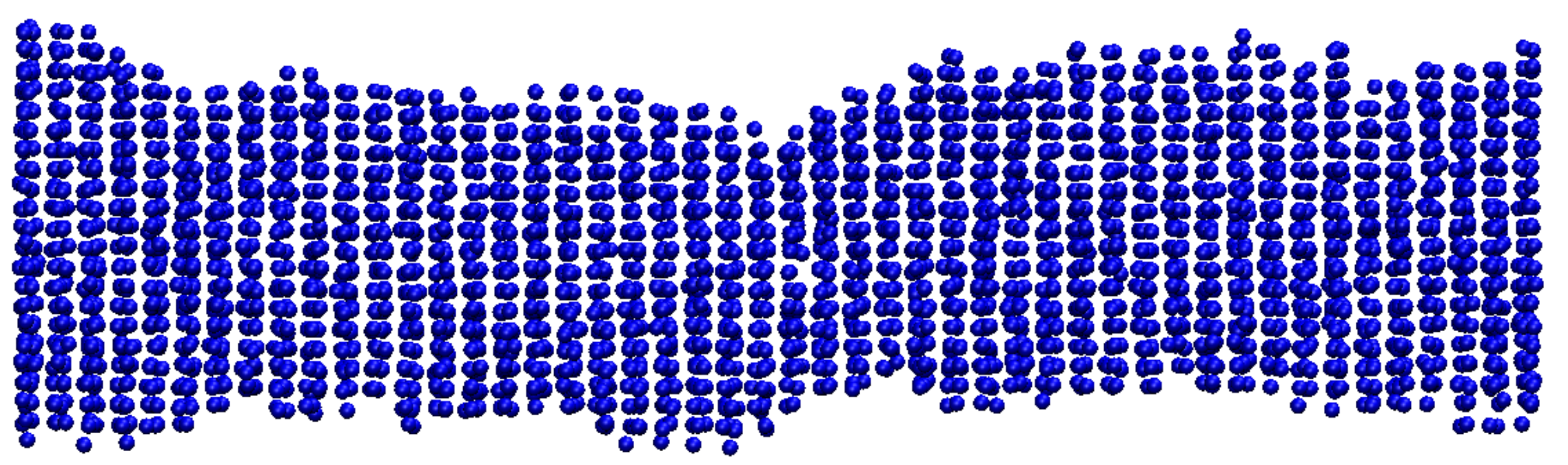}}}
%\caption{\scriptsize (pII)[pI] Ice/water}
\subfloat[\text{(pII)[pI] Ice/vapor}]{%
\resizebox{2.7in}{!}{\includegraphics{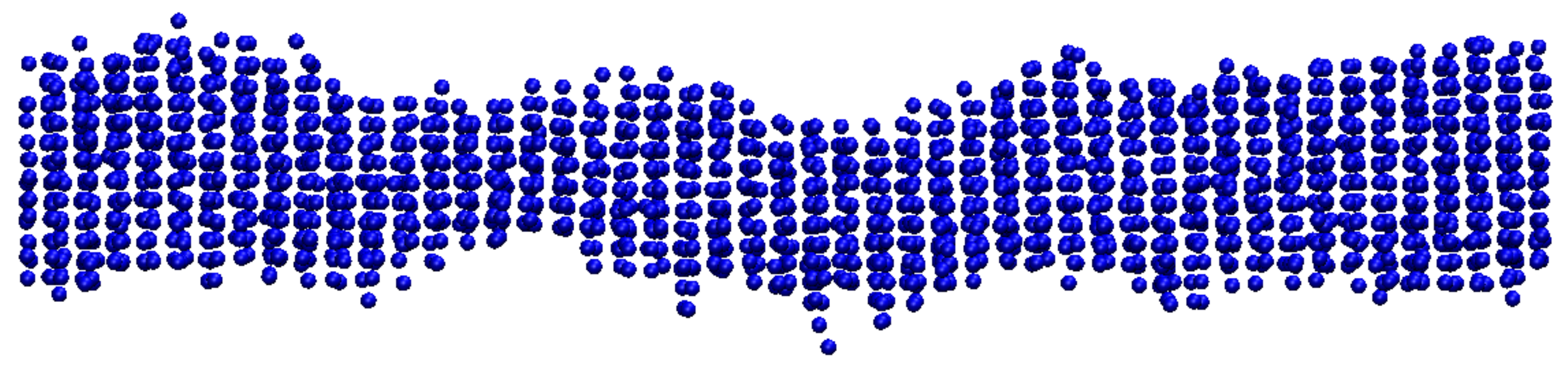}}}
%\caption{\scriptsize (pII)[pI] Ice/vapor}
 \caption{Ice structures for different interfaces.}
 \label{fig:estructuras}
\end{figure}

To see the difference in surface structure in a more intuitive way,
we accompany our results with snapshots taken from the simulations
in Fig. \ref{fig:estructuras}. The results seem to support clearly
the presence of terraces on the basal face, and appear clearly rough
for the pII face, with the pI facet exhibiting a structure somewhat
in between these limits.

Despite these indications, it must be noted that 
observation of a regime with apparently rough behavior 
does not allow to rule out the appearance of a smooth
behavior at length scales larger than the size of our
simulation box, as indicated in Table.\ref{limites}. Indeed, in the
event that the ice/water correlation length is much larger
than the water/vapor correlation length, it is possible to
observe a correlated rough behavior of the full surface before
the attainment of smooth behavior at wave-lengths larger than
$\xi_{iw}$. In fact,
Libbrecht has suggested that the ice interface remains
rough up to fairly large length scales of about 20 unit cells, but that could
eventually become smooth at larger length scales.\cite{libbrecht14}

%%%%%%%%%%%%%%%%%%%%%%%%%%%%%%%%%%%%%%%%%%%%%%%%%%%%%%%%%%%%%%%%%%%%%%%%%%%%%%
%%%%%%%%%%%%%%%%%%%%%%%%%%%%%%%%%%%%%%%%%%%%%%%%%%%%%%%%%%%%%%%%%%%%%%%%%%%%%%

\subsection{Structure at small wave-lengths}

In the previous section we have interpreted the spectrum of fluctuations 
at low wave-vectors on the
basis of the SG-CW model of section \ref{modelo}. We have 
concluded that
at a temperature two Kelvin below the triple point, the basal
facet is smooth, the pII facet is rough and the pI facet is likely very
close to a roughening transition.

We now test whether the model serves as a qualitative description of the
ice/vapor interface at large wave-vectors. We note that, according to 
the model, the effective stiffnesses
for the ice/film and film/vapor surfaces should become equal to
those of the ice/water and water/vapor surfaces, respectively, for large enough
wave vectors (c.f. Table \ref{limites}).

We test this hypothesis by comparing $\Gamma_{if}$ and $\Gamma_{fv}$ with
the stiffness coefficients, $\Gamma_{iw}$ and $\Gamma_{wv}$ obtained in our 
previous work for the ice/water and water/vapor interface,\cite{benet14}
as shown with open symbols in Fig.\ref{fig:hq}.

\begin{figure}
\centering
\resizebox{8cm}{!}{\includegraphics{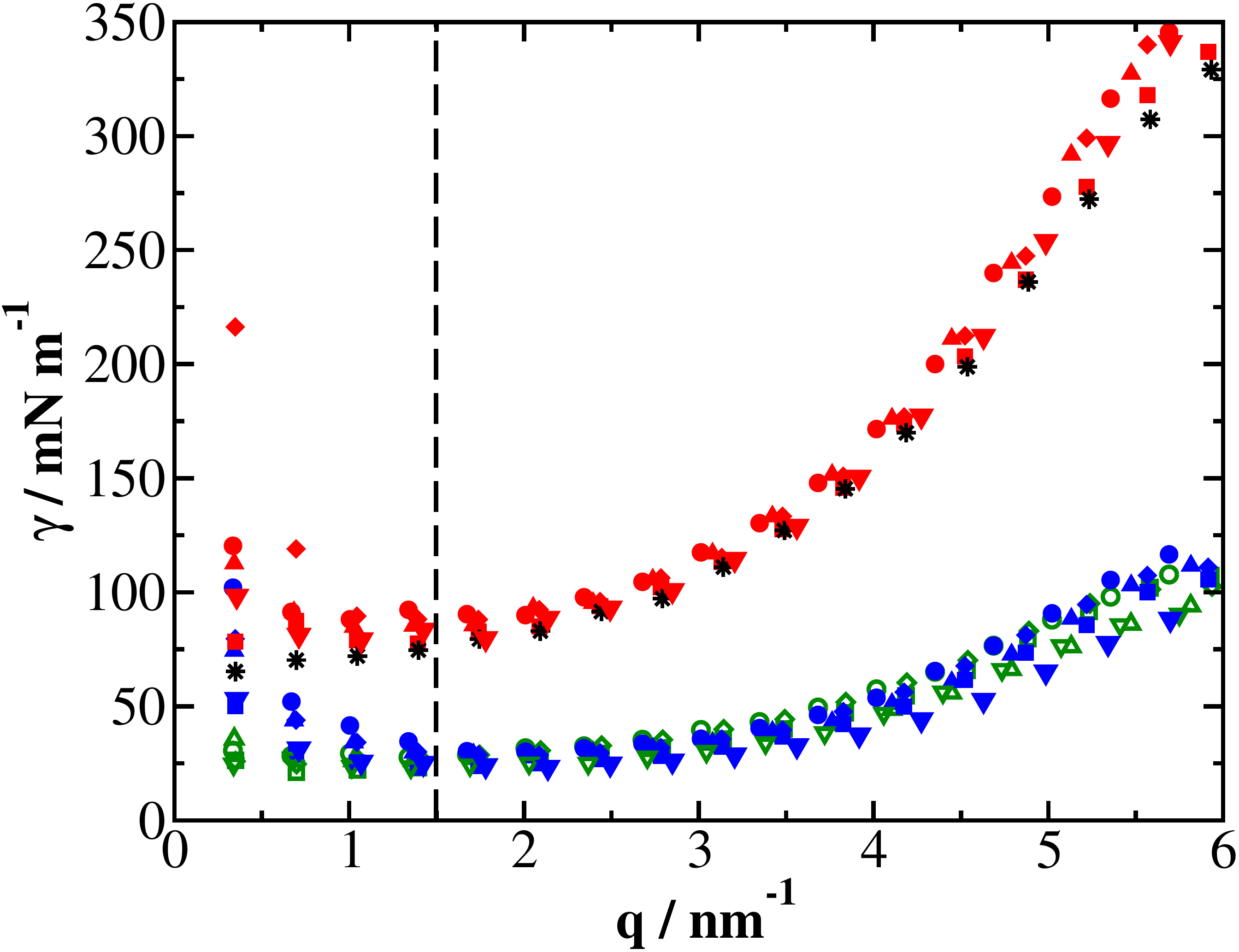}}
\caption{Stiffness of the ice/film (blue solid symbols), ice/water (green empty
symbols), film/vapor (red solid symbols)
and water/vapor (black stars) interfaces as a function of the 
wave-vector for all the
interfaces studied.
(Basal)[pII]: circles, (pI)[Basal]: diamonds, (pI)[pII]: triangles up,
(pII)[Basal]: squares and (pII)[pI]: triangles down. The vertical
dashed line defines the region where the surfaces are uncorrelated.
Note the round parenthesis indicate the plane studied, while
the square brackets correspond to the direction of capillary wave
propagation.
}
\label{fig:comp-stiffness}
\end{figure}

The results show that the stiffness coefficients  of the premelting
film  are indeed very similar to those of water for wave-vectors beyond 
about $q=1.5$~nm$^{-1}$, as summarized for all surface directions studied
in Fig. \ref{fig:comp-stiffness}.  Thus, for wavelengths smaller than
about $\lambda=4$~nm, the  ice surface cannot tell the difference between 
the bulk liquid
phase or the thin premelted film. Similarly, 
 the  liquid/vapor surface
cannot tell whether it limits a bulk liquid phase or a thin premelted film
in contact with bulk ice. Whence, at this length-scale, surface
properties of the bounding premelting film are hardly distinguishable
from those of water. This behavior is consistent with experimental
findings, which report clear signatures of bulk water on the premelting
film at small scales.\cite{smit17}

In terms of surface fluctuations, what this means is that for such wavelengths, the ice/film and
film/vapor surfaces behave independently from each other, and are therefore
uncorrelated. For that reason, the cross correlations
$\langle \left|h_{if}(q)h_{fv}^*\right|\rangle$ decay very fast
and the corresponding stiffness $\Gamma_{iv}(q)$ diverges, as is observed
in  Fig.\ref{fig:hq} (triangles) and is predicted from the analysis
of Eq.\ref{eq:h2surf} and Eq.\ref{eq:defg}
in Table.\ref{limites}

%%%%%%%%%%%%%%%%%%%%%%%%%%%%%%%%%%%%%%%%%%%%%%%%%%%%%%%%%%%%%%%%%%%%

%%%%%%%%%%%%%%%%%%%%%%%%%%%%%%%%%%%%%%%%%%%%%%%%%%%%%%%%%%%%%%%%%%%%
\begin{table}[h]
\footnotesize
\centering
\begin{tabular}{c c c c}
\hline
\hline
Orientation &
$\tilde{\gamma}_{iw} (mN/m)$ & $\tilde{\gamma}_{if} (mN/m)$ &
$\gamma_{fv} (mN/m)$ \\
\hline
(Basal)[pII] & 28.4 & 30.8 & 88.7 \\
(pI)[Basal]  & 26.1  & 30.3 & 87.1 \\
(pI)[pII]    & 27.2  & 29.2 & 84.6 \\
(pII)[Basal] & 22.8 & 24.2 & 79.3 \\
(pII)[pI]    & 24.8 & 25.3 & 79.2 \\
\hline
\hline
\end{tabular}
\caption{Stiffness and interfacial tension of the ice/water and ice/vapor
interfaces. The value of the water/vapor surface tension is
$\gamma_{wv}=72.3mN/m$}
\label{stiffnes-sl}
\end{table}

%%%%%%%%%%%%%%%%%%%%%%%%%%%%%%%%%%%%%%%%%%%%%%%%%%%%%%%%%%%%%%%%%%%%

In order to stress this point in a quantitative manner,
we fit $\Gamma_{if}(q)$ and $\Gamma_{fv}(q)$ for $q>1.5$~nm$^{-1}$
to the expression given by the
capillary wave theory \cite{meunier87,mecke99b,blokhuis09}:
\begin{equation}
\widetilde{\gamma}(q)=\widetilde{\gamma}_0 + \kappa q^2 + c q^4
\label{ajusteq}
\end{equation}
where $\kappa$ is known as the bending rigidity coefficient, and $c$
a fitting parameter with no particular physical significance.

Table \ref{stiffnes-sl} collects the data obtained this way, together with
the corresponding results obtained for the ice/water interface.\cite{benet15b}
Note that the results reported here correspond to direct
extrapolation of $\Gamma_{if}(q)$ to $q\to 0$. This provides  the
{\em interfacial stiffness coefficients} and not to the surface
tensions. The stiffness coefficients reported in the table are  suitably
analysed in Ref.\cite{benet15b} and show very good agreement with interface
 tension measurements from the mold integration technique.\cite{espinosa16}

The estimates obtained from the fluctuations of the premelting film  are in
reasonable agreement with the data for the independent interfaces, albeit
systematically too large.

A closer inspection shows that in fact this could
have been expected from our model. Indeed, since the stiffness coefficients are
actually not constants, but rather, are $q$ dependent, it follows that the
bending rigidity coefficients feed into the apparent stiffness 
so that, in fact, the fits yield the stiffness coefficients up to a
constant small factor of:
\begin{equation}
   1 + \frac{g''}{\gamma_{iw}\gamma_{wv}} ( \kappa_{iw}+\kappa_{wv})
\end{equation}
Since the bending rigidities are positive, and we expect $g''$ also to be small
but positive, this factor is larger than unity, and explains the somewhat larger
coefficients obtained in Table \ref{stiffnes-sl}. 

Additionally, this seems to be consistent with claims that
the film-vapor surface tension of an adsorbed film is expected to depend 
on the film thickness by a factor that is proportional to $\xi_b^2 g''$ (with
$\xi_b$ the bulk correlation length),
as shown recently,\cite{macdowell13,benet14c,macdowell14,macdowell18}.
Since $g''>0$, this term could 
account for an enhancement of
$\gamma_{fv}$ with respect to the water-vapor surface tension $\gamma_{wv}$
pertaining to semi-infinite amounts of bulk phase. Notice that
the need for a film thick dependent surface tension was also
suggested recently.\cite{murata16}

\section{Quantitative test}

In the two preceding sections we have shown that the qualitative results
of the phenomenological SG-CW model allow us to interpret the results 
obtained from simulations. 
A crucial issue as regards the surface structure is whether the
interface is overall smooth ($\upsilon$ finite) or rough ($\upsilon=0$),
which can be elucidated from the behavior of the effective stiffnesses
in the $q\to 0$ limit.
Unfortunately, the low wave-vector regime is
achieved very slowly, so there remains some uncertainty as to the
conclusions reached from visual extrapolation of the limited simulation
data. 

In our study, we have found that the SG-CW model not only provides a
qualitative framework to interpret the interfacial behavior;
it also provides a rather reasonable quantitative account of the
simulation results. This allows us to provide  estimates
of the phenomenological parameters $\upsilon$ and $g''$, and  also
allows us to exploit the intermediate wave-vector data in order
to determine whether the surface is rough or smooth in a consistent
and controlled manner. To show this, we performed  fits
of the stiffness coefficients to the SG-CW, with $\upsilon$
used either as a fitting parameter or fixed at $\upsilon=0$.
For (pI)[Basal] and (pII)[Basal], the constrained fit with $\upsilon=0$
provided the smallest squared deviations, while for (pI)[pII] and
(Basal)[pII], the smallest deviations were obtained using
a finite value for $\upsilon$.  Best fits to the simulation data are 
shown in Fig.\ref{fig:hq}. As illustrated in the figure (notice the
double logarithmic scale of the plot), the model provides a systematic
means for extrapolation of the results into the low $q$ region, and
supports the conclusions obtained previously by visual inspection.

Table \ref{tab:parametrosagua} collects the model parameters of the
best fits displayed on Fig.\ref{fig:hq}. The data can be exploited
to provide  estimates for very relevant surface properties that are
otherwise extremely difficult to obtain from computer simulation.
Firstly, we can trivially estimate the parallel correlation length of 
ice/film and film/vapor surfaces, since
$\xi_{if}=(\gamma_{iw}/\upsilon)^{1/2}$ and
$\xi_{fv}=(\gamma_{wv}/\upsilon)^{1/2}$.  
For smooth surfaces on a mono-crystal, growth  
proceeds by birth and spread of 2-d critical nuclei that are just one
lattice spacing higher than the flat facet. The parallel correlation length
$\xi_{if}$ dictates the range of decay of the ice/film profile, so that
the average slope of this profile at the edge of a pancake like nucleus
is given by $\approx b/\xi_{if}$. In order to clearly identify a terrace
in a molecular simulation, we would need a simulation box which is
several times larger than $\xi_{if}$ in the lateral direction. The table shows
that the Basal facet has a parallel correlation length of about
two nanometers, so that properly formed terraces can only be identified
in a simulation box with lateral dimensions of about 10 nanometers. This
explains why it appears to be so difficult to identify such structures
in computer simulations of the Basal facet.
Also $\xi_{fv}$ is a relevant property, since it determines the
length-scales over which defects on the ice-film surface are {\em healed}
by the premelting film.\cite{degennes04} Particularly, the liquid-film
heals irregularities of the ice surface with a length-scale smaller than
$\xi_{iv}$, but will adapt to irregularities which occur on a length-scale
larger than $\xi_{iv}$. For smooth realizations of the ice surface,
Table \ref{tab:parametrosagua} shows that $\xi_{if}$ is somewhat larger 
but of the same order of magnitude as $\xi_{fv}$, so we expect that
the profile of terraces formed on the ice-film surface will be followed
rather faithfully by the film-vapor surface. This explains why steps
are observed consistently on the basal surface of ice by optical microscopy,
despite the presence of a premelting 
film.\cite{sazaki12,inomata17,murata18,murata19}

Of even greater significance is the possibility to estimate
step free energies. Indeed, we expect that the step free energy, $\beta$,
 describing
the free energy cost of the edge on a terrace will be of the order
$u\xi_{if}$. For the sine-Gordon model, in fact,
$\beta=\frac{2b^2}{\pi^2}(\gamma_{iw}\upsilon)^{1/2}$.\cite{nozieres87}
As seen in Table  \ref{tab:parametrosagua}, estimates
using this formula   are found to be about $3\cdot 10^{-13}$~J/m, which
is the same order of magnitude as results reported by
Libbrecht from crystal growth measurements, 
$\beta\approx 8\cdot 10^{-13}$~J/m,\cite{libbrecht12} 
and a factor of 10 smaller than very recent estimates by
Murata et al. by direct observation of 2-d nucleation rates,
$\beta\approx 9\cdot 10^{-12}$~J/m.\cite{murata19}
\begin{table}
\footnotesize
\centering
\begin{tabular}{l c c c c c c}
\hline
\hline
Surface & $b$/nm & $\upsilon\cdot 10^{-15}$/J$\cdot$m$^{-4}$  
        &      $g''\cdot 10^{-15}$/J$\cdot$m$^{-4}$ 
        & $\xi_{if}$/nm
& $\beta\cdot 10^{12}$/Jm$^{-1}$ & $\xi_{fv}$/nm  \\
\hline
(Basal)[pII] & 0.37 & 6.5 & 6.0   & 2.1       & 0.37 & 3.5 \\
(pI)[Basal]  & 0.40 & 0   & 8.7   & $\infty$  & 0   & 2.9 \\
(pI)[pII]    & 0.40 & 3.3 & 7.3   & 2.9       & 0.30 & 3.2 \\
(pII)[Basal] & 0.22 & 0   & 3.6   & $\infty$  & 0   & 4.5 \\
\hline
\hline
\end{tabular}
\caption{Values of the structural properties of the ice/vapor interface
for the TIP4P/2005 model as extracted from fits to Eq.\ref{eq:h2surf} and
\ref{eq:defg}.}
\label{tab:parametrosagua}
\end{table}

\section{Surface Melting?}

Whereas it is quite clear from our simulations that there is a thin premelting
layer of about one nanometer thickness, it is far more difficult to confirm whether
this surface film diverges as the triple point is approached, i.e., whether the
system undergoes surface melting or not.  In principle, this could
be elucidated by measuring the premelting thickness as a function of
temperature, followed by suitable extrapolation.\cite{conde08,limmer14}
However, the test requires large systems, a
very precise knowledge of the triple point and also a fine control of the
temperature, which is difficult to achieve for inhomogeneous systems. 

In practice, the existence of a surface melting transition will depend 
essentially on the behavior of the interface potential, $g(h)$, at large 
film thickness. An interface potential decaying to zero with negative slope,
implies a repulsion of the film/vapor interface from the ice/film surface, and a propensity to surface
melt. On the contrary, if $g(h)$ decays with positive slope, then an increase in
film thickness is penalized, and the film will remain finite at
coexistence.\cite{schick90} 
In the absence of an explicit interface
potential calculated for this system, we can nevertheless make a discussion
based on general concepts of wetting and intermolecular forces.

Exactly how does $g(h)$ decay will
depend on the nature of the long range dispersion interactions. A precise 
account of such interactions in computer
simulations is actually  completely beyond present state of the
art, because they result from quantum
fluctuations of the electromagnetic field all the way from x-ray to infrared
frequencies, plus static contributions as well. Accordingly, an accurate representation of the van der Waals forces
would require to account for polarization effects in this whole range.
 A more appropriate theoretical framework is the
quantum field theory of
Dzyaloshinskii, Lifshitz and Pitaevskii (DLP), 
which allows us to describe these interactions from the
known dielectric response of the media
involved.\cite{dzyaloshinskii61,safran94,parsegian06}

Because the full expression of the DLP theory is fairly difficult to
interpret qualitatively, it is best to discuss the results separately
for film thickness that are either small (non-retarded interactions) or large
(retarded interactions) compared to the distance
traveled by light at ultraviolet frequencies.

For non-retarded interactions, the long range contribution to
the interface potential has the form:\cite{parsegian06,safran94,israelachvili91}
\begin{equation}
g_{lr}(h) = -\frac{A_{\omega=0}+A_{\omega>0}}{12\pi h^2}
\end{equation} 
where $A_{\omega=0}$ and $A_{\omega>0}$ are thermal and athermal contributions 
to the total
Hamaker constant, $A=A_{\omega=0}+A_{\omega>0}$. The first term depends only on the static
dielectric response of the media. 
It includes purely classical thermal averaging of the dipole fluctuations,
such as Keesom plus Debye type
interactions between the molecules. It  reads:
\begin{equation}
A_{\omega=0} = \frac{3 k_BT}{4}
 \frac{(\epsilon_i(0)-\epsilon_w(0))(\epsilon_v(0)-\epsilon_w(0))}
      {(\epsilon_i(0)+\epsilon_w(0))(\epsilon_v(0)+\epsilon_w(0))}
\end{equation} 
where $\epsilon_i(0)$, $\epsilon_w(0)$, $\epsilon_v(0)$ are the static 
dielectric
constants of the ice, water and vapor phases.

The second term stems from the frequency dependent dielectric response and has a
purely quantum mechanical origin:
\begin{equation}
A_ {\omega>0}=  \frac{3\hbar}{4\pi}
 \int_0^{\infty}  \frac{(\epsilon_i(i\omega)-\epsilon_w(i\omega))
                  (\epsilon_v(i\omega)-\epsilon_w(i\omega))}
           {(\epsilon_i(i\omega)+\epsilon_w(i\omega))(\epsilon_v(i\omega)+\epsilon_w(i\omega))}\,
d\omega
\end{equation} 
where know, the integrand involves the dielectric response as a function of
imaginary frequencies.

The static dielectric response of ice is larger than that of
water,\cite{israelachvili91}
$\epsilon_i(0)\approx 91.5$, $\epsilon_w(0)\approx 88.2$ and $\epsilon_v(0)\approx 1$, so that
$A_{\omega=0}$ is negative, and favors wetting of water on ice at the triple
point.\cite{elbaum91b}
At finite imaginary frequencies, the dielectric constant of
ice is smaller than that of water from the micro-wave to the
visible, but becomes larger again in the extreme-ultraviolet.\cite{elbaum91b}
In practice, the high frequency dielectric response dominates
the behavior of the Hamaker constant at small distances. Using
DLP theory, with dielectric data reported by Elbaum and
Schick as input, we find  $A_{\omega=0}+A_{\omega>0}=-4.1\cdot 10^{-22}$~J,
with a static contribution $A_{\omega=0}=-5.1\cdot 10^{-23}$~J that
accounts for about 10\% of the total Hamaker constant (c.f.
Ref.\citep{limmer14,limmer16}).

Actually, the above results are only
appropriate for film lengths that are smaller than the typical 
distance traveled by the electromagnetic field in the extreme ultraviolet 
region. 
For thicker films, at the nanometer range, retardation effects
cause the intermolecular forces to decay at a faster rate which
asymptotically is of the order $\approx h^{-3}$. More importantly, the constant
governing this decay depends on the visible and infra-red dielectric response
of the material and can have completely different sign from
$A_{\omega=0}$.\cite{dzyaloshinskii61} 
Elbaum and Schick carried a detailed analysis of the high frequency
dielectric response, and concluded that, because of retardation effects, the
interface potential at large distances takes a positive slope,
and therefore develops a very shallow minimum at about 36~\AA, so that, 
under the action of long range forces alone, water does not wet ice at the 
triple point.\cite{elbaum91b} 

Whence, contrary to erroneous claims by Limmer,\cite{limmer16} the origin
of incomplete surface melting of ice is not at all related to the
static Hamaker constant $A_{\omega=0}$, which is used with the
erroneous sing convention in Ref.\citep{limmer14,limmer16}, but rather, to
retarded interactions that are dominated by the visible and infrared
terms of the dielectric response and decay as $\approx h^{-3}$.

 Yet, these arguments leave completely aside the
possibility of pure Coulombic interactions that could result from a net charge
at the ice/water and water/vapor surface.\cite{wilen95} This can occur 
even in pure
water, because the broken symmetry of the interface promotes a small but
significant net charge transfer between molecules.\cite{lee12}

This discussion serves to understand how difficult it is to clarify the 
issue of
surface melting by using standard computer simulations. In all such studies to
date, rigid point charge models have been employed at
best.\cite{conde08,limmer14,pickering18}
Unfortunately, such models are not polarizable and do not properly account 
for the 
dielectric response, as they usually predict  a smaller static 
dielectric constant 
for ice than for water.\cite{rick03,rick05,macdowell10,aragones11} As a result, 
the sign of $A_{\omega=0}$ is inverted with respect to expectations based on the
experimental dielectric constant. 

This issue cannot be remedied by considering
the $A_{\omega>0}$ contribution. Indeed, that term, which stems from quantum
mechanical fluctuations of the electric field at extreme ultraviolet 
frequencies, 
is accounted in classical
simulations by the dispersive $-4\epsilon(\sigma/r)^{6}$ contribution of the Lennard-Jones
potential. Whence, $A_{\omega>0}$  depends only on the difference between the bulk
densities of the phases involved, and is given by:\cite{dietrich86,dietrich91}
\begin{equation}
 A_{\omega>0} =  4\pi^2 \sigma^6 \epsilon (\rho_i - \rho_w)(\rho_v-\rho_w)
\end{equation} 
For water, it is well known that $\rho_w>\rho_i>\rho_v$,
so that  $A_{\omega>0}$ is also positive and accordingly does not favor surface 
melting either. 

In fact, the latter contribution cannot be described accurately in computer
simulations, since, in practice, dispersive interactions are truncated beyond a
nanometer or so. Whence, long range dispersive interactions are replaced by 
short range truncated interactions, which produce an exponential rather than an
algebraic decay, and correspond to the complete neglect of $A_{\omega>0}$ beyond
the cutoff distance. 

A plausible assumption is that rigid point charge models will therefore not
yield surface melting, since they are dominated by the incorrect $A_{\omega=0}$
term stemming from the permanent dipoles. Unlike dispersive terms, such long
range contribution is indeed taken into account explicitly in simulations 
via the Ewald summation.  Possibly, however, the presence of a 
nanometer thick
premelting film close to the triple point growing as temperature is raised
towards the triple point,\cite{conde08} indicates that the short range
structural forces
in water do promote surface melting. This hypothesis is further supported by
recent simulations, where a coarse grained model of water that only enforces
short range tetrahedral correlations, but ignores charges and dispersion, has
shown convincingly to exhibit (short range) surface melting.\cite{limmer14} 
Assuming that 
point charge models exhibit short range structural forces of similar nature, one then
concludes that a thick premelting film must form at the triple point. Whether
such film exhibits a divergence or not in real water will be then dictated by 
the nature of
the long range dispersive interactions, which, at best, can be described within
the DLP framework and seem to indicate incomplete surface 
melting.\cite{elbaum91b} Computer simulations do seem to indicate
reliably from  the extent of
the prewetting films of our model,  that the range where short range structural
 forces
dominate over the Hamaker contribution is at least at the nanometer range. This
implies that the minimum estimated from consideration of long range forces
alone at about 36~\AA,\cite{elbaum91b} is most likely located at 
longer distances, and will  play a significant role at temperatures 
very close to the triple point.

This scenario is supported by recent experimental observations, which
indicate that thick but bound films of up to  9~nm can form on saturated
ice surfaces
close to the triple 
point.\cite{sazaki12,asakawa15,asakawa16,murata16} 
Unexpectedly, the experiments have also
revealed that these thick films grow discontinuously, as in a first
order  thin-to-thick surface phase transition, of the kind observed previously 
for alkanes adsorbed on water and polymer on silicium.\cite{shahidzadeh98,
bertrand00,mueller01,macdowell05}
This scenario implies the presence of two minima of the interface
potential,\cite{murata16} and currently, there is no known theoretical 
explanation to account for the origin of this additional minimum. 
Clearly, much greater efforts will be needed to fully elucidate the
intriguing surface behavior of ice close to the triple point.

\section{Conclusions}

In this paper we have studied the structure of the solid/vapor
interfaces of water in the neighborhood of the triple point for the TIP4P/2005
model. Our results show that the three most important planes of ice, basal,
primary prismatic and secondary prismatic exhibit a thin premelted liquid layer
of about 0.9 nm. This implies that the ice/vapor
interface may be described in terms of two additional surfaces, separating the
premelting layer from the bulk solid and bulk vapor phases. We have studied the
fluctuations of these two surfaces, and analyzed them in terms of a
simple phenomenological model of coupled surface fluctuations.

Our results suggest that the ice/film surface is smooth for the basal facet, it
is rough for the secondary prismatic facet, and very close to the  roughening
transition 2 K away from the triple point for the primary prismatic facet.
The basal facet has
a step free energy of about $10^{-13}$~$J/m$, in reasonable agreement
with experiments.\cite{libbrecht13,murata19} The film/vapor surface
at large wave-lengths is strongly correlated to the ice/film surface,
and exhibits qualitatively similar fluctuations. Our study reveals
that it is possible to have a highly disordered outer layer of premelted
molecules below a smooth and faceted surface. Furthermore, it shows
that the liquid like premelting film can adapt to the shape of the
steps on the surface, thus explaining why a premelting film can
be consistent with the observation of steps and terraces 
in experiments.\cite{sazaki12,asakawa15,asakawa16,murata16}

We test the phenomenological model by comparing the ice/film and film/vapor
fluctuations with those obtained from independent simulations of the
 ice/water and water/vapor interfaces. The results clearly demonstrate a
crossover from a microscopic regime of length scales smaller than about 4.5~nm,
where the premelted surfaces fluctuate independently, and nearly as those
of the ice/water and water/vapor interfaces. For larger wavelengths, on the
contrary, the two surfaces become correlated and behave as a single interface.
When the facet becomes rough, fluctuations become governed
by a stiffness which is close to, but somewhat larger
than the sum of the ice/water stiffness and the water/vapor surface tension,
whence, about $100$~$mJ/m^2$.

Our results lend support to a recent study, which indicates
that the crystal growth of ice crystallites in
 either bulk water or
bulk vapor follows a similar mechanism \cite{libbrecht14,murata18}. 
This implies, for the
crystal growth in the vapor phase, a process that is limited by the
crystallization of water molecules within the premelted liquid film.

On the contrary, we note that present state of the art computer simulations
have not reached the level of accuracy required to elucidate
accurately the problem of surface melting on ice.

\section{Acknowledgments}
The authors would like to thank A. Michaelides and B. Slater for providing
us a copy of Ref.\citep{slater19} prior to publication, as well as Eva G. Noya for helpful
discussions. We also wish to thank the Agencia Estatal de Investigaci\'on and
Fondo Europeo de Desarrollo Regional (FEDER) 
for research grant FIS-89361-C3-2-P. P.L. wishes
to acknowledge additional support  under grant FIS2015-72946-EXP. 

\bibliography{referenc}

%merlin.mbs apsrev4-1.bst 2010-07-25 4.21a (PWD, AO, DPC) hacked
%Control: key (0)
%Control: author (0) dotless jnrlst
%Control: editor formatted (1) identically to author
%Control: production of article title (0) allowed
%Control: page (1) range
%Control: year (0) verbatim
%Control: production of eprint (0) enabled
\begin{thebibliography}{108}%
\makeatletter
\providecommand \@ifxundefined [1]{%
 \@ifx{#1\undefined}
}%
\providecommand \@ifnum [1]{%
 \ifnum #1\expandafter \@firstoftwo
 \else \expandafter \@secondoftwo
 \fi
}%
\providecommand \@ifx [1]{%
 \ifx #1\expandafter \@firstoftwo
 \else \expandafter \@secondoftwo
 \fi
}%
\providecommand \natexlab [1]{#1}%
\providecommand \enquote  [1]{``#1''}%
\providecommand \bibnamefont  [1]{#1}%
\providecommand \bibfnamefont [1]{#1}%
\providecommand \citenamefont [1]{#1}%
\providecommand \href@noop [0]{\@secondoftwo}%
\providecommand \href [0]{\begingroup \@sanitize@url \@href}%
\providecommand \@href[1]{\@@startlink{#1}\@@href}%
\providecommand \@@href[1]{\endgroup#1\@@endlink}%
\providecommand \@sanitize@url [0]{\catcode `\\12\catcode `\$12\catcode
  `\&12\catcode `\#12\catcode `\^12\catcode `\_12\catcode `\%12\relax}%
\providecommand \@@startlink[1]{}%
\providecommand \@@endlink[0]{}%
\providecommand \url  [0]{\begingroup\@sanitize@url \@url }%
\providecommand \@url [1]{\endgroup\@href {#1}{\urlprefix }}%
\providecommand \urlprefix  [0]{URL }%
\providecommand \Eprint [0]{\href }%
\providecommand \doibase [0]{http://dx.doi.org/}%
\providecommand \selectlanguage [0]{\@gobble}%
\providecommand \bibinfo  [0]{\@secondoftwo}%
\providecommand \bibfield  [0]{\@secondoftwo}%
\providecommand \translation [1]{[#1]}%
\providecommand \BibitemOpen [0]{}%
\providecommand \bibitemStop [0]{}%
\providecommand \bibitemNoStop [0]{.\EOS\space}%
\providecommand \EOS [0]{\spacefactor3000\relax}%
\providecommand \BibitemShut  [1]{\csname bibitem#1\endcsname}%
\let\auto@bib@innerbib\@empty
%</preamble>
\bibitem [{\citenamefont {Libbrecht}(2005)}]{libbrecht05}%
  \BibitemOpen
  \bibfield  {author} {\bibinfo {author} {\bibfnamefont {K.~G.}\ \bibnamefont
  {Libbrecht}},\ }\bibfield  {title} {\enquote {\bibinfo {title} {The physics
  of snow crystals},}\ }\href@noop {} {\bibfield  {journal} {\bibinfo
  {journal} {Rep. Prog. Phys.}\ }\textbf {\bibinfo {volume} {68}},\ \bibinfo
  {pages} {855--895} (\bibinfo {year} {2005})}\BibitemShut {NoStop}%
\bibitem [{\citenamefont {Pfalzgraff}\ \emph {et~al.}(2010)\citenamefont
  {Pfalzgraff}, \citenamefont {Hulscher},\ and\ \citenamefont
  {Neshyba}}]{pfalzgraf10}%
  \BibitemOpen
  \bibfield  {author} {\bibinfo {author} {\bibfnamefont {W.~C.}\ \bibnamefont
  {Pfalzgraff}}, \bibinfo {author} {\bibfnamefont {R.~M.}\ \bibnamefont
  {Hulscher}}, \ and\ \bibinfo {author} {\bibfnamefont {S.~P.}\ \bibnamefont
  {Neshyba}},\ }\bibfield  {title} {\enquote {\bibinfo {title} {Scanning
  electron microscopy and molecular dynamics of surfaces of growing and
  ablating hexagonal ice crystals},}\ }\href@noop {} {\bibfield  {journal}
  {\bibinfo  {journal} {Atmos. Chem. Phys.}\ }\textbf {\bibinfo {volume}
  {10}},\ \bibinfo {pages} {2927--2935} (\bibinfo {year} {2010})}\BibitemShut
  {NoStop}%
\bibitem [{\citenamefont {Ball}(2016)}]{ball16}%
  \BibitemOpen
  \bibfield  {author} {\bibinfo {author} {\bibfnamefont {P.}~\bibnamefont
  {Ball}},\ }\bibfield  {title} {\enquote {\bibinfo {title} {Material witness:
  Close to the edge},}\ }\href@noop {} {\bibfield  {journal} {\bibinfo
  {journal} {Nature Mat.}\ }\textbf {\bibinfo {volume} {15}},\ \bibinfo {pages}
  {1060} (\bibinfo {year} {2016})}\BibitemShut {NoStop}%
\bibitem [{\citenamefont {Demange}\ \emph {et~al.}(2017)\citenamefont
  {Demange}, \citenamefont {Zapolsy}, \citenamefont {Patte},\ and\
  \citenamefont {Brunel}}]{demange17}%
  \BibitemOpen
  \bibfield  {author} {\bibinfo {author} {\bibfnamefont {G.}~\bibnamefont
  {Demange}}, \bibinfo {author} {\bibfnamefont {H.}~\bibnamefont {Zapolsy}},
  \bibinfo {author} {\bibfnamefont {R.}~\bibnamefont {Patte}}, \ and\ \bibinfo
  {author} {\bibfnamefont {M.}~\bibnamefont {Brunel}},\ }\bibfield  {title}
  {\enquote {\bibinfo {title} {A phase field model for snow crystal growth in
  three dimensions},}\ }\href@noop {} {\bibfield  {journal} {\bibinfo
  {journal} {npj Comp. Mat.}\ }\textbf {\bibinfo {volume} {3}},\ \bibinfo
  {pages} {15} (\bibinfo {year} {2017})}\BibitemShut {NoStop}%
\bibitem [{\citenamefont {Burton}\ \emph {et~al.}(1951)\citenamefont {Burton},
  \citenamefont {Cabrera},\ and\ \citenamefont {Frank}}]{burton51}%
  \BibitemOpen
  \bibfield  {author} {\bibinfo {author} {\bibfnamefont {W.~K.}\ \bibnamefont
  {Burton}}, \bibinfo {author} {\bibfnamefont {N.}~\bibnamefont {Cabrera}}, \
  and\ \bibinfo {author} {\bibfnamefont {F.~C.}\ \bibnamefont {Frank}},\
  }\bibfield  {title} {\enquote {\bibinfo {title} {The growth of crystals and
  the equilibrium structure of their surfaces},}\ }\href@noop {} {\bibfield
  {journal} {\bibinfo  {journal} {Phil. Trans. R. Soc. Lond. A. Math. Phys.}\
  }\textbf {\bibinfo {volume} {243}},\ \bibinfo {pages} {299--358} (\bibinfo
  {year} {1951})}\BibitemShut {NoStop}%
\bibitem [{\citenamefont {van Beijeren}(1977)}]{vanbeijeren77}%
  \BibitemOpen
  \bibfield  {author} {\bibinfo {author} {\bibfnamefont {Henk}\ \bibnamefont
  {van Beijeren}},\ }\bibfield  {title} {\enquote {\bibinfo {title} {Exactly
  solvable model for the roughening transition of a crystal surface},}\ }\href
  {\doibase 10.1103/PhysRevLett.38.993} {\bibfield  {journal} {\bibinfo
  {journal} {Phys. Rev. Lett.}\ }\textbf {\bibinfo {volume} {38}},\ \bibinfo
  {pages} {993--996} (\bibinfo {year} {1977})}\BibitemShut {NoStop}%
\bibitem [{\citenamefont {Abraham}(1980)}]{abraham80}%
  \BibitemOpen
  \bibfield  {author} {\bibinfo {author} {\bibfnamefont {D.~B.}\ \bibnamefont
  {Abraham}},\ }\bibfield  {title} {\enquote {\bibinfo {title} {Solvable model
  with a roughening transition for a planar ising ferromagnet},}\ }\href
  {\doibase 10.1103/PhysRevLett.44.1165} {\bibfield  {journal} {\bibinfo
  {journal} {Phys. Rev. Lett.}\ }\textbf {\bibinfo {volume} {44}},\ \bibinfo
  {pages} {1165--1168} (\bibinfo {year} {1980})}\BibitemShut {NoStop}%
\bibitem [{\citenamefont {Lipowsky}(1982)}]{lipowsky82}%
  \BibitemOpen
  \bibfield  {author} {\bibinfo {author} {\bibfnamefont {R.}~\bibnamefont
  {Lipowsky}},\ }\bibfield  {title} {\enquote {\bibinfo {title} {Critical
  surface phenomena at first-order bulk transitions},}\ }\href@noop {}
  {\bibfield  {journal} {\bibinfo  {journal} {Phys. Rev. Lett.}\ }\textbf
  {\bibinfo {volume} {49}},\ \bibinfo {pages} {1575--1578} (\bibinfo {year}
  {1982})}\BibitemShut {NoStop}%
\bibitem [{\citenamefont {Fisher}\ and\ \citenamefont
  {Weeks}(1983)}]{fisher83}%
  \BibitemOpen
  \bibfield  {author} {\bibinfo {author} {\bibfnamefont {Daniel~S.}\
  \bibnamefont {Fisher}}\ and\ \bibinfo {author} {\bibfnamefont {John~D.}\
  \bibnamefont {Weeks}},\ }\bibfield  {title} {\enquote {\bibinfo {title}
  {Shape of crystals at low temperatures: Absence of quantum roughening},}\
  }\href {\doibase 10.1103/PhysRevLett.50.1077} {\bibfield  {journal} {\bibinfo
   {journal} {Phys. Rev. Lett.}\ }\textbf {\bibinfo {volume} {50}},\ \bibinfo
  {pages} {1077--1080} (\bibinfo {year} {1983})}\BibitemShut {NoStop}%
\bibitem [{\citenamefont {Rommelse}\ and\ \citenamefont {den
  Nijs}(1987)}]{rommelse87}%
  \BibitemOpen
  \bibfield  {author} {\bibinfo {author} {\bibfnamefont {Koos}\ \bibnamefont
  {Rommelse}}\ and\ \bibinfo {author} {\bibfnamefont {Marcel}\ \bibnamefont
  {den Nijs}},\ }\bibfield  {title} {\enquote {\bibinfo {title} {Preroughening
  transitions in surfaces},}\ }\href {\doibase 10.1103/PhysRevLett.59.2578}
  {\bibfield  {journal} {\bibinfo  {journal} {Phys. Rev. Lett.}\ }\textbf
  {\bibinfo {volume} {59}},\ \bibinfo {pages} {2578--2581} (\bibinfo {year}
  {1987})}\BibitemShut {NoStop}%
\bibitem [{\citenamefont {Ball}\ and\ \citenamefont {Evans}(1988)}]{ball88}%
  \BibitemOpen
  \bibfield  {author} {\bibinfo {author} {\bibfnamefont {P.~C.}\ \bibnamefont
  {Ball}}\ and\ \bibinfo {author} {\bibfnamefont {R.}~\bibnamefont {Evans}},\
  }\bibfield  {title} {\enquote {\bibinfo {title} {Structure and adsorption at
  gas-solid interfaces: Layering transitions from a continuum theory},}\ }\href
  {\doibase 10.1063/1.454827} {\bibfield  {journal} {\bibinfo  {journal} {J.
  Chem. Phys.}\ }\textbf {\bibinfo {volume} {89}},\ \bibinfo {pages}
  {4412--4423} (\bibinfo {year} {1988})},\ \Eprint
  {http://arxiv.org/abs/https://doi.org/10.1063/1.454827}
  {https://doi.org/10.1063/1.454827} \BibitemShut {NoStop}%
\bibitem [{\citenamefont {Lipowsky}(1990)}]{lipowsky90}%
  \BibitemOpen
  \bibfield  {author} {\bibinfo {author} {\bibfnamefont {R.}~\bibnamefont
  {Lipowsky}},\ }\bibfield  {title} {\enquote {\bibinfo {title}
  {Surface-induced disorder and surface melting},}\ }in\ \href@noop {} {\emph
  {\bibinfo {booktitle} {Magnetic Properties of Low-Dimensional Systems II}}},\
  \bibinfo {series} {Springer Proceedings in Physics}, Vol.~\bibinfo {volume}
  {50},\ \bibinfo {editor} {edited by\ \bibinfo {editor} {\bibfnamefont
  {L.~M.}\ \bibnamefont {Falicov}}, \bibinfo {editor} {\bibfnamefont
  {F.}~\bibnamefont {Mejia-Lira}}, \ and\ \bibinfo {editor} {\bibfnamefont
  {J.~L.}\ \bibnamefont {Moran-Lopez}}}\ (\bibinfo  {publisher}
  {Springer-Verlag},\ \bibinfo {address} {Heidelberg},\ \bibinfo {year}
  {1990})\ pp.\ \bibinfo {pages} {158--166}\BibitemShut {NoStop}%
\bibitem [{\citenamefont {Evans}(1992)}]{evans92}%
  \BibitemOpen
  \bibfield  {author} {\bibinfo {author} {\bibfnamefont {R.}~\bibnamefont
  {Evans}},\ }\bibfield  {title} {\enquote {\bibinfo {title} {Density
  functionals in the theory of nonuniform fluids},}\ }in\ \href@noop {} {\emph
  {\bibinfo {booktitle} {Fundamentals of Inhomogenous Fluids}}},\ \bibinfo
  {editor} {edited by\ \bibinfo {editor} {\bibfnamefont {D.}~\bibnamefont
  {Henderson}}}\ (\bibinfo  {publisher} {Marcel Dekker},\ \bibinfo {address}
  {New York},\ \bibinfo {year} {1992})\ Chap.~\bibinfo {chapter} {3}, pp.\
  \bibinfo {pages} {85--175}\BibitemShut {NoStop}%
\bibitem [{\citenamefont {Herring}(1951)}]{herring51}%
  \BibitemOpen
  \bibfield  {author} {\bibinfo {author} {\bibfnamefont {Conyers}\ \bibnamefont
  {Herring}},\ }\bibfield  {title} {\enquote {\bibinfo {title} {Some theorems
  on the free energies of crystal surfaces},}\ }\href {\doibase
  10.1103/PhysRev.82.87} {\bibfield  {journal} {\bibinfo  {journal} {Phys.
  Rev.}\ }\textbf {\bibinfo {volume} {82}},\ \bibinfo {pages} {87--93}
  (\bibinfo {year} {1951})}\BibitemShut {NoStop}%
\bibitem [{\citenamefont {Rottman}\ and\ \citenamefont
  {Wortis}(1984)}]{rottman84}%
  \BibitemOpen
  \bibfield  {author} {\bibinfo {author} {\bibfnamefont {Craig}\ \bibnamefont
  {Rottman}}\ and\ \bibinfo {author} {\bibfnamefont {Michael}\ \bibnamefont
  {Wortis}},\ }\bibfield  {title} {\enquote {\bibinfo {title} {Equilibrium
  crystal shapes for lattice models with nearest-and next-nearest-neighbor
  interactions},}\ }\href {\doibase 10.1103/PhysRevB.29.328} {\bibfield
  {journal} {\bibinfo  {journal} {Phys. Rev. B}\ }\textbf {\bibinfo {volume}
  {29}},\ \bibinfo {pages} {328--339} (\bibinfo {year} {1984})}\BibitemShut
  {NoStop}%
\bibitem [{\citenamefont {Chaikin}\ and\ \citenamefont
  {Lubensky}(1995)}]{chaikin95}%
  \BibitemOpen
  \bibfield  {author} {\bibinfo {author} {\bibfnamefont {P.~M.}\ \bibnamefont
  {Chaikin}}\ and\ \bibinfo {author} {\bibfnamefont {T.~C.}\ \bibnamefont
  {Lubensky}},\ }\href@noop {} {\emph {\bibinfo {title} {Principles of
  Condensed Matter Physics}}}\ (\bibinfo  {publisher} {Cambridge University
  Press},\ \bibinfo {address} {Cambridge},\ \bibinfo {year} {1995})\BibitemShut
  {NoStop}%
\bibitem [{\citenamefont {Kuroda}\ and\ \citenamefont
  {Lacmann}(1982)}]{kuroda82}%
  \BibitemOpen
  \bibfield  {author} {\bibinfo {author} {\bibfnamefont {T.}~\bibnamefont
  {Kuroda}}\ and\ \bibinfo {author} {\bibfnamefont {R.}~\bibnamefont
  {Lacmann}},\ }\bibfield  {title} {\enquote {\bibinfo {title} {Growth kinetics
  of ice from the vapour phase and its growth forms},}\ }\href {\doibase
  http://dx.doi.org/10.1016/0022-0248(82)90028-8} {\bibfield  {journal}
  {\bibinfo  {journal} {J. Cryst. Growth}\ }\textbf {\bibinfo {volume} {56}},\
  \bibinfo {pages} {189--205} (\bibinfo {year} {1982})}\BibitemShut {NoStop}%
\bibitem [{\citenamefont {Jayaprakash}\ \emph {et~al.}(1983)\citenamefont
  {Jayaprakash}, \citenamefont {Saam},\ and\ \citenamefont
  {Teitel}}]{jayaprakash83}%
  \BibitemOpen
  \bibfield  {author} {\bibinfo {author} {\bibfnamefont {C.}~\bibnamefont
  {Jayaprakash}}, \bibinfo {author} {\bibfnamefont {W.~F.}\ \bibnamefont
  {Saam}}, \ and\ \bibinfo {author} {\bibfnamefont {S.}~\bibnamefont
  {Teitel}},\ }\bibfield  {title} {\enquote {\bibinfo {title} {Roughening and
  facet formation in crystals},}\ }\href {\doibase 10.1103/PhysRevLett.50.2017}
  {\bibfield  {journal} {\bibinfo  {journal} {Phys. Rev. Lett.}\ }\textbf
  {\bibinfo {volume} {50}},\ \bibinfo {pages} {2017--2020} (\bibinfo {year}
  {1983})}\BibitemShut {NoStop}%
\bibitem [{\citenamefont {Conde}\ \emph {et~al.}(2008)\citenamefont {Conde},
  \citenamefont {Vega},\ and\ \citenamefont {Patrykiejew}}]{conde08}%
  \BibitemOpen
  \bibfield  {author} {\bibinfo {author} {\bibfnamefont {M.~M.}\ \bibnamefont
  {Conde}}, \bibinfo {author} {\bibfnamefont {C.}~\bibnamefont {Vega}}, \ and\
  \bibinfo {author} {\bibfnamefont {A.}~\bibnamefont {Patrykiejew}},\
  }\bibfield  {title} {\enquote {\bibinfo {title} {The thickness of a liquid
  layer on the free surface of ice as obtained from computer simulation},}\
  }\href {\doibase 10.1063/1.2940195} {\bibfield  {journal} {\bibinfo
  {journal} {J. Chem. Phys.}\ }\textbf {\bibinfo {volume} {129}},\ \bibinfo
  {pages} {014702} (\bibinfo {year} {2008})},\ \Eprint
  {http://arxiv.org/abs/https://doi.org/10.1063/1.2940195}
  {https://doi.org/10.1063/1.2940195} \BibitemShut {NoStop}%
\bibitem [{\citenamefont {Yang}\ \emph {et~al.}(2013)\citenamefont {Yang},
  \citenamefont {Asta},\ and\ \citenamefont {Laird}}]{yang13}%
  \BibitemOpen
  \bibfield  {author} {\bibinfo {author} {\bibfnamefont {A.~J.}\ \bibnamefont
  {Yang}}, \bibinfo {author} {\bibfnamefont {M.}~\bibnamefont {Asta}}, \ and\
  \bibinfo {author} {\bibfnamefont {B.~B.}\ \bibnamefont {Laird}},\ }\bibfield
  {title} {\enquote {\bibinfo {title} {Solid--liquid interfacial premelting},}\
  }\href@noop {} {\bibfield  {journal} {\bibinfo  {journal} {Phys. Rev. Lett.}\
  }\textbf {\bibinfo {volume} {110}},\ \bibinfo {pages} {096102} (\bibinfo
  {year} {2013})}\BibitemShut {NoStop}%
\bibitem [{\citenamefont {Limmer}\ and\ \citenamefont
  {Chandler}(2014)}]{limmer14}%
  \BibitemOpen
  \bibfield  {author} {\bibinfo {author} {\bibfnamefont {David~T.}\
  \bibnamefont {Limmer}}\ and\ \bibinfo {author} {\bibfnamefont {David}\
  \bibnamefont {Chandler}},\ }\bibfield  {title} {\enquote {\bibinfo {title}
  {Premelting, fluctuations, and coarse-graining of water-ice interfaces},}\
  }\href {\doibase http://dx.doi.org/10.1063/1.4895399} {\bibfield  {journal}
  {\bibinfo  {journal} {J. Chem. Phys.}\ }\textbf {\bibinfo {volume} {141}},\
  \bibinfo {pages} {18C505} (\bibinfo {year} {2014})}\BibitemShut {NoStop}%
\bibitem [{\citenamefont {Kling}\ \emph {et~al.}(2018)\citenamefont {Kling},
  \citenamefont {Kling},\ and\ \citenamefont {Donadio}}]{kling18}%
  \BibitemOpen
  \bibfield  {author} {\bibinfo {author} {\bibfnamefont {Tanja}\ \bibnamefont
  {Kling}}, \bibinfo {author} {\bibfnamefont {Felix}\ \bibnamefont {Kling}}, \
  and\ \bibinfo {author} {\bibfnamefont {Davide}\ \bibnamefont {Donadio}},\
  }\bibfield  {title} {\enquote {\bibinfo {title} {Structure and dynamics of
  the quasi-liquid layer at the surface of ice from molecular simulations},}\
  }\href {\doibase 10.1021/acs.jpcc.8b07724} {\bibfield  {journal} {\bibinfo
  {journal} {J. Phys. Chem. C}\ }\textbf {\bibinfo {volume} {122}},\ \bibinfo
  {pages} {24780--24787} (\bibinfo {year} {2018})},\ \Eprint
  {http://arxiv.org/abs/https://doi.org/10.1021/acs.jpcc.8b07724}
  {https://doi.org/10.1021/acs.jpcc.8b07724} \BibitemShut {NoStop}%
\bibitem [{\citenamefont {Pickering}\ \emph {et~al.}(2018)\citenamefont
  {Pickering}, \citenamefont {Paleico}, \citenamefont {Sirkin}, \citenamefont
  {Scherlis},\ and\ \citenamefont {Factorovich}}]{pickering18}%
  \BibitemOpen
  \bibfield  {author} {\bibinfo {author} {\bibfnamefont {Ignacio}\ \bibnamefont
  {Pickering}}, \bibinfo {author} {\bibfnamefont {Martin}\ \bibnamefont
  {Paleico}}, \bibinfo {author} {\bibfnamefont {Yamila A.~Perez}\ \bibnamefont
  {Sirkin}}, \bibinfo {author} {\bibfnamefont {Damian~A.}\ \bibnamefont
  {Scherlis}}, \ and\ \bibinfo {author} {\bibfnamefont {Matías~H.}\
  \bibnamefont {Factorovich}},\ }\bibfield  {title} {\enquote {\bibinfo {title}
  {Grand canonical investigation of the quasi liquid layer of ice: Is it
  liquid?}}\ }\href {\doibase 10.1021/acs.jpcb.8b00784} {\bibfield  {journal}
  {\bibinfo  {journal} {J. Phys. Chem. B}\ }\textbf {\bibinfo {volume} {122}},\
  \bibinfo {pages} {4880--4890} (\bibinfo {year} {2018})},\ \bibinfo {note}
  {pMID: 29660281},\ \Eprint
  {http://arxiv.org/abs/https://doi.org/10.1021/acs.jpcb.8b00784}
  {https://doi.org/10.1021/acs.jpcb.8b00784} \BibitemShut {NoStop}%
\bibitem [{\citenamefont {Mohandesi}\ and\ \citenamefont
  {Kusalik}(2018)}]{mohandesi18}%
  \BibitemOpen
  \bibfield  {author} {\bibinfo {author} {\bibfnamefont {Ali}\ \bibnamefont
  {Mohandesi}}\ and\ \bibinfo {author} {\bibfnamefont {Peter~G.}\ \bibnamefont
  {Kusalik}},\ }\bibfield  {title} {\enquote {\bibinfo {title} {Probing ice
  growth from vapor phase: A molecular dynamics simulation approach},}\ }\href
  {\doibase https://doi.org/10.1016/j.jcrysgro.2017.11.022} {\bibfield
  {journal} {\bibinfo  {journal} {J. Cryst. Growth}\ }\textbf {\bibinfo
  {volume} {483}},\ \bibinfo {pages} {156 -- 163} (\bibinfo {year}
  {2018})}\BibitemShut {NoStop}%
\bibitem [{\citenamefont {Qiu}\ and\ \citenamefont {Molinero}(2018)}]{qiu18}%
  \BibitemOpen
  \bibfield  {author} {\bibinfo {author} {\bibfnamefont {Yuqing}\ \bibnamefont
  {Qiu}}\ and\ \bibinfo {author} {\bibfnamefont {Valeria}\ \bibnamefont
  {Molinero}},\ }\bibfield  {title} {\enquote {\bibinfo {title} {Why is it so
  difficult to identify the onset of ice premelting?}}\ }\href {\doibase
  10.1021/acs.jpclett.8b02244} {\bibfield  {journal} {\bibinfo  {journal} {J.
  Phys. Chem. Lett.}\ }\textbf {\bibinfo {volume} {9}},\ \bibinfo {pages}
  {5179--5182} (\bibinfo {year} {2018})},\ \Eprint
  {http://arxiv.org/abs/https://doi.org/10.1021/acs.jpclett.8b02244}
  {https://doi.org/10.1021/acs.jpclett.8b02244} \BibitemShut {NoStop}%
\bibitem [{\citenamefont {Schick}(1990)}]{schick90}%
  \BibitemOpen
  \bibfield  {author} {\bibinfo {author} {\bibfnamefont {M.}~\bibnamefont
  {Schick}},\ }\bibfield  {title} {\enquote {\bibinfo {title} {Introduction to
  wetting phenomena},}\ }in\ \href@noop {} {\emph {\bibinfo {booktitle}
  {Liquids at Interfaces}}},\ \bibinfo {series and number} {Les Houches Lecture
  Notes}\ (\bibinfo  {publisher} {Elsevier},\ \bibinfo {address} {Amsterdam},\
  \bibinfo {year} {1990})\ pp.\ \bibinfo {pages} {1--89}\BibitemShut {NoStop}%
\bibitem [{\citenamefont {Dosch}\ \emph {et~al.}(1995)\citenamefont {Dosch},
  \citenamefont {Lied},\ and\ \citenamefont {Bilgram}}]{dosch95}%
  \BibitemOpen
  \bibfield  {author} {\bibinfo {author} {\bibfnamefont {H.}~\bibnamefont
  {Dosch}}, \bibinfo {author} {\bibfnamefont {A.}~\bibnamefont {Lied}}, \ and\
  \bibinfo {author} {\bibfnamefont {J.~H.}\ \bibnamefont {Bilgram}},\
  }\bibfield  {title} {\enquote {\bibinfo {title} {Glancing angle x-ray
  scattering studies of the premelting of ice surfaces},}\ }\href@noop {}
  {\bibfield  {journal} {\bibinfo  {journal} {Surf. Sci.}\ }\textbf {\bibinfo
  {volume} {327}},\ \bibinfo {pages} {145--164} (\bibinfo {year}
  {1995})}\BibitemShut {NoStop}%
\bibitem [{\citenamefont {Cao}\ and\ \citenamefont {Conrad}(1990)}]{cao90}%
  \BibitemOpen
  \bibfield  {author} {\bibinfo {author} {\bibfnamefont {Yijian}\ \bibnamefont
  {Cao}}\ and\ \bibinfo {author} {\bibfnamefont {Edward~H.}\ \bibnamefont
  {Conrad}},\ }\bibfield  {title} {\enquote {\bibinfo {title} {Approach to
  thermal roughening of ni(110): A study by high-resolution low-energy electron
  diffraction},}\ }\href {\doibase 10.1103/PhysRevLett.64.447} {\bibfield
  {journal} {\bibinfo  {journal} {Phys. Rev. Lett.}\ }\textbf {\bibinfo
  {volume} {64}},\ \bibinfo {pages} {447--450} (\bibinfo {year}
  {1990})}\BibitemShut {NoStop}%
\bibitem [{\citenamefont {Gibbs}\ \emph {et~al.}(1988)\citenamefont {Gibbs},
  \citenamefont {Ocko}, \citenamefont {Zehner},\ and\ \citenamefont
  {Mochrie}}]{gibbs88}%
  \BibitemOpen
  \bibfield  {author} {\bibinfo {author} {\bibfnamefont {Doon}\ \bibnamefont
  {Gibbs}}, \bibinfo {author} {\bibfnamefont {B.}~\bibnamefont {Ocko}},
  \bibinfo {author} {\bibfnamefont {D.}~\bibnamefont {Zehner}}, \ and\ \bibinfo
  {author} {\bibfnamefont {S.}~\bibnamefont {Mochrie}},\ }\bibfield  {title}
  {\enquote {\bibinfo {title} {Absolute x-ray reflectivity study of the au(100)
  surface},}\ }\href@noop {} {\bibfield  {journal} {\bibinfo  {journal} {Phys.
  Rev. B}\ }\textbf {\bibinfo {volume} {38}},\ \bibinfo {pages} {7303--7310}
  (\bibinfo {year} {1988})}\BibitemShut {NoStop}%
\bibitem [{\citenamefont {Mei}\ and\ \citenamefont {Lu}(2007)}]{mei07}%
  \BibitemOpen
  \bibfield  {author} {\bibinfo {author} {\bibfnamefont {Q.~S.}\ \bibnamefont
  {Mei}}\ and\ \bibinfo {author} {\bibfnamefont {K.}~\bibnamefont {Lu}},\
  }\bibfield  {title} {\enquote {\bibinfo {title} {Melting and superheating of
  crystalline solids: From bulk to nanocrystals},}\ }\href@noop {} {\bibfield
  {journal} {\bibinfo  {journal} {Prog. Materials Sci.}\ }\textbf {\bibinfo
  {volume} {52}},\ \bibinfo {pages} {1175--1262} (\bibinfo {year}
  {2007})}\BibitemShut {NoStop}%
\bibitem [{\citenamefont {Jasnow}(1984)}]{jasnow84}%
  \BibitemOpen
  \bibfield  {author} {\bibinfo {author} {\bibfnamefont {D.}~\bibnamefont
  {Jasnow}},\ }\bibfield  {title} {\enquote {\bibinfo {title} {Critical
  phenomena at interfaces},}\ }\href@noop {} {\bibfield  {journal} {\bibinfo
  {journal} {Rep. Prog. Phys.}\ }\textbf {\bibinfo {volume} {47}},\ \bibinfo
  {pages} {1059} (\bibinfo {year} {1984})}\BibitemShut {NoStop}%
\bibitem [{\citenamefont {Chernov}\ and\ \citenamefont
  {Mikheev}(1988)}]{chernov88}%
  \BibitemOpen
  \bibfield  {author} {\bibinfo {author} {\bibfnamefont {A.~A.}\ \bibnamefont
  {Chernov}}\ and\ \bibinfo {author} {\bibfnamefont {L.~V.}\ \bibnamefont
  {Mikheev}},\ }\bibfield  {title} {\enquote {\bibinfo {title} {Wetting of
  solid surfaces by a structured simple liquid: Effect of fluctuations},}\
  }\href {\doibase 10.1103/PhysRevLett.60.2488} {\bibfield  {journal} {\bibinfo
   {journal} {Phys. Rev. Lett.}\ }\textbf {\bibinfo {volume} {60}},\ \bibinfo
  {pages} {2488--2491} (\bibinfo {year} {1988})}\BibitemShut {NoStop}%
\bibitem [{\citenamefont {Chernov}\ and\ \citenamefont
  {Mikheev}(1989)}]{chernov89}%
  \BibitemOpen
  \bibfield  {author} {\bibinfo {author} {\bibfnamefont {A.~A.}\ \bibnamefont
  {Chernov}}\ and\ \bibinfo {author} {\bibfnamefont {L.~V.}\ \bibnamefont
  {Mikheev}},\ }\bibfield  {title} {\enquote {\bibinfo {title} {Wetting and
  surface melting: Capillary fluctuations vs. layerwise short-range order},}\
  }\href@noop {} {\bibfield  {journal} {\bibinfo  {journal} {Physica. A}\
  }\textbf {\bibinfo {volume} {157}},\ \bibinfo {pages} {1042--1058} (\bibinfo
  {year} {1989})}\BibitemShut {NoStop}%
\bibitem [{\citenamefont {Henderson}(1994)}]{henderson94}%
  \BibitemOpen
  \bibfield  {author} {\bibinfo {author} {\bibfnamefont {J.~R.}\ \bibnamefont
  {Henderson}},\ }\bibfield  {title} {\enquote {\bibinfo {title} {Wetting
  phenomena and the decay of correlations at fluid interfaces},}\ }\href
  {\doibase 10.1103/PhysRevE.50.4836} {\bibfield  {journal} {\bibinfo
  {journal} {Phys. Rev. E}\ }\textbf {\bibinfo {volume} {50}},\ \bibinfo
  {pages} {4836--4846} (\bibinfo {year} {1994})}\BibitemShut {NoStop}%
\bibitem [{\citenamefont {Dash}\ \emph {et~al.}(2006)\citenamefont {Dash},
  \citenamefont {Rempel},\ and\ \citenamefont {Wettlaufer}}]{dash06}%
  \BibitemOpen
  \bibfield  {author} {\bibinfo {author} {\bibfnamefont {J.~G.}\ \bibnamefont
  {Dash}}, \bibinfo {author} {\bibfnamefont {A.~W.}\ \bibnamefont {Rempel}}, \
  and\ \bibinfo {author} {\bibfnamefont {J.~S.}\ \bibnamefont {Wettlaufer}},\
  }\bibfield  {title} {\enquote {\bibinfo {title} {The physics of premelted ice
  and its geophysical consequences},}\ }\href@noop {} {\bibfield  {journal}
  {\bibinfo  {journal} {Rev. Mod. Phys.}\ }\textbf {\bibinfo {volume} {78}},\
  \bibinfo {pages} {695--741} (\bibinfo {year} {2006})}\BibitemShut {NoStop}%
\bibitem [{\citenamefont {Elbaum}(1991)}]{elbaum91}%
  \BibitemOpen
  \bibfield  {author} {\bibinfo {author} {\bibfnamefont {Michael}\ \bibnamefont
  {Elbaum}},\ }\bibfield  {title} {\enquote {\bibinfo {title} {Roughening
  transition observed on the prism facet of ice},}\ }\href {\doibase
  10.1103/PhysRevLett.67.2982} {\bibfield  {journal} {\bibinfo  {journal}
  {Phys. Rev. Lett.}\ }\textbf {\bibinfo {volume} {67}},\ \bibinfo {pages}
  {2982--2985} (\bibinfo {year} {1991})}\BibitemShut {NoStop}%
\bibitem [{\citenamefont {Elbaum}\ and\ \citenamefont
  {Schick}(1991)}]{elbaum91b}%
  \BibitemOpen
  \bibfield  {author} {\bibinfo {author} {\bibfnamefont {Michael}\ \bibnamefont
  {Elbaum}}\ and\ \bibinfo {author} {\bibfnamefont {M.}~\bibnamefont
  {Schick}},\ }\bibfield  {title} {\enquote {\bibinfo {title} {Application of
  the theory of dispersion forces to the surface melting of ice},}\ }\href@noop
  {} {\bibfield  {journal} {\bibinfo  {journal} {Phys. Rev. Lett.}\ }\textbf
  {\bibinfo {volume} {66}},\ \bibinfo {pages} {1713--1716} (\bibinfo {year}
  {1991})}\BibitemShut {NoStop}%
\bibitem [{\citenamefont {Elbaum}\ \emph {et~al.}(1993)\citenamefont {Elbaum},
  \citenamefont {Lipson},\ and\ \citenamefont {Dash}}]{elbaum93}%
  \BibitemOpen
  \bibfield  {author} {\bibinfo {author} {\bibfnamefont {Michael}\ \bibnamefont
  {Elbaum}}, \bibinfo {author} {\bibfnamefont {S.~G.}\ \bibnamefont {Lipson}},
  \ and\ \bibinfo {author} {\bibfnamefont {J.~G.}\ \bibnamefont {Dash}},\
  }\bibfield  {title} {\enquote {\bibinfo {title} {Optical study of surface
  melting on ice},}\ }\href@noop {} {\bibfield  {journal} {\bibinfo  {journal}
  {J. Cryst. Growth}\ }\textbf {\bibinfo {volume} {129}},\ \bibinfo {pages}
  {491--505} (\bibinfo {year} {1993})}\BibitemShut {NoStop}%
\bibitem [{\citenamefont {Lied}\ \emph {et~al.}(1994)\citenamefont {Lied},
  \citenamefont {Dosch},\ and\ \citenamefont {Bilgram}}]{lied94}%
  \BibitemOpen
  \bibfield  {author} {\bibinfo {author} {\bibfnamefont {A.}~\bibnamefont
  {Lied}}, \bibinfo {author} {\bibfnamefont {H.}~\bibnamefont {Dosch}}, \ and\
  \bibinfo {author} {\bibfnamefont {J.~H.}\ \bibnamefont {Bilgram}},\
  }\bibfield  {title} {\enquote {\bibinfo {title} {Surface melting of ice
  {I$_h$} single crystals revealed by glancing angle x-ray scattering},}\
  }\href@noop {} {\bibfield  {journal} {\bibinfo  {journal} {Phys. Rev. Lett.}\
  }\textbf {\bibinfo {volume} {72}},\ \bibinfo {pages} {3554--3557} (\bibinfo
  {year} {1994})}\BibitemShut {NoStop}%
\bibitem [{\citenamefont {Bluhm}\ \emph {et~al.}(2002)\citenamefont {Bluhm},
  \citenamefont {Ogletree}, \citenamefont {Fadley}, \citenamefont {Hussain},\
  and\ \citenamefont {Salmeron}}]{bluhm02}%
  \BibitemOpen
  \bibfield  {author} {\bibinfo {author} {\bibfnamefont {H.}~\bibnamefont
  {Bluhm}}, \bibinfo {author} {\bibfnamefont {D.~F.}\ \bibnamefont {Ogletree}},
  \bibinfo {author} {\bibfnamefont {C.~S.}\ \bibnamefont {Fadley}}, \bibinfo
  {author} {\bibfnamefont {Z.}~\bibnamefont {Hussain}}, \ and\ \bibinfo
  {author} {\bibfnamefont {M.}~\bibnamefont {Salmeron}},\ }\bibfield  {title}
  {\enquote {\bibinfo {title} {The premelting of ice studied with photoelectron
  spectroscopy},}\ }\href@noop {} {\bibfield  {journal} {\bibinfo  {journal}
  {J. Phys.: Condens. Matter}\ }\textbf {\bibinfo {volume} {14}},\ \bibinfo
  {pages} {L227--L233} (\bibinfo {year} {2002})}\BibitemShut {NoStop}%
\bibitem [{\citenamefont {Sazaki}\ \emph {et~al.}(2012)\citenamefont {Sazaki},
  \citenamefont {Zepeda}, \citenamefont {Nakatsubo}, \citenamefont {Yokomine},\
  and\ \citenamefont {Furukawa}}]{sazaki12}%
  \BibitemOpen
  \bibfield  {author} {\bibinfo {author} {\bibfnamefont {Gen}\ \bibnamefont
  {Sazaki}}, \bibinfo {author} {\bibfnamefont {Salvador}\ \bibnamefont
  {Zepeda}}, \bibinfo {author} {\bibfnamefont {Shunichi}\ \bibnamefont
  {Nakatsubo}}, \bibinfo {author} {\bibfnamefont {Makoto}\ \bibnamefont
  {Yokomine}}, \ and\ \bibinfo {author} {\bibfnamefont {Yoshinori}\
  \bibnamefont {Furukawa}},\ }\bibfield  {title} {\enquote {\bibinfo {title}
  {Quasi-liquid layers on ice crystal surfaces are made up of two different
  phases},}\ }\href {\doibase 10.1073/pnas.1116685109} {\bibfield  {journal}
  {\bibinfo  {journal} {Proc. Natl. Acad. Sci. U.S.A.}\ }\textbf {\bibinfo
  {volume} {109}},\ \bibinfo {pages} {1052--1055} (\bibinfo {year} {2012})},\
  \Eprint
  {http://arxiv.org/abs/http://www.pnas.org/content/109/4/1052.full.pdf}
  {http://www.pnas.org/content/109/4/1052.full.pdf} \BibitemShut {NoStop}%
\bibitem [{\citenamefont {Asakawa}\ \emph {et~al.}(2015)\citenamefont
  {Asakawa}, \citenamefont {Sazaki}, \citenamefont {Nagashima}, \citenamefont
  {Nakatsubo},\ and\ \citenamefont {Furukawa}}]{asakawa15}%
  \BibitemOpen
  \bibfield  {author} {\bibinfo {author} {\bibfnamefont {Harutoshi}\
  \bibnamefont {Asakawa}}, \bibinfo {author} {\bibfnamefont {Gen}\ \bibnamefont
  {Sazaki}}, \bibinfo {author} {\bibfnamefont {Ken}\ \bibnamefont {Nagashima}},
  \bibinfo {author} {\bibfnamefont {Shunichi}\ \bibnamefont {Nakatsubo}}, \
  and\ \bibinfo {author} {\bibfnamefont {Yoshinori}\ \bibnamefont {Furukawa}},\
  }\bibfield  {title} {\enquote {\bibinfo {title} {Prism and other high-index
  faces of ice crystals exhibit two types of quasi-liquid layers},}\ }\href
  {\doibase 10.1021/acs.cgd.5b00438} {\bibfield  {journal} {\bibinfo  {journal}
  {Crystal Growth \& Design}\ }\textbf {\bibinfo {volume} {15}},\ \bibinfo
  {pages} {3339--3344} (\bibinfo {year} {2015})},\ \Eprint
  {http://arxiv.org/abs/http://dx.doi.org/10.1021/acs.cgd.5b00438}
  {http://dx.doi.org/10.1021/acs.cgd.5b00438} \BibitemShut {NoStop}%
\bibitem [{\citenamefont {Asakawa}\ \emph {et~al.}(2016)\citenamefont
  {Asakawa}, \citenamefont {Sazaki}, \citenamefont {Nagashima}, \citenamefont
  {Nakatsubo},\ and\ \citenamefont {Furukawa}}]{asakawa16}%
  \BibitemOpen
  \bibfield  {author} {\bibinfo {author} {\bibfnamefont {Harutoshi}\
  \bibnamefont {Asakawa}}, \bibinfo {author} {\bibfnamefont {Gen}\ \bibnamefont
  {Sazaki}}, \bibinfo {author} {\bibfnamefont {Ken}\ \bibnamefont {Nagashima}},
  \bibinfo {author} {\bibfnamefont {Shunichi}\ \bibnamefont {Nakatsubo}}, \
  and\ \bibinfo {author} {\bibfnamefont {Yoshinori}\ \bibnamefont {Furukawa}},\
  }\bibfield  {title} {\enquote {\bibinfo {title} {Two types of quasi-liquid
  layers on ice crystals are formed kinetically},}\ }\href {\doibase
  10.1073/pnas.1521607113} {\bibfield  {journal} {\bibinfo  {journal} {Proc.
  Natl. Acad. Sci. U.S.A.}\ }\textbf {\bibinfo {volume} {113}},\ \bibinfo
  {pages} {1749--1753} (\bibinfo {year} {2016})},\ \Eprint
  {http://arxiv.org/abs/http://www.pnas.org/content/113/7/1749.full.pdf}
  {http://www.pnas.org/content/113/7/1749.full.pdf} \BibitemShut {NoStop}%
\bibitem [{\citenamefont {Murata}\ \emph {et~al.}(2016)\citenamefont {Murata},
  \citenamefont {Asakawa}, \citenamefont {Nagashima}, \citenamefont
  {Furukawa},\ and\ \citenamefont {Sazaki}}]{murata16}%
  \BibitemOpen
  \bibfield  {author} {\bibinfo {author} {\bibfnamefont {Ken-ichiro}\
  \bibnamefont {Murata}}, \bibinfo {author} {\bibfnamefont {Harutoshi}\
  \bibnamefont {Asakawa}}, \bibinfo {author} {\bibfnamefont {Ken}\ \bibnamefont
  {Nagashima}}, \bibinfo {author} {\bibfnamefont {Yoshinori}\ \bibnamefont
  {Furukawa}}, \ and\ \bibinfo {author} {\bibfnamefont {Gen}\ \bibnamefont
  {Sazaki}},\ }\bibfield  {title} {\enquote {\bibinfo {title} {Thermodynamic
  origin of surface melting on ice crystals},}\ }\href@noop {} {\bibfield
  {journal} {\bibinfo  {journal} {Proc. Natl. Acad. Sci. U.S.A.}\ }\textbf
  {\bibinfo {volume} {113}},\ \bibinfo {pages} {E6741--E6748} (\bibinfo {year}
  {2016})}\BibitemShut {NoStop}%
\bibitem [{\citenamefont {S{\'a}nchez}\ \emph {et~al.}(2017)\citenamefont
  {S{\'a}nchez}, \citenamefont {Kling}, \citenamefont {Ishiyama}, \citenamefont
  {van Zadel}, \citenamefont {Bisson}, \citenamefont {Mezger}, \citenamefont
  {Jochum}, \citenamefont {Cyran}, \citenamefont {Smit}, \citenamefont
  {Bakker}, \citenamefont {Shultz}, \citenamefont {Morita}, \citenamefont
  {Donadio}, \citenamefont {Nagata}, \citenamefont {Bonn},\ and\ \citenamefont
  {Backus}}]{sanchez17}%
  \BibitemOpen
  \bibfield  {author} {\bibinfo {author} {\bibfnamefont {M.~Alejandra}\
  \bibnamefont {S{\'a}nchez}}, \bibinfo {author} {\bibfnamefont {Tanja}\
  \bibnamefont {Kling}}, \bibinfo {author} {\bibfnamefont {Tatsuya}\
  \bibnamefont {Ishiyama}}, \bibinfo {author} {\bibfnamefont {Marc-Jan}\
  \bibnamefont {van Zadel}}, \bibinfo {author} {\bibfnamefont {Patrick~J.}\
  \bibnamefont {Bisson}}, \bibinfo {author} {\bibfnamefont {Markus}\
  \bibnamefont {Mezger}}, \bibinfo {author} {\bibfnamefont {Mara~N.}\
  \bibnamefont {Jochum}}, \bibinfo {author} {\bibfnamefont {Jenée~D.}\
  \bibnamefont {Cyran}}, \bibinfo {author} {\bibfnamefont {Wilbert~J.}\
  \bibnamefont {Smit}}, \bibinfo {author} {\bibfnamefont {Huib~J.}\
  \bibnamefont {Bakker}}, \bibinfo {author} {\bibfnamefont {Mary~Jane}\
  \bibnamefont {Shultz}}, \bibinfo {author} {\bibfnamefont {Akihiro}\
  \bibnamefont {Morita}}, \bibinfo {author} {\bibfnamefont {Davide}\
  \bibnamefont {Donadio}}, \bibinfo {author} {\bibfnamefont {Yuki}\
  \bibnamefont {Nagata}}, \bibinfo {author} {\bibfnamefont {Mischa}\
  \bibnamefont {Bonn}}, \ and\ \bibinfo {author} {\bibfnamefont {Ellen H.~G.}\
  \bibnamefont {Backus}},\ }\bibfield  {title} {\enquote {\bibinfo {title}
  {Experimental and theoretical evidence for bilayer-by-bilayer surface melting
  of crystalline ice},}\ }\href {\doibase 10.1073/pnas.1612893114} {\bibfield
  {journal} {\bibinfo  {journal} {Proc. Natl. Acad. Sci. U.S.A.}\ }\textbf
  {\bibinfo {volume} {114}},\ \bibinfo {pages} {227--232} (\bibinfo {year}
  {2017})},\ \Eprint
  {http://arxiv.org/abs/http://www.pnas.org/content/114/2/227.full.pdf}
  {http://www.pnas.org/content/114/2/227.full.pdf} \BibitemShut {NoStop}%
\bibitem [{\citenamefont {Slater}\ and\ \citenamefont
  {Michaelides}(2019)}]{slater19}%
  \BibitemOpen
  \bibfield  {author} {\bibinfo {author} {\bibfnamefont {B.}~\bibnamefont
  {Slater}}\ and\ \bibinfo {author} {\bibfnamefont {A.}~\bibnamefont
  {Michaelides}},\ }\bibfield  {title} {\enquote {\bibinfo {title} {Surface
  premelting of water ice},}\ }\href@noop {} {\bibfield  {journal} {\bibinfo
  {journal} {Nat. Rev. Chem}\ }\textbf {\bibinfo {volume} {3}},\ \bibinfo
  {pages} {172--188} (\bibinfo {year} {2019})}\BibitemShut {NoStop}%
\bibitem [{\citenamefont {Sadtchenko}\ and\ \citenamefont
  {Ewing}(2002)}]{sadtchenko02}%
  \BibitemOpen
  \bibfield  {author} {\bibinfo {author} {\bibfnamefont {Vlad}\ \bibnamefont
  {Sadtchenko}}\ and\ \bibinfo {author} {\bibfnamefont {George~E.}\
  \bibnamefont {Ewing}},\ }\bibfield  {title} {\enquote {\bibinfo {title}
  {Interfacial melting of thin ice films: An infrared study},}\ }\href
  {\doibase 10.1063/1.1449947} {\bibfield  {journal} {\bibinfo  {journal} {J.
  Chem. Phys.}\ }\textbf {\bibinfo {volume} {116}},\ \bibinfo {pages}
  {4686--4697} (\bibinfo {year} {2002})},\ \Eprint
  {http://arxiv.org/abs/https://doi.org/10.1063/1.1449947}
  {https://doi.org/10.1063/1.1449947} \BibitemShut {NoStop}%
\bibitem [{\citenamefont {Smit}\ and\ \citenamefont {Bakker}(2017)}]{smit17}%
  \BibitemOpen
  \bibfield  {author} {\bibinfo {author} {\bibfnamefont {Wilbert~J.}\
  \bibnamefont {Smit}}\ and\ \bibinfo {author} {\bibfnamefont {Huib~J.}\
  \bibnamefont {Bakker}},\ }\bibfield  {title} {\enquote {\bibinfo {title} {The
  surface of ice is like supercooled liquid water},}\ }\href {\doibase
  10.1002/anie.201707530} {\bibfield  {journal} {\bibinfo  {journal} {Angew.
  Chem. Int. Ed. Engl.}\ }\textbf {\bibinfo {volume} {56}},\ \bibinfo {pages}
  {15540--15544} (\bibinfo {year} {2017})},\ \Eprint
  {http://arxiv.org/abs/https://onlinelibrary.wiley.com/doi/pdf/10.1002/anie.201707530}
  {https://onlinelibrary.wiley.com/doi/pdf/10.1002/anie.201707530} \BibitemShut
  {NoStop}%
\bibitem [{\citenamefont {Michaelides}\ and\ \citenamefont
  {Slater}(2017)}]{michaelides17}%
  \BibitemOpen
  \bibfield  {author} {\bibinfo {author} {\bibfnamefont {Angelos}\ \bibnamefont
  {Michaelides}}\ and\ \bibinfo {author} {\bibfnamefont {Ben}\ \bibnamefont
  {Slater}},\ }\bibfield  {title} {\enquote {\bibinfo {title} {Melting the ice
  one layer at a time},}\ }\href {\doibase 10.1073/pnas.1619259114} {\bibfield
  {journal} {\bibinfo  {journal} {Proc. Natl. Acad. Sci. U.S.A.}\ }\textbf
  {\bibinfo {volume} {114}},\ \bibinfo {pages} {195--197} (\bibinfo {year}
  {2017})},\ \Eprint
  {http://arxiv.org/abs/http://www.pnas.org/content/early/2017/01/01/1619259114.full.pdf}
  {http://www.pnas.org/content/early/2017/01/01/1619259114.full.pdf}
  \BibitemShut {NoStop}%
\bibitem [{\citenamefont {Gonda}\ \emph {et~al.}(1999)\citenamefont {Gonda},
  \citenamefont {Arai},\ and\ \citenamefont {Sei}}]{gonda99}%
  \BibitemOpen
  \bibfield  {author} {\bibinfo {author} {\bibfnamefont {T.}~\bibnamefont
  {Gonda}}, \bibinfo {author} {\bibfnamefont {T.}~\bibnamefont {Arai}}, \ and\
  \bibinfo {author} {\bibfnamefont {T.}~\bibnamefont {Sei}},\ }\bibfield
  {title} {\enquote {\bibinfo {title} {Experimental study on the melting
  process of ice crystals just below the melting point.}}\ }\href@noop {}
  {\bibfield  {journal} {\bibinfo  {journal} {Polar Meteorol. Glaciol.}\
  }\textbf {\bibinfo {volume} {13}},\ \bibinfo {pages} {38--42} (\bibinfo
  {year} {1999})}\BibitemShut {NoStop}%
\bibitem [{\citenamefont {Wettlaufer}(1999)}]{wettlaufer99}%
  \BibitemOpen
  \bibfield  {author} {\bibinfo {author} {\bibfnamefont {J.}~\bibnamefont
  {Wettlaufer}},\ }\bibfield  {title} {\enquote {\bibinfo {title} {Impurity
  effects in the premelting of ice},}\ }\href {\doibase
  10.1103/PhysRevLett.82.2516} {\bibfield  {journal} {\bibinfo  {journal}
  {Phys. Rev. Lett.}\ }\textbf {\bibinfo {volume} {82}},\ \bibinfo {pages}
  {2516--2519} (\bibinfo {year} {1999})}\BibitemShut {NoStop}%
\bibitem [{\citenamefont {Mitsui}\ and\ \citenamefont {Aoki}(2019)}]{mitsui19}%
  \BibitemOpen
  \bibfield  {author} {\bibinfo {author} {\bibfnamefont {Takahisa}\
  \bibnamefont {Mitsui}}\ and\ \bibinfo {author} {\bibfnamefont {Kenichiro}\
  \bibnamefont {Aoki}},\ }\bibfield  {title} {\enquote {\bibinfo {title}
  {Fluctuation spectroscopy of surface melting of ice with and without
  impurities},}\ }\href {\doibase 10.1103/PhysRevE.99.010801} {\bibfield
  {journal} {\bibinfo  {journal} {Phys. Rev. E}\ }\textbf {\bibinfo {volume}
  {99}},\ \bibinfo {pages} {010801} (\bibinfo {year} {2019})}\BibitemShut
  {NoStop}%
\bibitem [{\citenamefont {Libbrecht}(2013)}]{libbrecht13}%
  \BibitemOpen
  \bibfield  {author} {\bibinfo {author} {\bibfnamefont {K.~G.}\ \bibnamefont
  {Libbrecht}},\ }\bibfield  {title} {\enquote {\bibinfo {title} {On the
  equilibrium shape of an ice crystal},}\ }\href@noop {} {\bibfield  {journal}
  {\bibinfo  {journal} {arXiv:1205.1452 [cond-mat.mtrl-sci]}\ } (\bibinfo
  {year} {2013})}\BibitemShut {NoStop}%
\bibitem [{\citenamefont {Maruyama}\ \emph {et~al.}(1997)\citenamefont
  {Maruyama}, \citenamefont {Nishida},\ and\ \citenamefont
  {Sawada}}]{maruyama97}%
  \BibitemOpen
  \bibfield  {author} {\bibinfo {author} {\bibfnamefont {Minoru}\ \bibnamefont
  {Maruyama}}, \bibinfo {author} {\bibfnamefont {Takehito}\ \bibnamefont
  {Nishida}}, \ and\ \bibinfo {author} {\bibfnamefont {Tsutomu}\ \bibnamefont
  {Sawada}},\ }\bibfield  {title} {\enquote {\bibinfo {title} {Crystal shape of
  high-pressure ice ih in water and roughening transition of the (10-10)
  plane},}\ }\href {\doibase 10.1021/jp9631745} {\bibfield  {journal} {\bibinfo
   {journal} {The Journal of Physical Chemistry B}\ }\textbf {\bibinfo {volume}
  {101}},\ \bibinfo {pages} {6151--6153} (\bibinfo {year} {1997})},\ \Eprint
  {http://arxiv.org/abs/http://pubs.acs.org/doi/pdf/10.1021/jp9631745}
  {http://pubs.acs.org/doi/pdf/10.1021/jp9631745} \BibitemShut {NoStop}%
\bibitem [{\citenamefont {Gonda}\ and\ \citenamefont
  {Yamakazi}(1978)}]{gonda78}%
  \BibitemOpen
  \bibfield  {author} {\bibinfo {author} {\bibfnamefont {T.}~\bibnamefont
  {Gonda}}\ and\ \bibinfo {author} {\bibfnamefont {T.}~\bibnamefont
  {Yamakazi}},\ }\bibfield  {title} {\enquote {\bibinfo {title} {Morphology of
  ice droxtals grown from supercooled water droplets},}\ }\href@noop {}
  {\bibfield  {journal} {\bibinfo  {journal} {J. Cryst. Growth}\ }\textbf
  {\bibinfo {volume} {45}},\ \bibinfo {pages} {66--69} (\bibinfo {year}
  {1978})}\BibitemShut {NoStop}%
\bibitem [{\citenamefont {Colbeck}(1983)}]{colbeck83}%
  \BibitemOpen
  \bibfield  {author} {\bibinfo {author} {\bibfnamefont {S.~C.}\ \bibnamefont
  {Colbeck}},\ }\bibfield  {title} {\enquote {\bibinfo {title} {Ice crystal
  morphology and growth rates at low supersaturations and high temperatures},}\
  }\href {\doibase http://dx.doi.org/10.1063/1.332290} {\bibfield  {journal}
  {\bibinfo  {journal} {J. App. Phys.}\ }\textbf {\bibinfo {volume} {54}},\
  \bibinfo {pages} {2677--2682} (\bibinfo {year} {1983})}\BibitemShut {NoStop}%
\bibitem [{\citenamefont {Benet}\ \emph
  {et~al.}(2014{\natexlab{a}})\citenamefont {Benet}, \citenamefont
  {MacDowell},\ and\ \citenamefont {Sanz}}]{benet14c}%
  \BibitemOpen
  \bibfield  {author} {\bibinfo {author} {\bibfnamefont {Jorge}\ \bibnamefont
  {Benet}}, \bibinfo {author} {\bibfnamefont {Luis~G.}\ \bibnamefont
  {MacDowell}}, \ and\ \bibinfo {author} {\bibfnamefont {Eduardo}\ \bibnamefont
  {Sanz}},\ }\bibfield  {title} {\enquote {\bibinfo {title} {A study of the
  ice-water interface using the tip4p/2005 water model},}\ }\href {\doibase
  10.1039/C4CP03398A} {\bibfield  {journal} {\bibinfo  {journal} {Phys. Chem.
  Chem. Phys.}\ }\textbf {\bibinfo {volume} {16}},\ \bibinfo {pages}
  {22159--22166} (\bibinfo {year} {2014}{\natexlab{a}})}\BibitemShut {NoStop}%
\bibitem [{\citenamefont {Furukawa}\ and\ \citenamefont
  {Nada}(1997)}]{furukawa97}%
  \BibitemOpen
  \bibfield  {author} {\bibinfo {author} {\bibfnamefont {Yoshinori}\
  \bibnamefont {Furukawa}}\ and\ \bibinfo {author} {\bibfnamefont {Hiroki}\
  \bibnamefont {Nada}},\ }\bibfield  {title} {\enquote {\bibinfo {title}
  {Anisotropic surface melting of an ice crystal and its relationship to growth
  forms},}\ }\href {\doibase 10.1021/jp9631700} {\bibfield  {journal} {\bibinfo
   {journal} {J. Phys. Chem. B}\ }\textbf {\bibinfo {volume} {101}},\ \bibinfo
  {pages} {6167--6170} (\bibinfo {year} {1997})},\ \Eprint
  {http://arxiv.org/abs/http://dx.doi.org/10.1021/jp9631700}
  {http://dx.doi.org/10.1021/jp9631700} \BibitemShut {NoStop}%
\bibitem [{\citenamefont {Bishop}\ \emph {et~al.}(2009)\citenamefont {Bishop},
  \citenamefont {Pan}, \citenamefont {Liu}, \citenamefont {Tribello},
  \citenamefont {Michaelides}, \citenamefont {Wang},\ and\ \citenamefont
  {Slater}}]{bishop08}%
  \BibitemOpen
  \bibfield  {author} {\bibinfo {author} {\bibfnamefont {C.~L.}\ \bibnamefont
  {Bishop}}, \bibinfo {author} {\bibfnamefont {D.}~\bibnamefont {Pan}},
  \bibinfo {author} {\bibfnamefont {L.~M.}\ \bibnamefont {Liu}}, \bibinfo
  {author} {\bibfnamefont {G.~A.}\ \bibnamefont {Tribello}}, \bibinfo {author}
  {\bibfnamefont {A.}~\bibnamefont {Michaelides}}, \bibinfo {author}
  {\bibfnamefont {E.~G.}\ \bibnamefont {Wang}}, \ and\ \bibinfo {author}
  {\bibfnamefont {B.}~\bibnamefont {Slater}},\ }\bibfield  {title} {\enquote
  {\bibinfo {title} {On thin ice: surface order and disorder during
  pre-melting},}\ }\href {\doibase 10.1039/B807377P} {\bibfield  {journal}
  {\bibinfo  {journal} {Faraday Discuss.}\ }\textbf {\bibinfo {volume} {141}},\
  \bibinfo {pages} {277--292} (\bibinfo {year} {2009})}\BibitemShut {NoStop}%
\bibitem [{\citenamefont {Pereyra}\ and\ \citenamefont
  {Carignano}(2009)}]{pereyra09}%
  \BibitemOpen
  \bibfield  {author} {\bibinfo {author} {\bibfnamefont {Rodolfo~G.}\
  \bibnamefont {Pereyra}}\ and\ \bibinfo {author} {\bibfnamefont {Marcelo~A.}\
  \bibnamefont {Carignano}},\ }\bibfield  {title} {\enquote {\bibinfo {title}
  {Ice nanocolumns: A molecular dynamics study},}\ }\href {\doibase
  10.1021/jp903404n} {\bibfield  {journal} {\bibinfo  {journal} {J. Phys. Chem.
  C}\ }\textbf {\bibinfo {volume} {113}},\ \bibinfo {pages} {12699--12705}
  (\bibinfo {year} {2009})},\ \Eprint
  {http://arxiv.org/abs/http://dx.doi.org/10.1021/jp903404n}
  {http://dx.doi.org/10.1021/jp903404n} \BibitemShut {NoStop}%
\bibitem [{\citenamefont {Pan}\ \emph {et~al.}(2011)\citenamefont {Pan},
  \citenamefont {Liu}, \citenamefont {Slater}, \citenamefont {Michaelides},\
  and\ \citenamefont {Wang}}]{pan11}%
  \BibitemOpen
  \bibfield  {author} {\bibinfo {author} {\bibfnamefont {D.}~\bibnamefont
  {Pan}}, \bibinfo {author} {\bibfnamefont {L-M.}\ \bibnamefont {Liu}},
  \bibinfo {author} {\bibfnamefont {B.}~\bibnamefont {Slater}}, \bibinfo
  {author} {\bibfnamefont {A.}~\bibnamefont {Michaelides}}, \ and\ \bibinfo
  {author} {\bibfnamefont {E.}~\bibnamefont {Wang}},\ }\bibfield  {title}
  {\enquote {\bibinfo {title} {Melting the ice: On the relation between
  temerature and size for nanoscale ice crystals},}\ }\href@noop {} {\bibfield
  {journal} {\bibinfo  {journal} {ACS nano}\ }\textbf {\bibinfo {volume} {5}},\
  \bibinfo {pages} {4562--4569} (\bibinfo {year} {2011})}\BibitemShut {NoStop}%
\bibitem [{\citenamefont {Benet}(2015)}]{benet15b}%
  \BibitemOpen
  \bibfield  {author} {\bibinfo {author} {\bibfnamefont {Jorge}\ \bibnamefont
  {Benet}},\ }\emph {\bibinfo {title} {Estudio por Simulaci\'on de
  Fluctuaciones Capilares: Interfases Fluidas, Adsorbidas y S\'olidas}},\
  \href@noop {} {Ph.D. thesis},\ \bibinfo  {school} {Universidad Complutense de
  Madrid} (\bibinfo {year} {2015})\BibitemShut {NoStop}%
\bibitem [{\citenamefont {Benet}\ \emph {et~al.}(2016)\citenamefont {Benet},
  \citenamefont {Llombart}, \citenamefont {Sanz},\ and\ \citenamefont
  {MacDowell}}]{benet16}%
  \BibitemOpen
  \bibfield  {author} {\bibinfo {author} {\bibfnamefont {Jorge}\ \bibnamefont
  {Benet}}, \bibinfo {author} {\bibfnamefont {Pablo}\ \bibnamefont {Llombart}},
  \bibinfo {author} {\bibfnamefont {Eduardo}\ \bibnamefont {Sanz}}, \ and\
  \bibinfo {author} {\bibfnamefont {Luis~G.}\ \bibnamefont {MacDowell}},\
  }\bibfield  {title} {\enquote {\bibinfo {title} {Premelting-induced
  smoothening of the ice-vapor interface},}\ }\href {\doibase
  10.1103/PhysRevLett.117.096101} {\bibfield  {journal} {\bibinfo  {journal}
  {Phys. Rev. Lett.}\ }\textbf {\bibinfo {volume} {117}},\ \bibinfo {pages}
  {096101} (\bibinfo {year} {2016})}\BibitemShut {NoStop}%
\bibitem [{\citenamefont {Abascal}\ \emph {et~al.}(2005)\citenamefont
  {Abascal}, \citenamefont {Sanz}, \citenamefont {Fernandez},\ and\
  \citenamefont {Vega}}]{abascal05}%
  \BibitemOpen
  \bibfield  {author} {\bibinfo {author} {\bibfnamefont {J.~L.~F.}\
  \bibnamefont {Abascal}}, \bibinfo {author} {\bibfnamefont {E.}~\bibnamefont
  {Sanz}}, \bibinfo {author} {\bibfnamefont {R.~G.}\ \bibnamefont {Fernandez}},
  \ and\ \bibinfo {author} {\bibfnamefont {C.}~\bibnamefont {Vega}},\
  }\bibfield  {title} {\enquote {\bibinfo {title} {A potential model for the
  study of ices and amorphous water: {TIP4P/Ice}},}\ }\href@noop {} {\bibfield
  {journal} {\bibinfo  {journal} {J. Chem. Phys.}\ }\textbf {\bibinfo {volume}
  {122}},\ \bibinfo {pages} {234511} (\bibinfo {year} {2005})}\BibitemShut
  {NoStop}%
\bibitem [{\citenamefont {Vega}\ and\ \citenamefont
  {Abascal}(2011)}]{abascal11}%
  \BibitemOpen
  \bibfield  {author} {\bibinfo {author} {\bibfnamefont {Carlos}\ \bibnamefont
  {Vega}}\ and\ \bibinfo {author} {\bibfnamefont {Jose L.~F.}\ \bibnamefont
  {Abascal}},\ }\bibfield  {title} {\enquote {\bibinfo {title} {Simulating
  water with rigid non-polarizable models: a general perspective},}\ }\href
  {\doibase 10.1039/C1CP22168J} {\bibfield  {journal} {\bibinfo  {journal}
  {Phys. Chem. Chem. Phys.}\ }\textbf {\bibinfo {volume} {13}},\ \bibinfo
  {pages} {19663--19688} (\bibinfo {year} {2011})}\BibitemShut {NoStop}%
\bibitem [{\citenamefont {MacDowell}\ and\ \citenamefont
  {Vega}(2010)}]{macdowell10}%
  \BibitemOpen
  \bibfield  {author} {\bibinfo {author} {\bibfnamefont {L.~G.}\ \bibnamefont
  {MacDowell}}\ and\ \bibinfo {author} {\bibfnamefont {C.}~\bibnamefont
  {Vega}},\ }\bibfield  {title} {\enquote {\bibinfo {title} {Dielectric
  constant of ice ih and ice v: A computer simulation study},}\ }\href@noop {}
  {\bibfield  {journal} {\bibinfo  {journal} {J. Phys. Chem. B}\ }\textbf
  {\bibinfo {volume} {114}},\ \bibinfo {pages} {6089--6098} (\bibinfo {year}
  {2010})}\BibitemShut {NoStop}%
\bibitem [{\citenamefont {MacDowell}\ \emph {et~al.}(2014)\citenamefont
  {MacDowell}, \citenamefont {Benet}, \citenamefont {Katcho},\ and\
  \citenamefont {Palanco}}]{macdowell14}%
  \BibitemOpen
  \bibfield  {author} {\bibinfo {author} {\bibfnamefont {Luis~G.}\ \bibnamefont
  {MacDowell}}, \bibinfo {author} {\bibfnamefont {Jorge}\ \bibnamefont
  {Benet}}, \bibinfo {author} {\bibfnamefont {Nebil~A.}\ \bibnamefont
  {Katcho}}, \ and\ \bibinfo {author} {\bibfnamefont {Jose~M.G.}\ \bibnamefont
  {Palanco}},\ }\bibfield  {title} {\enquote {\bibinfo {title} {Disjoining
  pressure and the film-height-dependent surface tension of thin liquid films:
  New insight from capillary wave fluctuations},}\ }\href {\doibase
  http://dx.doi.org/10.1016/j.cis.2013.11.003} {\bibfield  {journal} {\bibinfo
  {journal} {Adv. Colloid Interface Sci.}\ }\textbf {\bibinfo {volume} {206}},\
  \bibinfo {pages} {150--171} (\bibinfo {year} {2014})}\BibitemShut {NoStop}%
\bibitem [{\citenamefont {Benet}\ \emph
  {et~al.}(2014{\natexlab{b}})\citenamefont {Benet}, \citenamefont {Palanco},
  \citenamefont {Sanz},\ and\ \citenamefont {MacDowell}}]{benet14b}%
  \BibitemOpen
  \bibfield  {author} {\bibinfo {author} {\bibfnamefont {J.}~\bibnamefont
  {Benet}}, \bibinfo {author} {\bibfnamefont {J.~G.}\ \bibnamefont {Palanco}},
  \bibinfo {author} {\bibfnamefont {E.}~\bibnamefont {Sanz}}, \ and\ \bibinfo
  {author} {\bibfnamefont {L.~G.}\ \bibnamefont {MacDowell}},\ }\bibfield
  {title} {\enquote {\bibinfo {title} {Disjoining pressure, healing distance,
  and film height dependent surface tension of thin wetting films},}\
  }\href@noop {} {\bibfield  {journal} {\bibinfo  {journal} {J. Phys. Chem. C}\
  }\textbf {\bibinfo {volume} {118}},\ \bibinfo {pages} {22079--22089}
  (\bibinfo {year} {2014}{\natexlab{b}})}\BibitemShut {NoStop}%
\bibitem [{\citenamefont {Nelson}\ \emph {et~al.}(2004)\citenamefont {Nelson},
  \citenamefont {Piran},\ and\ \citenamefont {Weinberg}}]{nelson04}%
  \BibitemOpen
  \bibfield  {author} {\bibinfo {author} {\bibfnamefont {D.}~\bibnamefont
  {Nelson}}, \bibinfo {author} {\bibfnamefont {T.}~\bibnamefont {Piran}}, \
  and\ \bibinfo {author} {\bibfnamefont {S.}~\bibnamefont {Weinberg}},\
  }\href@noop {} {\emph {\bibinfo {title} {Statistical Mechanics of Membranes
  and Surfaces}}}\ (\bibinfo  {publisher} {Word Scientific, Singapore},\
  \bibinfo {year} {2004})\BibitemShut {NoStop}%
\bibitem [{\citenamefont {Safran}(1994)}]{safran94}%
  \BibitemOpen
  \bibfield  {author} {\bibinfo {author} {\bibfnamefont {Samuel~A.}\
  \bibnamefont {Safran}},\ }\href@noop {} {\emph {\bibinfo {title} {Statistical
  Thermodynamics of Surfaces, Interfaces and Membranes}}},\ \bibinfo {edition}
  {1st}\ ed.\ (\bibinfo  {publisher} {Addison-Wesley},\ \bibinfo {address}
  {Reading},\ \bibinfo {year} {1994})\BibitemShut {NoStop}%
\bibitem [{\citenamefont {Li}\ \emph {et~al.}(2001)\citenamefont {Li},
  \citenamefont {Tikhonov}, \citenamefont {Chaiko},\ and\ \citenamefont
  {Schlossman}}]{li01}%
  \BibitemOpen
  \bibfield  {author} {\bibinfo {author} {\bibfnamefont {Ming}\ \bibnamefont
  {Li}}, \bibinfo {author} {\bibfnamefont {Aleksey~M.}\ \bibnamefont
  {Tikhonov}}, \bibinfo {author} {\bibfnamefont {David~J.}\ \bibnamefont
  {Chaiko}}, \ and\ \bibinfo {author} {\bibfnamefont {Mark~L.}\ \bibnamefont
  {Schlossman}},\ }\bibfield  {title} {\enquote {\bibinfo {title} {Coupled
  capillary wave fluctuations in thin aqueous films on an aqueous subphase},}\
  }\href {\doibase 10.1103/PhysRevLett.86.5934} {\bibfield  {journal} {\bibinfo
   {journal} {Phys. Rev. Lett.}\ }\textbf {\bibinfo {volume} {86}},\ \bibinfo
  {pages} {5934--5937} (\bibinfo {year} {2001})}\BibitemShut {NoStop}%
\bibitem [{\citenamefont {Fukuto}\ \emph {et~al.}(2006)\citenamefont {Fukuto},
  \citenamefont {Gang}, \citenamefont {Alvine},\ and\ \citenamefont
  {Pershan}}]{fukuto06}%
  \BibitemOpen
  \bibfield  {author} {\bibinfo {author} {\bibfnamefont {Masafumi}\
  \bibnamefont {Fukuto}}, \bibinfo {author} {\bibfnamefont {Oleg}\ \bibnamefont
  {Gang}}, \bibinfo {author} {\bibfnamefont {Kyle~J.}\ \bibnamefont {Alvine}},
  \ and\ \bibinfo {author} {\bibfnamefont {Peter~S.}\ \bibnamefont {Pershan}},\
  }\bibfield  {title} {\enquote {\bibinfo {title} {Capillary wave fluctuations
  and intrinsic widths of coupled fluid-fluid interfaces: An x-ray scattering
  study of a wetting film on bulk liquid},}\ }\href {\doibase
  10.1103/PhysRevE.74.031607} {\bibfield  {journal} {\bibinfo  {journal} {Phys.
  Rev. E}\ }\textbf {\bibinfo {volume} {74}},\ \bibinfo {pages} {031607}
  (\bibinfo {year} {2006})}\BibitemShut {NoStop}%
\bibitem [{\citenamefont {Pershan}\ and\ \citenamefont
  {Schlossman}(2012)}]{pershan12}%
  \BibitemOpen
  \bibfield  {author} {\bibinfo {author} {\bibfnamefont {P.~S.}\ \bibnamefont
  {Pershan}}\ and\ \bibinfo {author} {\bibfnamefont {M.}~\bibnamefont
  {Schlossman}},\ }\href@noop {} {\emph {\bibinfo {title} {Liquid Surfaces and
  Interfaces: Synchrotron {X-ray} Methods}}}\ (\bibinfo  {publisher} {Cambridge
  University Press},\ \bibinfo {address} {Cambridge},\ \bibinfo {year} {2012})\
  pp.\ \bibinfo {pages} {1--311}\BibitemShut {NoStop}%
\bibitem [{\citenamefont {Lechner}\ and\ \citenamefont
  {Dellago}(2008)}]{lechner08}%
  \BibitemOpen
  \bibfield  {author} {\bibinfo {author} {\bibfnamefont {Wolfgang}\
  \bibnamefont {Lechner}}\ and\ \bibinfo {author} {\bibfnamefont {Christoph}\
  \bibnamefont {Dellago}},\ }\bibfield  {title} {\enquote {\bibinfo {title}
  {Accurate determination of crystal structures based on averaged local bond
  order parameters},}\ }\href {\doibase 10.1063/1.2977970} {\bibfield
  {journal} {\bibinfo  {journal} {J. Chem. Phys.}\ }\textbf {\bibinfo {volume}
  {129}},\ \bibinfo {eid} {114707} (\bibinfo {year} {2008})}\BibitemShut
  {NoStop}%
\bibitem [{\citenamefont {Benet}\ \emph
  {et~al.}(2014{\natexlab{c}})\citenamefont {Benet}, \citenamefont
  {MacDowell},\ and\ \citenamefont {Sanz}}]{benet14}%
  \BibitemOpen
  \bibfield  {author} {\bibinfo {author} {\bibfnamefont {J.}~\bibnamefont
  {Benet}}, \bibinfo {author} {\bibfnamefont {L.~G.}\ \bibnamefont
  {MacDowell}}, \ and\ \bibinfo {author} {\bibfnamefont {E.}~\bibnamefont
  {Sanz}},\ }\bibfield  {title} {\enquote {\bibinfo {title} {Computer
  simulation study of surface wave dynamics at the crystal--melt interface},}\
  }\href@noop {} {\bibfield  {journal} {\bibinfo  {journal} {J. Chem. Phys.}\
  }\textbf {\bibinfo {volume} {141}},\ \bibinfo {pages} {034701} (\bibinfo
  {year} {2014}{\natexlab{c}})}\BibitemShut {NoStop}%
\bibitem [{\citenamefont {Rozmanov}\ and\ \citenamefont
  {Kusalik}(2011)}]{rozmanov11}%
  \BibitemOpen
  \bibfield  {author} {\bibinfo {author} {\bibfnamefont {Dmitri}\ \bibnamefont
  {Rozmanov}}\ and\ \bibinfo {author} {\bibfnamefont {Peter~G.}\ \bibnamefont
  {Kusalik}},\ }\bibfield  {title} {\enquote {\bibinfo {title} {Temperature
  dependence of crystal growth of hexagonal ice (ih)},}\ }\href {\doibase
  10.1039/C1CP21210A} {\bibfield  {journal} {\bibinfo  {journal} {Phys. Chem.
  Chem. Phys}\ }\textbf {\bibinfo {volume} {13}},\ \bibinfo {pages}
  {15501--15511} (\bibinfo {year} {2011})}\BibitemShut {NoStop}%
\bibitem [{\citenamefont {Conde}\ \emph {et~al.}(2017)\citenamefont {Conde},
  \citenamefont {Rovere},\ and\ \citenamefont {Gallo}}]{conde17}%
  \BibitemOpen
  \bibfield  {author} {\bibinfo {author} {\bibfnamefont {M.~M.}\ \bibnamefont
  {Conde}}, \bibinfo {author} {\bibfnamefont {M.}~\bibnamefont {Rovere}}, \
  and\ \bibinfo {author} {\bibfnamefont {P.}~\bibnamefont {Gallo}},\ }\bibfield
   {title} {\enquote {\bibinfo {title} {High precision determination of the
  melting points of water tip4p/2005 and water tip4p/ice models by the direct
  coexistence technique},}\ }\href {\doibase 10.1063/1.5008478} {\bibfield
  {journal} {\bibinfo  {journal} {The Journal of Chemical Physics}\ }\textbf
  {\bibinfo {volume} {147}},\ \bibinfo {pages} {244506} (\bibinfo {year}
  {2017})},\ \Eprint {http://arxiv.org/abs/https://doi.org/10.1063/1.5008478}
  {https://doi.org/10.1063/1.5008478} \BibitemShut {NoStop}%
\bibitem [{\citenamefont {Bussi}\ \emph {et~al.}(2007)\citenamefont {Bussi},
  \citenamefont {Donadio},\ and\ \citenamefont {Parrinello}}]{bussi07}%
  \BibitemOpen
  \bibfield  {author} {\bibinfo {author} {\bibfnamefont {Giovanni}\
  \bibnamefont {Bussi}}, \bibinfo {author} {\bibfnamefont {Davide}\
  \bibnamefont {Donadio}}, \ and\ \bibinfo {author} {\bibfnamefont {Michele}\
  \bibnamefont {Parrinello}},\ }\bibfield  {title} {\enquote {\bibinfo {title}
  {Canonical sampling through velocity rescaling},}\ }\href {\doibase
  http://dx.doi.org/10.1063/1.2408420} {\bibfield  {journal} {\bibinfo
  {journal} {J. Chem. Phys.}\ }\textbf {\bibinfo {volume} {126}},\ \bibinfo
  {eid} {014101} (\bibinfo {year} {2007})}\BibitemShut {NoStop}%
\bibitem [{\citenamefont {Mastny}\ and\ \citenamefont
  {de~Pablo}(2007)}]{mastny07}%
  \BibitemOpen
  \bibfield  {author} {\bibinfo {author} {\bibfnamefont {Ethan~A.}\
  \bibnamefont {Mastny}}\ and\ \bibinfo {author} {\bibfnamefont {Juan~J.}\
  \bibnamefont {de~Pablo}},\ }\bibfield  {title} {\enquote {\bibinfo {title}
  {Melting line of the lennard-jones system, infinite size, and full poten
  tial},}\ }\href {\doibase 10.1063/1.2753149} {\bibfield  {journal} {\bibinfo
  {journal} {J. Chem. Phys.}\ }\textbf {\bibinfo {volume} {127}},\ \bibinfo
  {pages} {104504} (\bibinfo {year} {2007})}\BibitemShut {NoStop}%
\bibitem [{\citenamefont {Libbrecht}(2014)}]{libbrecht14}%
  \BibitemOpen
  \bibfield  {author} {\bibinfo {author} {\bibfnamefont {K.~G.}\ \bibnamefont
  {Libbrecht}},\ }\bibfield  {title} {\enquote {\bibinfo {title} {Towards a
  comprehensive model of snow crystal growth: 3. the correspondence between ice
  growth from water vapor and ice growth from liquid water},}\ }\href@noop {}
  {\bibfield  {journal} {\bibinfo  {journal} {arXiv:1407.0740
  [cond-mat.mtrl-sci]}\ } (\bibinfo {year} {2014})}\BibitemShut {NoStop}%
\bibitem [{\citenamefont {Murata}\ \emph {et~al.}(2018)\citenamefont {Murata},
  \citenamefont {Nagashima},\ and\ \citenamefont {Sazaki}}]{murata18}%
  \BibitemOpen
  \bibfield  {author} {\bibinfo {author} {\bibfnamefont {Ken-ichiro}\
  \bibnamefont {Murata}}, \bibinfo {author} {\bibfnamefont {Ken}\ \bibnamefont
  {Nagashima}}, \ and\ \bibinfo {author} {\bibfnamefont {Gen}\ \bibnamefont
  {Sazaki}},\ }\bibfield  {title} {\enquote {\bibinfo {title} {In situ
  observations of spiral growth on ice crystal surfaces},}\ }\href {\doibase
  10.1103/PhysRevMaterials.2.093402} {\bibfield  {journal} {\bibinfo  {journal}
  {Phys. Rev. Materials}\ }\textbf {\bibinfo {volume} {2}},\ \bibinfo {pages}
  {093402} (\bibinfo {year} {2018})}\BibitemShut {NoStop}%
\bibitem [{\citenamefont {Inomata}\ \emph {et~al.}(2018)\citenamefont
  {Inomata}, \citenamefont {Murata}, \citenamefont {Asakawa}, \citenamefont
  {Nagashima}, \citenamefont {Nakatsubo}, \citenamefont {Furukawa},\ and\
  \citenamefont {Sazaki}}]{inomata17}%
  \BibitemOpen
  \bibfield  {author} {\bibinfo {author} {\bibfnamefont {Masahiro}\
  \bibnamefont {Inomata}}, \bibinfo {author} {\bibfnamefont {Ken-ichiro}\
  \bibnamefont {Murata}}, \bibinfo {author} {\bibfnamefont {Harutoshi}\
  \bibnamefont {Asakawa}}, \bibinfo {author} {\bibfnamefont {Ken}\ \bibnamefont
  {Nagashima}}, \bibinfo {author} {\bibfnamefont {Shunichi}\ \bibnamefont
  {Nakatsubo}}, \bibinfo {author} {\bibfnamefont {Yoshinori}\ \bibnamefont
  {Furukawa}}, \ and\ \bibinfo {author} {\bibfnamefont {Gen}\ \bibnamefont
  {Sazaki}},\ }\bibfield  {title} {\enquote {\bibinfo {title} {Temperature
  dependence of the growth kinetics of elementary spiral steps on ice basal
  faces grown from water vapor},}\ }\href {\doibase 10.1021/acs.cgd.7b01251}
  {\bibfield  {journal} {\bibinfo  {journal} {Cryst. Growth Des.}\ }\textbf
  {\bibinfo {volume} {18}},\ \bibinfo {pages} {786--793} (\bibinfo {year}
  {2018})},\ \Eprint
  {http://arxiv.org/abs/https://doi.org/10.1021/acs.cgd.7b01251}
  {https://doi.org/10.1021/acs.cgd.7b01251} \BibitemShut {NoStop}%
\bibitem [{\citenamefont {Murata}\ \emph {et~al.}(2019)\citenamefont {Murata},
  \citenamefont {Nagashima},\ and\ \citenamefont {Sazaki}}]{murata19}%
  \BibitemOpen
  \bibfield  {author} {\bibinfo {author} {\bibfnamefont {Ken-ichiro}\
  \bibnamefont {Murata}}, \bibinfo {author} {\bibfnamefont {Ken}\ \bibnamefont
  {Nagashima}}, \ and\ \bibinfo {author} {\bibfnamefont {Gen}\ \bibnamefont
  {Sazaki}},\ }\bibfield  {title} {\enquote {\bibinfo {title} {How do ice
  crystals grow inside quasiliquid layers?}}\ }\href {\doibase
  10.1103/PhysRevLett.122.026102} {\bibfield  {journal} {\bibinfo  {journal}
  {Phys. Rev. Lett.}\ }\textbf {\bibinfo {volume} {122}},\ \bibinfo {pages}
  {026102} (\bibinfo {year} {2019})}\BibitemShut {NoStop}%
\bibitem [{\citenamefont {Beckmann}\ and\ \citenamefont
  {Lacmann}(1982)}]{beckmann82}%
  \BibitemOpen
  \bibfield  {author} {\bibinfo {author} {\bibfnamefont {W.}~\bibnamefont
  {Beckmann}}\ and\ \bibinfo {author} {\bibfnamefont {R.}~\bibnamefont
  {Lacmann}},\ }\bibfield  {title} {\enquote {\bibinfo {title} {Interface
  kinetics of the growth and evaporation of ice single crystals from the vapour
  phase: Ii. measurements in a pure water vapour environment},}\ }\href
  {\doibase http://dx.doi.org/10.1016/0022-0248(82)90292-5} {\bibfield
  {journal} {\bibinfo  {journal} {J. Cryst. Growth}\ }\textbf {\bibinfo
  {volume} {58}},\ \bibinfo {pages} {433--442} (\bibinfo {year}
  {1982})}\BibitemShut {NoStop}%
\bibitem [{\citenamefont {Meunier}(1987)}]{meunier87}%
  \BibitemOpen
  \bibfield  {author} {\bibinfo {author} {\bibfnamefont {J.}~\bibnamefont
  {Meunier}},\ }\bibfield  {title} {\enquote {\bibinfo {title} {Liquid
  interfaces: Role of the fluctuations and analysis of ellipsometry and
  reflectivity measurements},}\ }\href@noop {} {\bibfield  {journal} {\bibinfo
  {journal} {J. Phys.(Paris)}\ }\textbf {\bibinfo {volume} {48}},\ \bibinfo
  {pages} {1819--1831} (\bibinfo {year} {1987})}\BibitemShut {NoStop}%
\bibitem [{\citenamefont {Mecke}\ and\ \citenamefont
  {Dietrich}(1999)}]{mecke99b}%
  \BibitemOpen
  \bibfield  {author} {\bibinfo {author} {\bibfnamefont {K.~R.}\ \bibnamefont
  {Mecke}}\ and\ \bibinfo {author} {\bibfnamefont {S.}~\bibnamefont
  {Dietrich}},\ }\bibfield  {title} {\enquote {\bibinfo {title} {Effective
  hamiltonian for liquid-vapor interfaces},}\ }\href {\doibase
  10.1103/PhysRevE.59.6766} {\bibfield  {journal} {\bibinfo  {journal} {Phys.
  Rev. E}\ }\textbf {\bibinfo {volume} {59}},\ \bibinfo {pages} {6766--6784}
  (\bibinfo {year} {1999})}\BibitemShut {NoStop}%
\bibitem [{\citenamefont {Blokhuis}(2009)}]{blokhuis09}%
  \BibitemOpen
  \bibfield  {author} {\bibinfo {author} {\bibfnamefont {Edgar~M.}\
  \bibnamefont {Blokhuis}},\ }\bibfield  {title} {\enquote {\bibinfo {title}
  {On the spectrum of fluctuations of a liquid surface: From the molecular
  scale to the macroscopic scale},}\ }\href@noop {} {\bibfield  {journal}
  {\bibinfo  {journal} {J. Chem. Phys.}\ }\textbf {\bibinfo {volume} {130}},\
  \bibinfo {eid} {014706} (\bibinfo {year} {2009})}\BibitemShut {NoStop}%
\bibitem [{\citenamefont {Espinosa}\ \emph {et~al.}(2016)\citenamefont
  {Espinosa}, \citenamefont {Vega},\ and\ \citenamefont {Sanz}}]{espinosa16}%
  \BibitemOpen
  \bibfield  {author} {\bibinfo {author} {\bibfnamefont {Jorge~R.}\
  \bibnamefont {Espinosa}}, \bibinfo {author} {\bibfnamefont {Carlos}\
  \bibnamefont {Vega}}, \ and\ \bibinfo {author} {\bibfnamefont {Eduardo}\
  \bibnamefont {Sanz}},\ }\bibfield  {title} {\enquote {\bibinfo {title}
  {Ice-water interfacial free energy for the tip4p, tip4p/2005, tip4p/ice, and
  mw models as obtained from the mold integration technique},}\ }\href
  {\doibase 10.1021/acs.jpcc.5b11221} {\bibfield  {journal} {\bibinfo
  {journal} {J. Phys. Chem. C}\ }\textbf {\bibinfo {volume} {120}},\ \bibinfo
  {pages} {8068--8075} (\bibinfo {year} {2016})},\ \Eprint
  {http://arxiv.org/abs/https://doi.org/10.1021/acs.jpcc.5b11221}
  {https://doi.org/10.1021/acs.jpcc.5b11221} \BibitemShut {NoStop}%
\bibitem [{\citenamefont {MacDowell}\ \emph {et~al.}(2013)\citenamefont
  {MacDowell}, \citenamefont {Benet},\ and\ \citenamefont
  {Katcho}}]{macdowell13}%
  \BibitemOpen
  \bibfield  {author} {\bibinfo {author} {\bibfnamefont {Luis~G.}\ \bibnamefont
  {MacDowell}}, \bibinfo {author} {\bibfnamefont {Jorge}\ \bibnamefont
  {Benet}}, \ and\ \bibinfo {author} {\bibfnamefont {Nebil~A.}\ \bibnamefont
  {Katcho}},\ }\bibfield  {title} {\enquote {\bibinfo {title} {Capillary
  fluctuations and film-height-dependent surface tension of an adsorbed liquid
  film},}\ }\href {\doibase 10.1103/PhysRevLett.111.047802} {\bibfield
  {journal} {\bibinfo  {journal} {Phys. Rev. Lett.}\ }\textbf {\bibinfo
  {volume} {111}},\ \bibinfo {pages} {047802} (\bibinfo {year}
  {2013})}\BibitemShut {NoStop}%
\bibitem [{\citenamefont {MacDowell}\ \emph {et~al.}(2018)\citenamefont
  {MacDowell}, \citenamefont {Llombart}, \citenamefont {Benet}, \citenamefont
  {Palanco},\ and\ \citenamefont {Guerrero-Martinez}}]{macdowell18}%
  \BibitemOpen
  \bibfield  {author} {\bibinfo {author} {\bibfnamefont {Luis~G.}\ \bibnamefont
  {MacDowell}}, \bibinfo {author} {\bibfnamefont {Pablo}\ \bibnamefont
  {Llombart}}, \bibinfo {author} {\bibfnamefont {Jorge}\ \bibnamefont {Benet}},
  \bibinfo {author} {\bibfnamefont {Jose~G.}\ \bibnamefont {Palanco}}, \ and\
  \bibinfo {author} {\bibfnamefont {Andrés}\ \bibnamefont
  {Guerrero-Martinez}},\ }\bibfield  {title} {\enquote {\bibinfo {title}
  {Nanocapillarity and liquid bridge-mediated force between colloidal
  nanoparticles},}\ }\href {\doibase 10.1021/acsomega.7b01650} {\bibfield
  {journal} {\bibinfo  {journal} {ACS Omega}\ }\textbf {\bibinfo {volume}
  {3}},\ \bibinfo {pages} {112--123} (\bibinfo {year} {2018})},\ \Eprint
  {http://arxiv.org/abs/http://dx.doi.org/10.1021/acsomega.7b01650}
  {http://dx.doi.org/10.1021/acsomega.7b01650} \BibitemShut {NoStop}%
\bibitem [{\citenamefont {de~Gennes}\ \emph {et~al.}(2004)\citenamefont
  {de~Gennes}, \citenamefont {Brochard-Wyart},\ and\ \citenamefont
  {Qu{\'e}r{\'e}}}]{degennes04}%
  \BibitemOpen
  \bibfield  {author} {\bibinfo {author} {\bibfnamefont {P.~G.}\ \bibnamefont
  {de~Gennes}}, \bibinfo {author} {\bibfnamefont {F.}~\bibnamefont
  {Brochard-Wyart}}, \ and\ \bibinfo {author} {\bibfnamefont {D.}~\bibnamefont
  {Qu{\'e}r{\'e}}},\ }\href@noop {} {\emph {\bibinfo {title} {Capillarity and
  Wetting Phenomena}}}\ (\bibinfo  {publisher} {Springer},\ \bibinfo {address}
  {New York},\ \bibinfo {year} {2004})\ pp.\ \bibinfo {pages}
  {1--292}\BibitemShut {NoStop}%
\bibitem [{\citenamefont {Nozi\'eres}\ and\ \citenamefont
  {Gallet}(1987)}]{nozieres87}%
  \BibitemOpen
  \bibfield  {author} {\bibinfo {author} {\bibfnamefont {P.}~\bibnamefont
  {Nozi\'eres}}\ and\ \bibinfo {author} {\bibfnamefont {F.}~\bibnamefont
  {Gallet}},\ }\bibfield  {title} {\enquote {\bibinfo {title} {The roughening
  transition of crystal surfaces. i. static and dynamic renormalization theory,
  crystal shape and facet grwoth},}\ }\href@noop {} {\bibfield  {journal}
  {\bibinfo  {journal} {J. Phys.(Paris)}\ }\textbf {\bibinfo {volume} {48}},\
  \bibinfo {pages} {353--367} (\bibinfo {year} {1987})}\BibitemShut {NoStop}%
\bibitem [{\citenamefont {Libbrecht}\ and\ \citenamefont
  {Rickerby}(2013)}]{libbrecht12}%
  \BibitemOpen
  \bibfield  {author} {\bibinfo {author} {\bibfnamefont {K.~G.}\ \bibnamefont
  {Libbrecht}}\ and\ \bibinfo {author} {\bibfnamefont {M.~E.}\ \bibnamefont
  {Rickerby}},\ }\bibfield  {title} {\enquote {\bibinfo {title} {Meassurements
  of surface attachment kinetics for faceted ice crystal growth},}\ }\href@noop
  {} {\bibfield  {journal} {\bibinfo  {journal} {J. Cryst. Growth}\ }\textbf
  {\bibinfo {volume} {377}},\ \bibinfo {pages} {1--8} (\bibinfo {year}
  {2013})}\BibitemShut {NoStop}%
\bibitem [{\citenamefont {Dzyaloshinskii}\ \emph {et~al.}(1961)\citenamefont
  {Dzyaloshinskii}, \citenamefont {Lifshitz},\ and\ \citenamefont
  {Pitaevskii}}]{dzyaloshinskii61}%
  \BibitemOpen
  \bibfield  {author} {\bibinfo {author} {\bibfnamefont {I.~E.}\ \bibnamefont
  {Dzyaloshinskii}}, \bibinfo {author} {\bibfnamefont {E.~M.}\ \bibnamefont
  {Lifshitz}}, \ and\ \bibinfo {author} {\bibfnamefont {Lev.~P.}\ \bibnamefont
  {Pitaevskii}},\ }\bibfield  {title} {\enquote {\bibinfo {title} {General
  theory of van der waals forces},}\ }\href
  {http://stacks.iop.org/0038-5670/4/i=2/a=R01} {\bibfield  {journal} {\bibinfo
   {journal} {Soviet Physics Uspekhi}\ }\textbf {\bibinfo {volume} {4}},\
  \bibinfo {pages} {153.175} (\bibinfo {year} {1961})}\BibitemShut {NoStop}%
\bibitem [{\citenamefont {Parsegian}(2006)}]{parsegian06}%
  \BibitemOpen
  \bibfield  {author} {\bibinfo {author} {\bibfnamefont {V.~A.}\ \bibnamefont
  {Parsegian}},\ }\href@noop {} {\emph {\bibinfo {title} {Van der Waals
  Forces}}}\ (\bibinfo  {publisher} {Cambridge University Press},\ \bibinfo
  {address} {Cambridge},\ \bibinfo {year} {2006})\ pp.\ \bibinfo {pages}
  {1--311}\BibitemShut {NoStop}%
\bibitem [{\citenamefont {Israelachvili}(1991)}]{israelachvili91}%
  \BibitemOpen
  \bibfield  {author} {\bibinfo {author} {\bibfnamefont {J.~N.}\ \bibnamefont
  {Israelachvili}},\ }\href@noop {} {\emph {\bibinfo {title} {Intermolecular
  and Surfaces Forces}}},\ \bibinfo {edition} {2nd}\ ed.\ (\bibinfo
  {publisher} {Academic Press},\ \bibinfo {address} {London},\ \bibinfo {year}
  {1991})\BibitemShut {NoStop}%
\bibitem [{\citenamefont {Limmer}(2016)}]{limmer16}%
  \BibitemOpen
  \bibfield  {author} {\bibinfo {author} {\bibfnamefont {David~T.}\
  \bibnamefont {Limmer}},\ }\bibfield  {title} {\enquote {\bibinfo {title}
  {Closer look at the surface of ice},}\ }\href@noop {} {\bibfield  {journal}
  {\bibinfo  {journal} {Proc. Natl. Acad. Sci. U.S.A.}\ }\textbf {\bibinfo
  {volume} {113}},\ \bibinfo {pages} {12347--12349} (\bibinfo {year}
  {2016})}\BibitemShut {NoStop}%
\bibitem [{\citenamefont {Wilen}\ \emph {et~al.}(1995)\citenamefont {Wilen},
  \citenamefont {Wettlaufer}, \citenamefont {Elbaum},\ and\ \citenamefont
  {Schick}}]{wilen95}%
  \BibitemOpen
  \bibfield  {author} {\bibinfo {author} {\bibfnamefont {L.~A.}\ \bibnamefont
  {Wilen}}, \bibinfo {author} {\bibfnamefont {J.~S.}\ \bibnamefont
  {Wettlaufer}}, \bibinfo {author} {\bibfnamefont {M.}~\bibnamefont {Elbaum}},
  \ and\ \bibinfo {author} {\bibfnamefont {M.}~\bibnamefont {Schick}},\
  }\bibfield  {title} {\enquote {\bibinfo {title} {Dispersion-force effects in
  interfacial premelting of ice},}\ }\href {\doibase 10.1103/PhysRevB.52.12426}
  {\bibfield  {journal} {\bibinfo  {journal} {Phys. Rev. B}\ }\textbf {\bibinfo
  {volume} {52}},\ \bibinfo {pages} {12426--12433} (\bibinfo {year}
  {1995})}\BibitemShut {NoStop}%
\bibitem [{\citenamefont {Lee}\ and\ \citenamefont {Rick}(2012)}]{lee12}%
  \BibitemOpen
  \bibfield  {author} {\bibinfo {author} {\bibfnamefont {Alexis~J.}\
  \bibnamefont {Lee}}\ and\ \bibinfo {author} {\bibfnamefont {Steven~W.}\
  \bibnamefont {Rick}},\ }\bibfield  {title} {\enquote {\bibinfo {title}
  {Characterizing charge transfer at water ice interfaces},}\ }\href {\doibase
  10.1021/jz301411q} {\bibfield  {journal} {\bibinfo  {journal} {J. Phys. Chem.
  Lett.}\ }\textbf {\bibinfo {volume} {3}},\ \bibinfo {pages} {3199--3203}
  (\bibinfo {year} {2012})},\ \Eprint
  {http://arxiv.org/abs/http://dx.doi.org/10.1021/jz301411q}
  {http://dx.doi.org/10.1021/jz301411q} \BibitemShut {NoStop}%
\bibitem [{\citenamefont {Rick}\ and\ \citenamefont {Haymet}(2003)}]{rick03}%
  \BibitemOpen
  \bibfield  {author} {\bibinfo {author} {\bibfnamefont {S.~W.}\ \bibnamefont
  {Rick}}\ and\ \bibinfo {author} {\bibfnamefont {A.~D.~J.}\ \bibnamefont
  {Haymet}},\ }\bibfield  {title} {\enquote {\bibinfo {title} {Dielectric
  constant and proton order and disorder in ice {Ih}: {M}onte {C}arlo computer
  simulations},}\ }\href@noop {} {\bibfield  {journal} {\bibinfo  {journal} {J.
  Chem. Phys.}\ }\textbf {\bibinfo {volume} {118}},\ \bibinfo {pages}
  {9291--9296} (\bibinfo {year} {2003})}\BibitemShut {NoStop}%
\bibitem [{\citenamefont {Rick}(2005)}]{rick05}%
  \BibitemOpen
  \bibfield  {author} {\bibinfo {author} {\bibfnamefont {S.~W.}\ \bibnamefont
  {Rick}},\ }\href@noop {} {\bibfield  {journal} {\bibinfo  {journal} {J. Chem.
  Phys.}\ }\textbf {\bibinfo {volume} {122}},\ \bibinfo {pages} {094504}
  (\bibinfo {year} {2005})}\BibitemShut {NoStop}%
\bibitem [{\citenamefont {Aragones}\ \emph {et~al.}(2011)\citenamefont
  {Aragones}, \citenamefont {MacDowell}, \citenamefont {Siepmann},\ and\
  \citenamefont {Vega}}]{aragones11}%
  \BibitemOpen
  \bibfield  {author} {\bibinfo {author} {\bibfnamefont {J.~L.}\ \bibnamefont
  {Aragones}}, \bibinfo {author} {\bibfnamefont {L.~G.}\ \bibnamefont
  {MacDowell}}, \bibinfo {author} {\bibfnamefont {J.~I.}\ \bibnamefont
  {Siepmann}}, \ and\ \bibinfo {author} {\bibfnamefont {C.}~\bibnamefont
  {Vega}},\ }\bibfield  {title} {\enquote {\bibinfo {title} {Phase diagram of
  water under an applied electric field},}\ }\href {\doibase
  10.1103/PhysRevLett.107.155702} {\bibfield  {journal} {\bibinfo  {journal}
  {Phys. Rev. Lett.}\ }\textbf {\bibinfo {volume} {107}},\ \bibinfo {pages}
  {155702} (\bibinfo {year} {2011})}\BibitemShut {NoStop}%
\bibitem [{\citenamefont {Dietrich}\ and\ \citenamefont
  {Schick}(1986)}]{dietrich86}%
  \BibitemOpen
  \bibfield  {author} {\bibinfo {author} {\bibfnamefont {S.}~\bibnamefont
  {Dietrich}}\ and\ \bibinfo {author} {\bibfnamefont {M.}~\bibnamefont
  {Schick}},\ }\bibfield  {title} {\enquote {\bibinfo {title} {Order of wetting
  transitions},}\ }\href {\doibase 10.1103/PhysRevB.33.4952} {\bibfield
  {journal} {\bibinfo  {journal} {Phys. Rev. B}\ }\textbf {\bibinfo {volume}
  {33}},\ \bibinfo {pages} {4952--4968} (\bibinfo {year} {1986})}\BibitemShut
  {NoStop}%
\bibitem [{\citenamefont {Dietrich}\ and\ \citenamefont
  {Napi\'orkowski}(1991)}]{dietrich91}%
  \BibitemOpen
  \bibfield  {author} {\bibinfo {author} {\bibfnamefont {S.}~\bibnamefont
  {Dietrich}}\ and\ \bibinfo {author} {\bibfnamefont {M.}~\bibnamefont
  {Napi\'orkowski}},\ }\bibfield  {title} {\enquote {\bibinfo {title} {Analytic
  results for wetting transitions in the presence of van der waals tails},}\
  }\href {\doibase 10.1103/PhysRevA.43.1861} {\bibfield  {journal} {\bibinfo
  {journal} {Phys. Rev. A}\ }\textbf {\bibinfo {volume} {43}},\ \bibinfo
  {pages} {1861--1885} (\bibinfo {year} {1991})}\BibitemShut {NoStop}%
\bibitem [{\citenamefont {Shahidzadeh}\ \emph {et~al.}(1998)\citenamefont
  {Shahidzadeh}, \citenamefont {Bonn}, \citenamefont {Ragil}, \citenamefont
  {Broseta},\ and\ \citenamefont {Meunier}}]{shahidzadeh98}%
  \BibitemOpen
  \bibfield  {author} {\bibinfo {author} {\bibfnamefont {N.}~\bibnamefont
  {Shahidzadeh}}, \bibinfo {author} {\bibfnamefont {D.}~\bibnamefont {Bonn}},
  \bibinfo {author} {\bibfnamefont {K.}~\bibnamefont {Ragil}}, \bibinfo
  {author} {\bibfnamefont {D.}~\bibnamefont {Broseta}}, \ and\ \bibinfo
  {author} {\bibfnamefont {J.}~\bibnamefont {Meunier}},\ }\bibfield  {title}
  {\enquote {\bibinfo {title} {Sequence of two wetting transitions induced by
  tuning the hamaker constant},}\ }\href@noop {} {\bibfield  {journal}
  {\bibinfo  {journal} {Phys. Rev. Lett.}\ }\textbf {\bibinfo {volume} {80}},\
  \bibinfo {pages} {3992--3995} (\bibinfo {year} {1998})}\BibitemShut {NoStop}%
\bibitem [{\citenamefont {Bertrand}\ \emph {et~al.}(2000)\citenamefont
  {Bertrand}, \citenamefont {Dobbs}, \citenamefont {Broseta}, \citenamefont
  {Indekeu}, \citenamefont {Bonn},\ and\ \citenamefont {Meunier}}]{bertrand00}%
  \BibitemOpen
  \bibfield  {author} {\bibinfo {author} {\bibfnamefont {E.}~\bibnamefont
  {Bertrand}}, \bibinfo {author} {\bibfnamefont {H.}~\bibnamefont {Dobbs}},
  \bibinfo {author} {\bibfnamefont {D.}~\bibnamefont {Broseta}}, \bibinfo
  {author} {\bibfnamefont {J.}~\bibnamefont {Indekeu}}, \bibinfo {author}
  {\bibfnamefont {D.}~\bibnamefont {Bonn}}, \ and\ \bibinfo {author}
  {\bibfnamefont {J.}~\bibnamefont {Meunier}},\ }\bibfield  {title} {\enquote
  {\bibinfo {title} {First--order and critical wetting of alkanes on water},}\
  }\href@noop {} {\bibfield  {journal} {\bibinfo  {journal} {Phys. Rev. Lett.}\
  }\textbf {\bibinfo {volume} {85}},\ \bibinfo {pages} {1282--1285} (\bibinfo
  {year} {2000})}\BibitemShut {NoStop}%
\bibitem [{\citenamefont {M{\"u}ller}\ \emph {et~al.}(2001)\citenamefont
  {M{\"u}ller}, \citenamefont {MacDowell}, \citenamefont
  {M{\"u}ller-Buschbaum}, \citenamefont {Wunnike},\ and\ \citenamefont
  {Stamm}}]{mueller01}%
  \BibitemOpen
  \bibfield  {author} {\bibinfo {author} {\bibfnamefont {M.}~\bibnamefont
  {M{\"u}ller}}, \bibinfo {author} {\bibfnamefont {L.~G.}\ \bibnamefont
  {MacDowell}}, \bibinfo {author} {\bibfnamefont {P.}~\bibnamefont
  {M{\"u}ller-Buschbaum}}, \bibinfo {author} {\bibfnamefont {O.}~\bibnamefont
  {Wunnike}}, \ and\ \bibinfo {author} {\bibfnamefont {M.}~\bibnamefont
  {Stamm}},\ }\bibfield  {title} {\enquote {\bibinfo {title} {Nano-dewtting:
  Interplay between van-der-{W}aals and short range interactions},}\
  }\href@noop {} {\bibfield  {journal} {\bibinfo  {journal} {J. Chem. Phys.}\
  }\textbf {\bibinfo {volume} {115}},\ \bibinfo {pages} {9960--9969} (\bibinfo
  {year} {2001})}\BibitemShut {NoStop}%
\bibitem [{\citenamefont {MacDowell}\ and\ \citenamefont
  {M{\"u}ller}(2005)}]{macdowell05}%
  \BibitemOpen
  \bibfield  {author} {\bibinfo {author} {\bibfnamefont {L.~G.}\ \bibnamefont
  {MacDowell}}\ and\ \bibinfo {author} {\bibfnamefont {M.}~\bibnamefont
  {M{\"u}ller}},\ }\bibfield  {title} {\enquote {\bibinfo {title} {Observation
  of autophobic dewetting on polymer brushes from computer simulation},}\
  }\href@noop {} {\bibfield  {journal} {\bibinfo  {journal} {J. Phys.: Condens.
  Matter}\ }\textbf {\bibinfo {volume} {17}},\ \bibinfo {pages} {S3523--S3528}
  (\bibinfo {year} {2005})}\BibitemShut {NoStop}%
\end{thebibliography}%

\end{document}